\documentclass[preprint]{elsarticle}
\usepackage{appendix}
\usepackage{amssymb}
\usepackage{amsmath}
\usepackage[final]{epsfig}
\usepackage{color}
\usepackage{graphicx}
\usepackage{lineno}
\usepackage{bm}
\usepackage{comment}
\usepackage{braket}
\usepackage{dsfont}
\usepackage{placeins} 
\usepackage{slashed}

\newcommand{\beq}{\begin{equation}}
\newcommand{\eeq}{\end{equation}}
\newcommand{\beqs}{\begin{equation*}}
\newcommand{\eeqs}{\end{equation*}}
\newcommand{\bdi}{\begin{displaymath}}
\newcommand{\edi}{\end{displaymath}}

\newcommand{\bea}{\begin{eqnarray}}
\newcommand{\eea}{\end{eqnarray}}
\newcommand{\beas}{\begin{eqnarray*}}
\newcommand{\eeas}{\end{eqnarray*}}

\newcommand{\crg}{\hat{a}^{\dagger}}
\newcommand{\cre}{\hat{a}_e^{\dagger}}
\newcommand{\crh}{\hat{a}_h^{\dagger}}
\newcommand{\ang}{\hat{a}}
\newcommand{\ane}{\hat{a}_e}
\newcommand{\anh}{\hat{a}_h}

\newcommand{\Numg}{\hat{N}}
\newcommand{\Nume}{\hat{N}_e}
\newcommand{\Numh}{\hat{N}_h}

\newcommand{\nume}{\hat{n}_e}
\newcommand{\numh}{\hat{n}_h}

\newcommand{\transe}{\hat{T}_e}
\newcommand{\transh}{\hat{T}_h}

\newcommand{\opg}{\hat{\mathcal{O}}}

\newcommand{\reg}{\mathcal{R}}
\newcommand{\indicator}{\mathbf{1}}

\newcommand{\avgbra}{\bra{\mathrm{avg}}}
\newcommand{\avgbraket}[1]{\braket{\mathrm{avg}|#1}}
\newcommand{\avg}[1]{\left\langle #1 \right\rangle}

\newcommand{\cov}{\mathrm{cov}}

\newcommand{\tret}{t_{\mathrm{ret}}}

\newcommand{\Iind}{I^{\mathrm{ind}}}

\newcommand{\uhat}{\hat{U}}
\newcommand{\ham}{\hat{H}}

\newcommand{\ttrans}{T_{\mathrm{transit}}}

\newcommand{\Nth}{N_{\mathrm{th}}}
\newcommand{\Ith}{I_{\mathrm{th}}}
\newcommand{\tth}{t_{\mathrm{th}}}
\newcommand{\sth}{\sigma_{\mathrm{th}}}
\newcommand{\sthinf}{\sigma_{\mathrm{th},\infty}}
\newcommand{\Aeff}{A_{\mathrm{eff}}}

\newcommand{\dNeh}{K_{eh}}

\newcommand{\realpart}{\mathrm{Re}}
\newcommand{\imagpart}{\mathrm{Im}}

\begin{document}

\begin{frontmatter}

\title{The statistics of electron-hole avalanches}

\author[Oxford]{P. Windischhofer}
\author[CERN]{W. Riegler}

\address[Oxford]{University of Oxford}
\address[CERN]{CERN}

\begin{abstract}
  Charge multiplication through avalanche processes is commonly employed in the detection of single photons or charged particles in high-energy physics and beyond. In this report, we provide a detailed discussion of the properties of avalanches driven by two species of charge carriers, e.g.~electrons and holes in a semiconductor exposed to an electric field.
  We derive equations that describe the general case of avalanches developing in position-dependent electric fields and give their analytical solutions for constant fields. We discuss consequences for the time resolution achievable with detectors that operate above the breakdown limit, e.g.~single-photon avalanche diodes (SPADs) and silicon photomultipliers (SiPMs). Our results also describe avalanches that achieve finite gain and are important for avalanche photodiodes (APDs) and low-gain avalanche detectors (LGADs).
\end{abstract}

\end{frontmatter}

\clearpage

\section{Introduction}
\label{Sec:introduction}

Ever since the discovery of electron ionisation processes in gases by Townsend \cite{avalanches_townsend}, avalanche processes have been used as powerful mechanisms to provide high gain in situations where a small amount of initial charge needs to be detected.
In recent decades, advances in the manufacturing of semiconductors have made it possible to exploit avalanche multiplication occurring in the solid state. This has led to the development of different types of semiconductor devices that make use of avalanche charge multiplication in various ways. Detectors based on this principle include avalanche photodiodes (APDs), low-gain avalanche detectors (LGADs), single-photon avalanche detectors (SPADs), and silicon photomultipliers (SiPMs).

Contrary to avalanches developing in gases, where solely electrons drive the avalanche, charge carriers of both polarities (i.e.~electrons and holes) can be important in semiconductors.
Many properties of electron-hole avalanches are already well understood, including the condition for breakdown \cite{mcintyre_breakdown, oldham_breakdown}, fluctuations of the avalanche gain \cite{mcintyre_avalanche_gain} and some aspects concerning the average development of the avalanche \cite{average_avalanche_solution, average_avalanche_solution_2}. In this report, we provide a detailed discussion of the statistical properties of electron-hole avalanches that complements the available literature.

\subsection{Description of the problem}
\label{Sec:definition_avalanche_process}

\begin{figure}[tph]
  \centering
  a)\quad\raisebox{0.2\height}{\includegraphics[width=2.8cm]{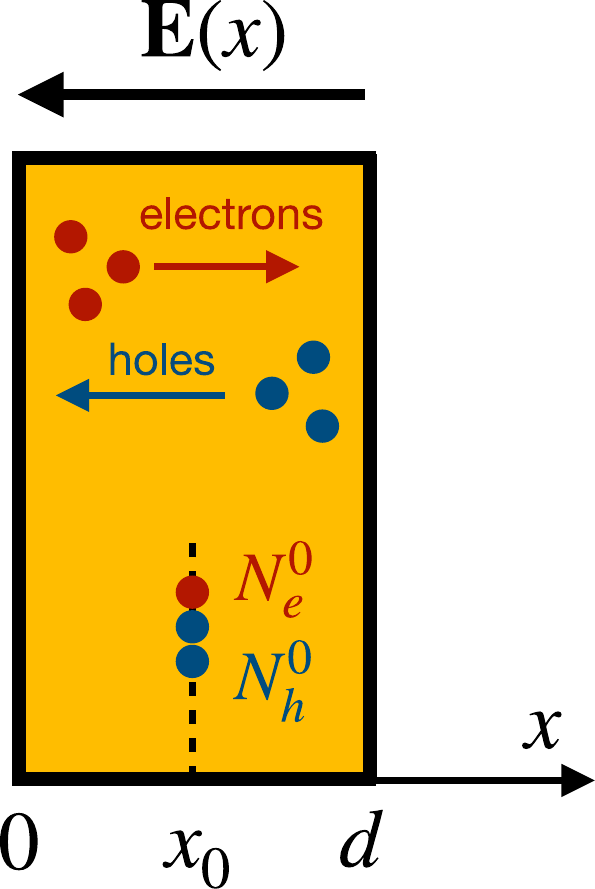}}\qquad\qquad
  b)\includegraphics[width=8.5cm]{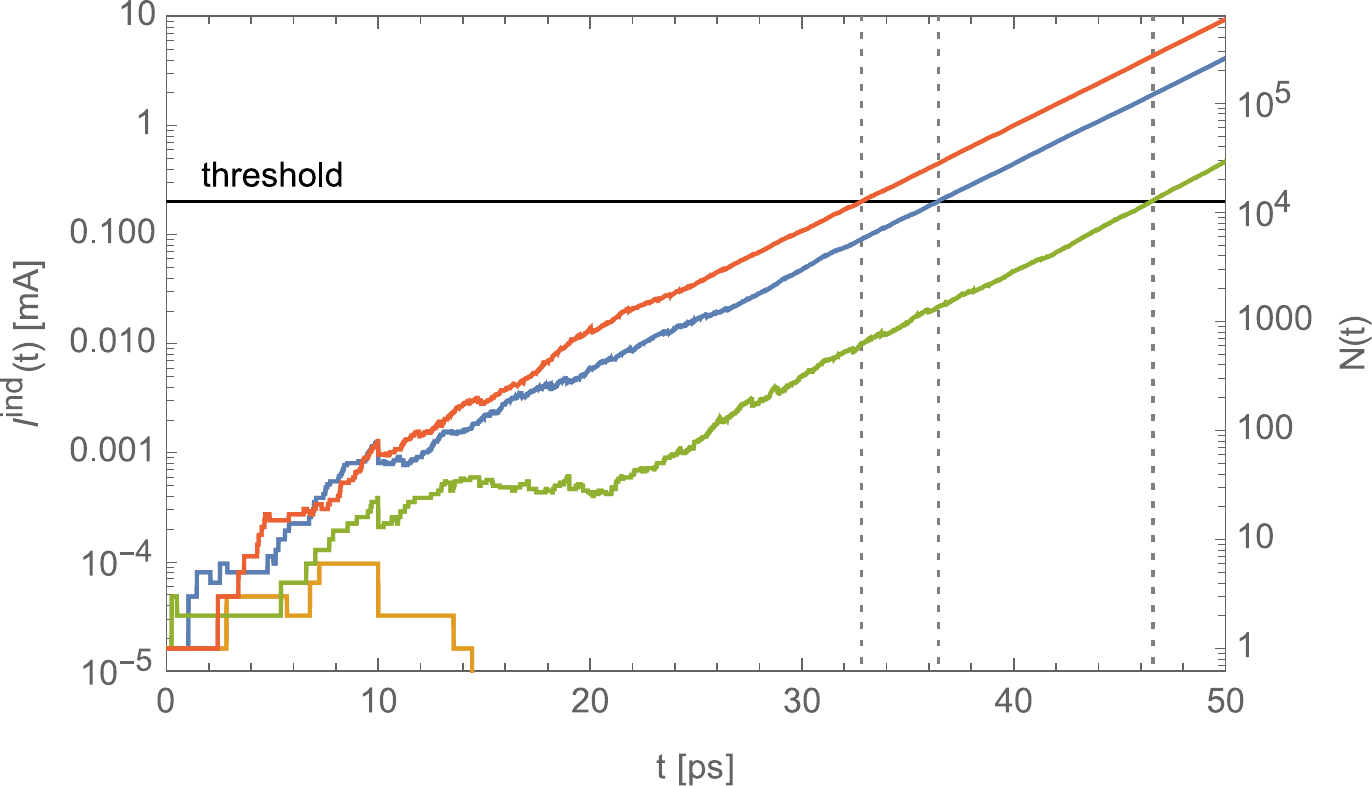}
  \caption{a) An electron-hole avalanche in a one-dimensional solid of thickness $d$ initiated by $N_e^0$ electrons and $N_h^0$ holes at $x = x_0$. In the applied electric field, electrons move in the positive $x$-direction and holes move in the negative $x$-direction. b) (Colour online.) The current induced by a series of electron-hole avalanches in a thin layer of silicon with $d=1\,\mu$m, obtained from a Monte Carlo simulation. The instantaneous number of charges $N(t)$ in the avalanche determines the current $\Iind(t)$ induced on the readout electrodes. The time at which the current crosses the applied threshold is indicated by a vertical dashed line for each avalanche. One avalanche fails to diverge and does not cross the threshold.}
  \label{Fig:practical_situation}
\end{figure}

Figure \ref{Fig:practical_situation}a shows the setup discussed in this report. A number of electrons and holes are deposited at $x = x_0$ in a solid state material of thickness $d$ exposed to a strong electric field $\mathbf{E}(x)$ ranging from 20-60\,V/$\mu$m. This configuration can be created in a reverse-biased p-n junction with suitable dopant concentrations. The thickness $d$ of this ``avalanche region'' is typically of the order of 1-2\,$\mu$m. Both types of charge carriers then multiply in the (generally position-dependent) electric field, which leads to the formation of an avalanche and the multiplication of the initial charge.
In SPADs or SiPMs, the lateral extent $D$ of the avalanche region is of the order of 5-15\,$\mu$m \cite{SPAD_magnitude}. The high aspect ratio $D/d$ allows to treat the field configuration as (effectively) one-dimensional, i.e.~to take all electric field lines to be parallel. 

The instantaneous number of charge carriers in the avalanche determines the current induced on the readout electrodes of a particle detector. This current forms the ``signal'' that is processed by the readout electronics. Detectors such as SPADs and SiPMs operate above the breakdown limit, where the initial charge triggers a diverging avalanche with a certain efficiency. A threshold can be applied to the induced current, as shown in Figure \ref{Fig:practical_situation}b. The threshold-crossing time fluctuates due to the stochasticity of the avalanche, and these variations determine the time resolution of the device.

\begin{figure}[tph]
  \centering
  \includegraphics[width=7cm]{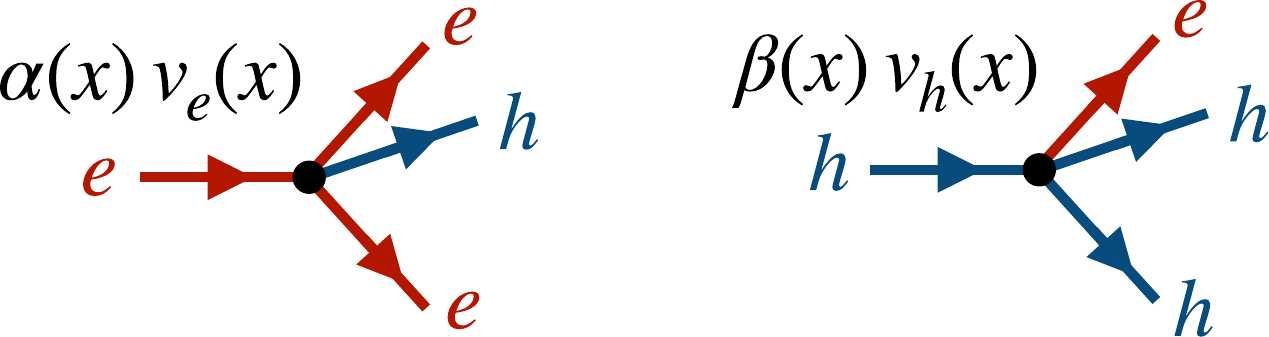}
  \caption{Multiplication processes occurring in the material. Electrons ($e$) move with a drift velocity $v_e(x)$ and can create additional electron-hole pairs as described by the impact ionisation coefficient $\alpha(x)$. Holes ($h$) move with a drift velocity $v_h(x)$ and can create additional electron-hole pairs as described by the impact ionisation coefficient $\beta(x)$. The rate with which these processes occur depends on the strength of the local electric field $\mathbf{E}(x)$, and thus on the position $x$.}
  \label{Fig:introduction_avalanche_description}
\end{figure}

\subsection{Outline}

This report consists of four main parts. Section \ref{Sec:assumptions} outlines the assumptions we use to model the electron-hole avalanche. Section \ref{Sec:results_unbounded} discusses the development of such an avalanche in an infinitely extended semiconductor to which a constant electric field is applied. This allows us to identify the basic characteristics of electron-hole avalanches in isolation. In Section \ref{Sec:results_bounded}, we then return to the practical situation mentioned above and in Figure \ref{Fig:practical_situation}a.
For both scenarios, we give a complete treatment of the average development of the avalanche as well as its fluctuations around this average evolution. Taken together, this allows us to quantify the achievable time resolution in the sense of Figure \ref{Fig:practical_situation}b. All of our results are phrased in terms of explicit analytic formulas, or, in situations of greater complexity, efficient numerical procedures.

Our discussions have as their starting point a series of differential equations that encode the dynamics of the avalanche process. These equations are most easily derived from the principles of stochastic mechanics, to which we give a self-contained introduction, along with a list of references, in Appendix \ref{Sec:formalism}. However, the material presented there is purely supplementary and our results, and all calculations presented in the main body, can be appreciated without consulting this appendix.

We apply the results presented in this paper to the case of charge avalanches in SPADs and SiPMs, and provide a detailed discussion of the time resolution and efficiency of these devices in Ref.~\cite{SPAD_paper}, which is published alongside this article.

\section{Avalanche model and assumptions}
\label{Sec:assumptions}

We model the development of the avalanche as shown in Figure \ref{Fig:introduction_avalanche_description}. Both electrons and holes can undergo multiplication reactions and produce additional electron-hole pairs. The probability for an electron to multiply in the small spatial interval $dx$ is taken to be $\alpha(x) dx$. Analogously, the probability for a hole to multiply is given by $\beta(x) dx$.
We assume that electrons and holes move with drift velocities $v_e(x)$ and $v_h(x)$ through the material. These drift velocities depend on the strength of the local electric field $\mathbf{E}(x)$ in the medium, and therefore on the position $x$. The probabilities for a multiplication to occur in a small time interval $dt$ are then $\alpha(x) v_e(x) dt$ and $\beta(x) v_h(x) dt$.

The impact ionisation coefficients $\alpha$ and $\beta$ encode the details of the underlying scattering processes in the material, and as such are also strongly dependent on the applied electric field $\mathbf{E}(x)$. In analogy to avalanches in gases, they are sometimes referred to as ``Townsend coefficients''. We treat them here as external input parameters and refer to parametrisations for specific materials that are available in the literature.

Note that we assume the Townsend coefficients to be local quantities: they prescribe the probability for impact ionisation at a certain position $x$. We neglect any dependence on the history of the charge carriers. In particular, our model does not include ``dark space'' effects in the sense of Ref.~\cite{nonlocal_impact_coeffs} that cause the ionisation probability to also depend on the distance travelled by the charge carrier since the last ionisation event. 

In accordance with Figure \ref{Fig:practical_situation}a, we model the avalanche as a one-dimensional system. In a physical device, the avalanche discharge forms a cylindrical ``microplasma tube'' with a nontrivial phenomenology \cite{microplasma_discharge}. In a parallel field geometry, its radial extent is determined by the diffusion of the participating charge carriers. However, this does not affect the statistics of the avalanche process and is thus irrelevant for the phenomena discussed here.

Furthermore, we do not consider space charge effects, i.e~assume all charge carriers in the avalanche to be independent of each other, irrespective of the total number of charges already present in the avalanche. These effects become important only for very large avalanches. The space charge field $\mathbf{E}_{\mathrm{sc}}$ becomes comparable to the externally applied electric field $\mathbf{E}$ for avalanches containing more than $10^4-10^5$ charges, depending on the radial evolution of the discharge tube.
As Figure \ref{Fig:practical_situation}b shows (and as will be discussed in detail throughout this report), fluctuations of charge avalanches are generated predominantly in the early stages of their evolution. In this regime, $|\mathbf{E}_{\mathrm{sc}}| \ll |\mathbf{E}|$ and the avalanche statistics can be treated in excellent approximation without considering the backreaction caused by $\mathbf{E}_{\mathrm{sc}}$.
Space charge effects remain of course very important for the understanding of other characteristics of the avalanche, such as the onset of the ``quenching'' of the discharge, which are not discussed here.



\section{Electron-hole avalanche in an infinite semiconductor with a constant electric field}
\label{Sec:results_unbounded}

\begin{figure}[ht]
  \centering
  a) \quad \includegraphics[height=4cm]{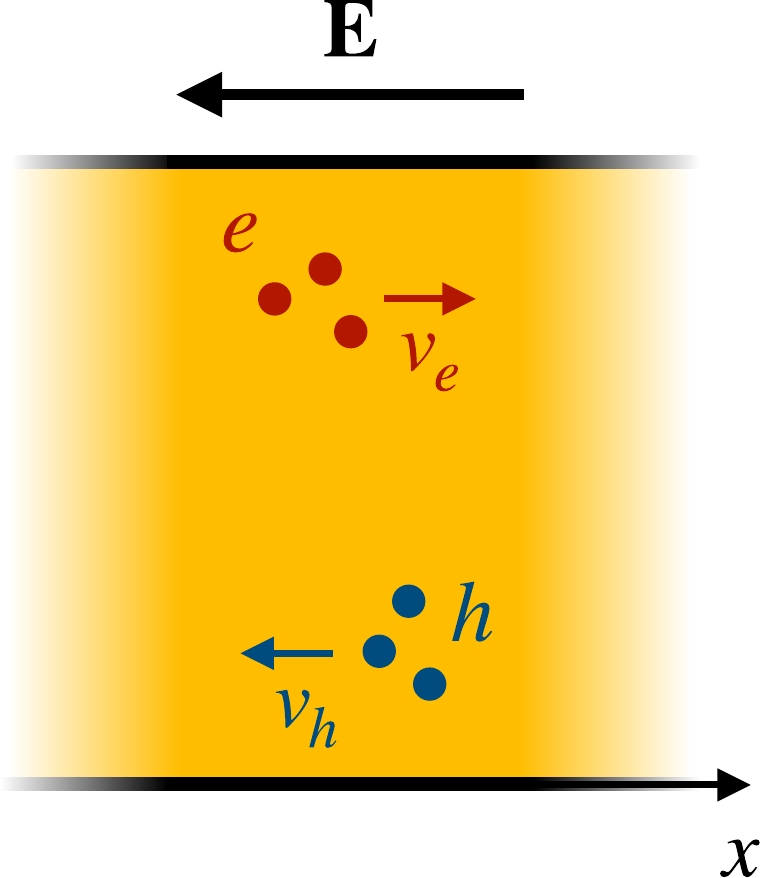}\qquad\qquad
  b) \quad \includegraphics[height=4cm]{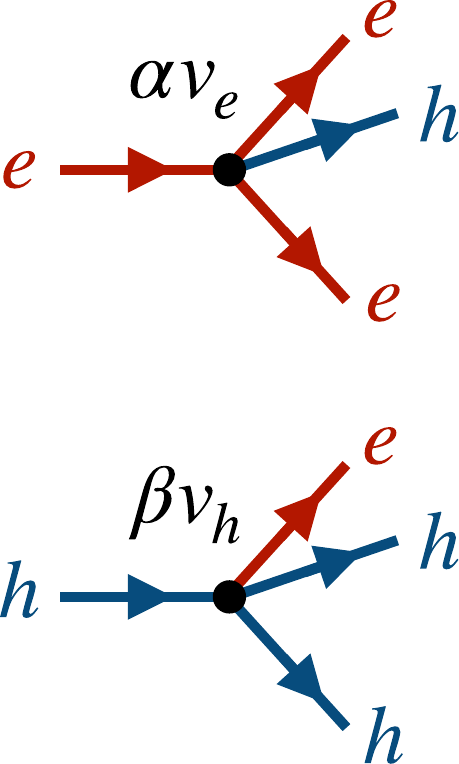}
  \caption{a) A one-dimensional semiconductor of infinite spatial extent, exposed to a constant electric field $\mathbf{E}$. Electrons ($e$) move to the right with a constant drift velocity $v_e$ and holes ($h$) move to the left with a constant drift velocity $v_h$. b) Multiplication processes occur in the material at a rate independent of the position $x$.}
  \label{Fig:unbounded_domain}
\end{figure}

We first consider the idealised scenario of an infinitely extended semiconductor exposed to a constant electric field $\mathbf{E}$, as shown in Figure \ref{Fig:unbounded_domain}. Then, the Townsend coefficients $\alpha$ and $\beta$ and the drift velocities $v_e$ and $v_h$ become independent of the position $x$. 

We are interested in describing the evolution of the instantaneous total number of electrons and holes in the medium, $N_e$ and $N_h$, as the avalanche progresses. The formation of the avalanche is a stochastic process, i.e.~$N_e$ and $N_h$ at a given time $t$ are understood as random variables. We denote the probability to find an avalanche containing $N_e$ electrons and $N_h$ at time $t$, starting from certain initial conditions, as $p(N_e, N_h, t)$. This probability satisfies the following differential equation, which determines its time evolution,
\bea
\frac{d}{dt} p(N_e, N_h, t) = p(N_e - 1, N_h - 1, t) \left[\alpha v_e (N_e - 1) + \beta v_h (N_h - 1) \right] - p(N_e, N_h, t) \left[\alpha v_e N_e + \beta v_h N_h \right].
\label{evolution_equation_p_Ne_Nh}
\eea

The content of this equation can be appreciated intuitively. There are two ways in which the probability $p(N_e, N_h, t)$ can be modified in a small time interval $dt$. First, any charge carrier in an avalanche with $N_e - 1$ electrons and $N_h - 1$ holes can multiply and add an additional electron-hole pair, thereby increasing $p(N_e, N_h, t)$. The probability (per unit time) for this to happen is proportional to $p(N_e-1, N_h-1, t)$ and given by the first, positive, term on the right-hand side. Second, any multiplication in an avalanche with $N_e$ electrons and $N_h$ holes reduces $p(N_e, N_h, t)$ by an amount proportional to itself. It is given by the second, negative, term.
In the physics and statistics literature, Equation \ref{evolution_equation_p_Ne_Nh} is sometimes referred to as a ``master equation'' for a Markov process. Its structure is a direct consequence of our independence and locality assumptions. We give a formal derivation in Appendix \ref{Sec:derivation_evolution_equation_p_Ne_Nh}.

\subsection{Exact expression for $p(N_e, N_h, t)$}
\label{probability_distribution_unbounded}
To solve Equation \ref{evolution_equation_p_Ne_Nh}, we consider an avalanche that is initiated by $N_e^0$ electrons and $N_h^0$ holes at $t = 0$, i.e.~$p(N_e, N_h, t=0) = \delta_{N_e, N_e^0} \, \delta_{N_h, N_h^0}$. The master equation only couples the probabilities $p(N_e, N_h, t)$ and $p(N_e + 1, N_h + 1, t)$, i.e.~it relates configurations that differ only by some number of electron-hole pairs. As a first step, it thus makes sense to consider the evolution of the number of additional electron-hole pairs created during the evolution of the avalanche. We label this quantity by the random variable $\dNeh$. Then, the probability $p(N_e, N_h, t)$ can be expressed in terms of $\dNeh$ as
\beq
p(N_e, N_h, t) = \sum_{\dNeh = 0}^{\infty} p(\dNeh, t) \, \delta_{N_e, N_e(\dNeh)} \, \delta_{N_h, N_h(\dNeh)},
\label{p_Ne_Nh_vs_p_Neh}
\eeq
where $N_e(\dNeh) = N_e^0 + \dNeh$ and $N_h(\dNeh) = N_h^0 + \dNeh$ are the numbers of electrons and holes present in the avalanche after the creation of $\dNeh$ additional pairs. Inserting this relation into Equation \ref{evolution_equation_p_Ne_Nh} yields the following evolution equation for $p(\dNeh, t)$,
\bea
\frac{d}{dt} p(\dNeh, t) = (\dNeh - 1) \left(\alpha v_e + \beta v_h \right) p(\dNeh - 1, t) - \dNeh \left(\alpha v_e + \beta v_h \right) p(\dNeh, t) + \nonumber\\
+ (\alpha v_e N_e^0 + \beta v_h N_h^0) \left[ p(\dNeh- 1, t) - p(\dNeh, t) \right].
\label{evolution_equation_p_Neh}
\eea
We now express this equation in terms of the $z$-transform of $p(\dNeh, t)$, which is defined as
\beqs
P(z, t) := \sum_{\dNeh = 0}^{\infty} \frac{p(\dNeh, t)}{z^{\dNeh}}.
\eeqs
In the $z$-domain, Equation \ref{evolution_equation_p_Neh} becomes
\beq
\frac{\partial}{\partial t} P(z, t) = (\alpha v_e + \beta v_h) (z - 1) \frac{\partial}{\partial z} P(z, t) + (\alpha v_e N_e^0 + \beta v_h N_h^0) \left(\frac{1}{z} - 1\right) P(z, t).
\label{evolution_equation_p_Neh_z_domain}
\eeq
Since $\dNeh$ is defined as the number of electron-hole pairs created during the evolution of the avalanche, the initial condition for Equation \ref{evolution_equation_p_Neh_z_domain} is $p(\dNeh, t = 0) = \delta_{\dNeh, 0}$. In the $z$-domain, this becomes $P(z, t = 0) = 1$. The evolution equation can be solved with the method of characteristics and the solution compatible with the above initial condition is
\beq
P(z, t) = \left(\frac{z}{1 + (z-1) \nu(t)} \right)^A,
\label{solution_p_Neh_z_domain}
\eeq
with
\beqs
\nu(t) = e^{(\alpha v_e + \beta v_h) t}, \qquad\qquad\qquad\qquad A = \frac{\alpha v_e N_e^0 + \beta v_h N_h^0}{\alpha v_e + \beta v_h}.
\eeqs
The parameter $A$ characterises the avalanche, and as we shall see below, many of its properties can be expressed in terms of it. We refer to $A$ as the ``avalanche parameter''. The quantity $\nu(t)$ sets the absolute scale for the (exponential) growth of the avalanche.

\begin{figure}[ht]
  \centering
  \includegraphics[width=10cm]{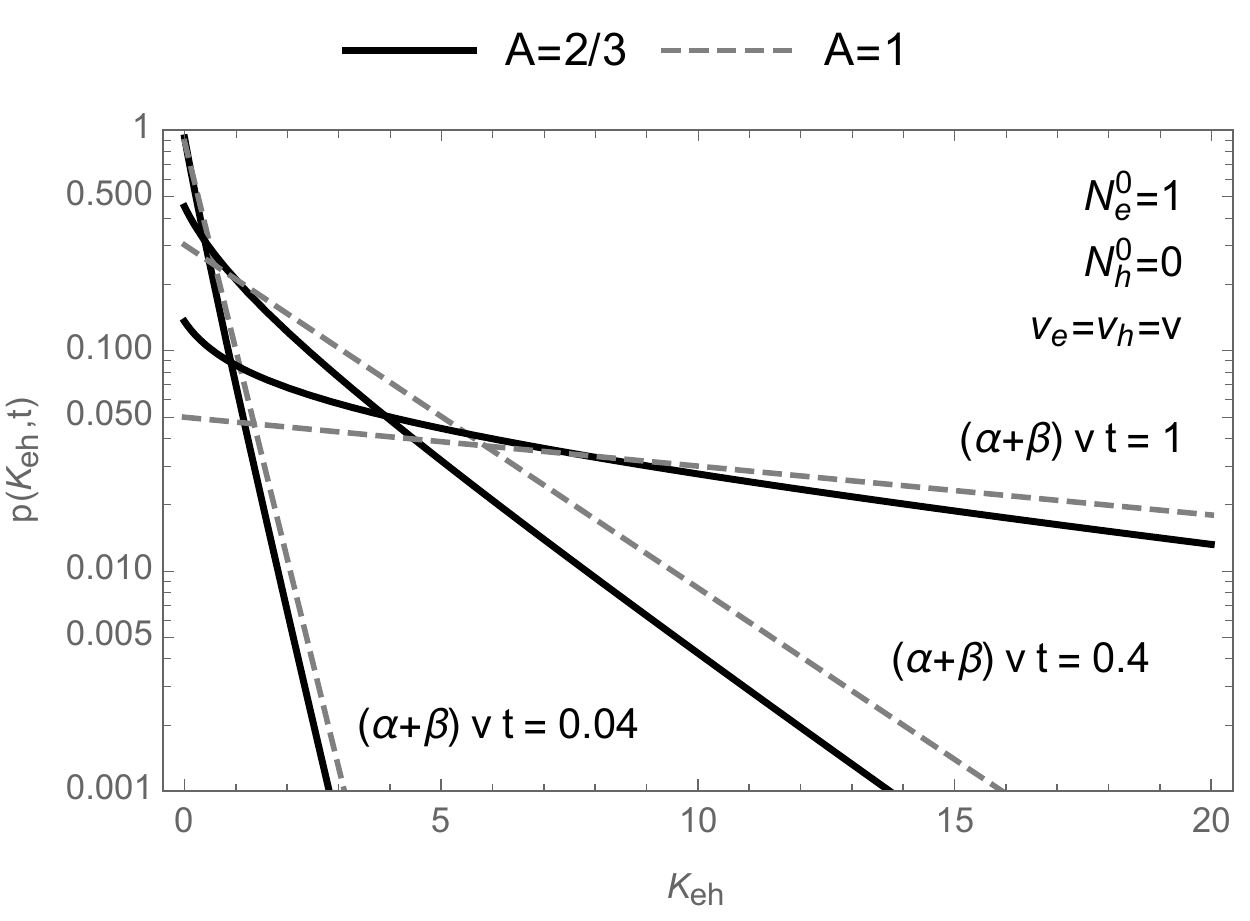}
  \caption{Probability $p(\dNeh, t)$ to find $\dNeh$ additional electron-hole pairs at time $t$ in an avalanche triggered by a single electron. The probability $p(\dNeh, t)$ is defined only for integer $\dNeh$, but for visual clarity is drawn here as a continuous function. The drift velocities of electrons and holes are taken to be equal, $v_e=v_h=v$. The dashed grey line corresponds to $\beta = 0$, i.e.~an avalanche in which only electrons multiply. The solid black line shows an electron-hole avalanche with identical $\alpha+\beta$, but $\beta\neq 0$, i.e.~a non-integer value for the avalanche parameter $A$.}
  \label{Fig:p_N_comparison}
\end{figure}

To perform the inverse $z$-transformation of Equation \ref{solution_p_Neh_z_domain}, we make use of the relation
\beqs
p(\dNeh, t) = \frac{1}{\Gamma(1+\dNeh)} \frac{d^{\dNeh}}{dz^{\dNeh}}\, P\left(1/z, t\right) \Big\rvert_{z=0},
\eeqs
where $\Gamma(\cdot)$ is the gamma function. This gives
\beq
p(\dNeh, t) = \frac{\Gamma(A+\dNeh)}{\Gamma(A) \Gamma(1+\dNeh)} \left(\frac{1}{\nu(t)}\right)^A \, \left(1 - \frac{1}{\nu(t)} \right)^{\dNeh} \approx \frac{1}{\Gamma(A)} \left(\frac{1}{\nu(t)} \right)^A \dNeh^{A-1} e^{-\frac{\dNeh}{\nu(t)}},
\label{result_p_Neh}
\eeq
where the approximation holds in the limit of late times, i.e.~large avalanches. The avalanche parameter $A$ effectively controls the shape of the distribution, as illustrated in Figure \ref{Fig:p_N_comparison}. Via Equation \ref{p_Ne_Nh_vs_p_Neh}, this immediately yields the solution for $p(N_e, N_h, t)$.

For many practical applications, the total number $N$ of charge carriers is relevant, $N = N_e + N_h$, e.g.~for the computation of the current induced on the readout electrodes in a particle detector. Its distribution $p(N, t)$ is related to $p(N_e, N_h, t)$ via
\beqs
p(N, t) = \sum_{N_e + N_h = N} p(N_e, N_h, t) = p\left( \dNeh(N), t \right),
\eeqs
where $\dNeh(N) = (N-N^0)/2$ is the number of additional electron-hole pairs that are required to reach $N$ charges from $N^0 = N_e^0 + N_h^0$.
Inserting the result obtained for $p(\dNeh, t)$, we find
\beq
p(N, t) = \begin{cases}
  p\left(\dNeh = \frac{N-N^0}{2}, t \right) = \frac{\Gamma(A+\frac{N-N^0}{2})}{\Gamma(A) \Gamma(1+\frac{N-N^0}{2})} \left(\frac{1}{\nu(t)}\right)^A \, \left(1 - \frac{1}{\nu(t)} \right)^{\frac{N-N^0}{2}} &\text{if $\frac{N-N^0}{2}\in\mathbb{N}$,} \\
  0 & \mathrm{otherwise.}
\end{cases}
\label{result_p_N}
\eeq

\subsection{Moments}
\label{Sec:moments_unbounded}
The probability $p(\dNeh, t)$ determines the $m$-th moment $\avg{\dNeh^m}$, which is defined as
\beq
\avg{\dNeh^m} := \sum_{\dNeh = 0}^{\infty} \dNeh^m \, p(\dNeh, t).
\label{moments_explicit_sum}
\eeq
In particular the first two moments $\avg{\dNeh}$ and $\avg{\dNeh^2}$ are important, since they contain information about the average evolution of the avalanche as well as its fluctuations around this average. Through the identities $N = N^0 + 2 \dNeh$, $N_e = N_e^0 + \dNeh$ and $N_h = N_h^0 + \dNeh$, the moments $\avg{\dNeh^m}$ also give access to the expectations $\avg{N_e^m N_h^n}$ and $\avg{N^m}$.
We just note here the mean and the variance of $\dNeh$, which are the most important ones in practice,
\bea
\avg{\dNeh} &=& A \left(\nu(t) - 1 \right), \label{eq:first_moment}\\
\sigma(\dNeh)^2 &=& \avg{\dNeh^2} - \avg{\dNeh}^2 = A\, \nu(t) \left(\nu(t) - 1\right). \label{eq:variance}
\eea

Summing Equation \ref{moments_explicit_sum} explicitly for large $m$ can be difficult. An alternative way to obtain the moments $\avg{N_{e}^m N_{h}^n}$ directly is to consider them as explicit functions of time. As is shown in Appendix \ref{Sec:derivation_evolution_equation_moments}, the moment $\avg{N_{e}^m N_{h}^n}$ satisfies the following differential equation,
\beq
\frac{d}{dt}\avg{N_e^m N_h^n} = \sum_{(l, k) \in S} \binom{m}{l}\binom{n}{k} \left(\alpha v_e \avg{N_e^{l+1} N_h^k} + \beta v_h \avg{N_e^l N_h^{k+1}} \right).
\label{differential_equation_general_moment}
\eeq
The sum runs over the set $S = \{(l, k) | 0 \leq l \leq m, 0 \leq k \leq n \} \setminus (m, n)$, i.e.~contains all combinations except the one where $l=m$ and $k=n$ simultaneously. An inspection of the right-hand side of Equation \ref{differential_equation_general_moment} shows that it contains only terms $\avg{N_e^l N_h^k}$ where $l + k \leq m + n$. Iterating this expression thus gives rise to a closed system of differential equations that can be solved systematically. The relevant initial conditions are $\avg{N_e^m N_h^n} = \left(N_e^0\right)^m \left(N_h^0\right)^n$.

For the first moments $\avg{N_e}$ and $\avg{N_h}$, these equations read
\beq
\frac{d}{dt}\avg{N_e} = \alpha v_e \avg{N_e} + \beta v_h \avg{N_h}, \qquad\qquad \frac{d}{dt}\avg{N_h} = \alpha v_e \avg{N_e} + \beta v_h \avg{N_h},
\label{equations_first_moment_unbounded}
\eeq
while for the three second moments $\avg{N_e^2}$, $\avg{N_h^2}$, and $\avg{N_e N_h}$ they are
\bea
\frac{d}{dt}\avg{N_e^2} &=& \alpha v_e \avg{N_e} + \beta v_h \avg{N_h} + 2 \alpha v_e \avg{N_e^2} + 2 \beta v_h \avg{N_e N_h},\label{equations_second_moment_unbounded_start}\\
\frac{d}{dt}\avg{N_h^2} &=& \alpha v_e \avg{N_e} + \beta v_h \avg{N_h} + 2 \beta v_h \avg{N_h^2} + 2 \alpha v_e \avg{N_e N_h},\\
\frac{d}{dt}\avg{N_e N_h} &=& \alpha v_e \avg{N_e} + \beta v_h \avg{N_h} + \alpha v_e \avg{N_e^2} + \beta v_h \avg{N_h^2} + (\alpha v_e + \beta v_h) \avg{N_e N_h}\label{equations_second_moment_unbounded_end}.
\eea

As a convenient way to quantify the fluctuations of the avalanche around its average evolution, we introduce the ratio of the variance of $N$ and the square of its mean, $\sigma(N)^2/\avg{N}^2$. We will refer to this quantity as measuring the ``relative fluctuations'' of $N$. Similar measures can be defined for other observables such as $N_e$ or $N_h$, and we shall make extensive use of them in what follows.
The solutions of Equations \ref{equations_second_moment_unbounded_start}-\ref{equations_second_moment_unbounded_end} then reveal the following relation between the avalanche parameter $A$ and the relative fluctuations of the number of charges at late times
\beq
\lim_{t\rightarrow\infty} \frac{\sigma(N)^2}{\avg{N}^2} = \lim_{t\rightarrow\infty} \frac{\sigma(N_e)^2}{\avg{N_e}^2} = \lim_{t\rightarrow\infty} \frac{\sigma(N_h)^2}{\avg{N_h}^2} = \frac{1}{A}.
\label{relfluc_late_times_unbounded}
\eeq
Using Equations \ref{eq:first_moment} and \ref{eq:variance}, we see that the timescale on which the relative fluctuations approach this limiting value is given by the ``saturation time'' $T_{\mathrm{sat}} = (\alpha v_e + \beta v_h)^{-1}$.

Equations \ref{equations_second_moment_unbounded_start}-\ref{equations_second_moment_unbounded_end} also imply that $N_e$ and $N_h$ are maximally Pearson correlated at all times,
\beqs
\rho(N_e, N_h) = \frac{\cov[N_e, N_h]}{\sigma(N_e)\sigma(N_h)} = 1.
\eeqs

\subsection{Avalanches driven by a single species: the case of integer-valued avalanche parameter}
The results given above have important special cases. In case the avalanche parameter $A$ is an integer, the dependence on $\dNeh$ in Equation \ref{result_p_Neh} simplifies. We get
\beq
p(\dNeh, t) = \binom{\dNeh + A - 1}{\dNeh} \left(\frac{1}{\nu(t)}\right)^{A} \left(1 - \frac{1}{\nu(t)} \right)^{\dNeh}.
\eeq
This covers the case of avalanches that are driven by electrons only ($\beta = 0$), where we have $A = N_e^0$ and
\beq
p(N_e, t) = \binom{N_e - 1}{N_e - N_e^0} \left(\frac{N_e^0}{\avg{N_e}}\right)^{N_e^0} \left(1 - \frac{N_e^0}{\avg{N_e}} \right)^{N_e - N_e^0},
\eeq
where $\avg{N_e} = N_e^0 e^{\alpha v_e t}$. This is the well-known Yule-Furry law \cite{furry_law, yule_law} for electron avalanches.

Another important special case concerns avalanches that are initiated by a certain number of electron hole pairs, i.e.~$N_e^0 = N_h^0 = N_{eh}^0$. In this case, the avalanche parameter becomes $A = N_{eh}^0$. The avalanche is then conveniently described in terms of the total number of electron-hole pairs, $N_{eh} = N_{eh}^0 + \dNeh$. The statistics of these avalanches is identical to those driven by electrons only,
\beq
p(N_{eh}, t) = \binom{N_{eh} - 1}{N_{eh} - N_{eh}^0} \left(\frac{N_{eh}^0}{\avg{N_{eh}}}\right)^{N_{eh}^0} \left(1 - \frac{N_{eh}^0}{\avg{N_{eh}}} \right)^{N_{eh} - N_{eh}^0},
\eeq
where $\avg{N_{eh}} = N_{eh}^0 e^{(\alpha v_e + \beta v_h) t}$.

\subsection{Time response function}
\label{Sec:time_response_function}
The current $\Iind(t)$ induced on the readout electrodes in a particle detector can be computed with the Ramo-Shockley theorem \cite{shockley, ramo}. If we assume a constant weighting field $E_w/V_w$ throughout the semiconductor, the current produced by the avalanche is
\beq
\Iind(t) = e_0 \frac{E_w}{V_w} \left(v_e N_e + v_h N_h \right),
\label{current_random_variable}
\eeq
where $e_0$ is the elementary charge. We define $\tth$ as the time at which the current crosses an applied threshold of $\Ith$. The threshold-crossing time is a random variable whose distribution contains information about the achievable time resolution. The current in Equation \ref{current_random_variable} can also be expressed as
\beqs
\Iind(t) = e_0 \frac{E_w}{V_w} \frac{v_e + v_h}{2} \left(\frac{v_e-v_h}{v_e+v_h} (N_e^0 - N_h^0) + N \right).
\eeqs
Making use of this relation, a threshold $\Ith$ applied to the current can be converted into an equivalent threshold $\Nth$ on the total number of charges in the avalanche (which is not generally experimentally accessible). It is thus sufficient to study the statistics of the threshold-crossing time for this case.

We define the time response function $\rho(\Nth, t) dt$ to give the probability that the avalanche crosses a threshold of $\Nth$ charges in the small time interval $[t, t + dt]$. It is given by
\beqs
\rho(\Nth, t) dt = p(\Nth, t) \left[\alpha v_e N_e(\Nth) + \beta v_h N_h(\Nth)\right] dt,
\eeqs
where the additional factor $\left[\alpha v_e N_e(\Nth) + \beta v_h N_h(\Nth)\right] dt$ gives the probability that an avalanche with $\Nth$ charges crosses this threshold in the time $dt$ through the multiplication of an electron or hole. The time response function is properly normalised such that $\int_0^{\infty} dt \, \rho(\Nth, t) = 1$.

\begin{figure}[ht]
  \centering
  \includegraphics[width=10cm]{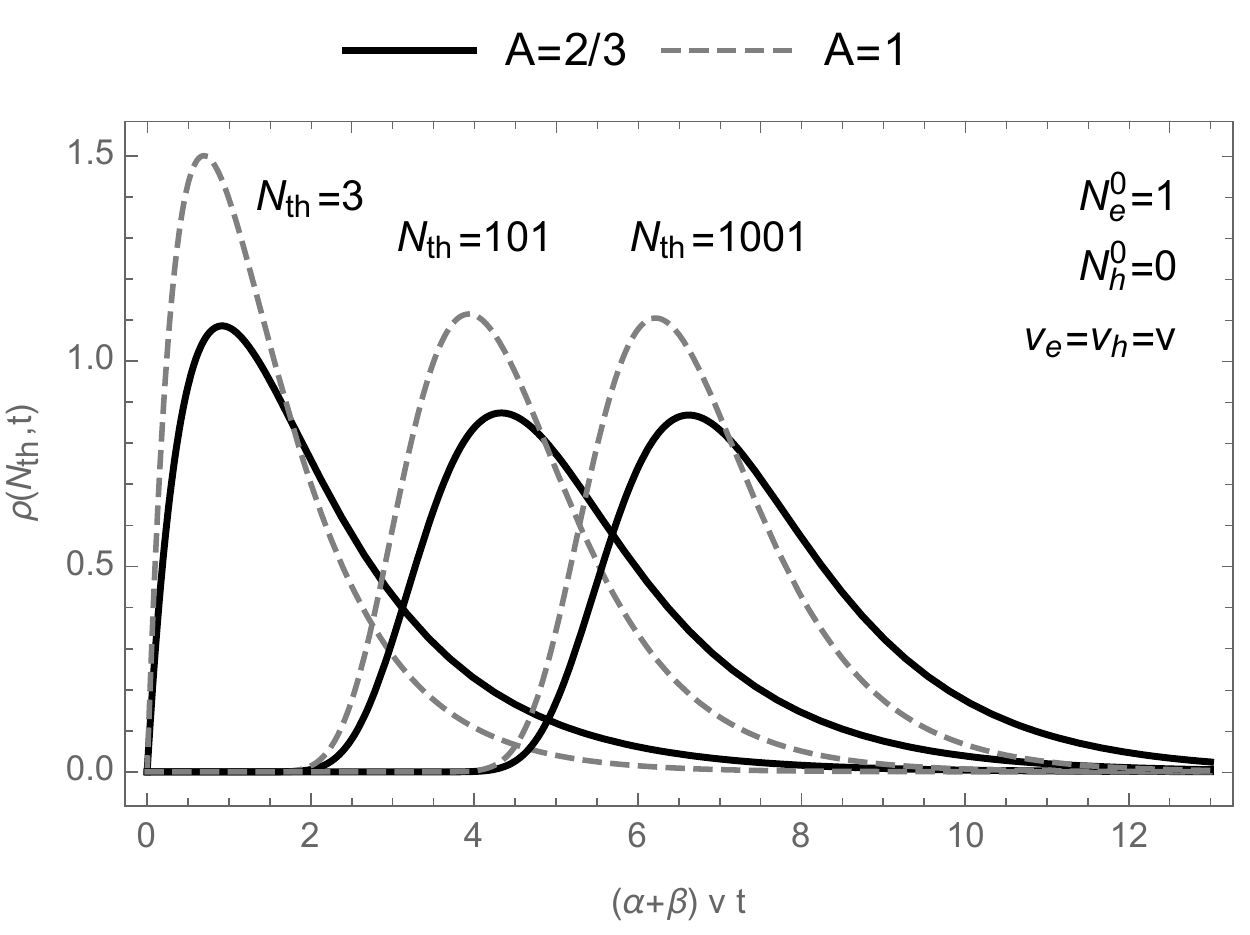}
  \caption{Time response function $\rho(\Nth, t)$ for an avalanche initiated by a single electron for thresholds $\Nth = 3, 101, 1001$. The dashed grey line corresponds to $\beta = 0$, i.e.~a pure electron avalanche. The solid black line shows an electron-hole avalanche with identical $\alpha+\beta$, but $\beta\neq 0$, i.e.~a non-integer value for the avalanche parameter $A$.}
  \label{Fig:time_response_function_comparison}
\end{figure}

Making use of Equation \ref{result_p_N}, it can be written as
\beq
\rho(\Nth, t) = (\alpha v_e + \beta v_h) \, \frac{\Gamma\left(1 + \frac{\Nth-N^0}{2} + A \right)}{\Gamma(A) \Gamma\left(1 +\frac{\Nth-N^0}{2} \right)} \left(\frac{1}{\nu(t)}\right)^{A} \left(1 - \frac{1}{\nu(t)} \right)^{\frac{\Nth-N^0}{2}},
\eeq
where it is again understood that $(\Nth-N^0)/2 \in \mathbb{N}$. For large thresholds (and late times), the time response function is approximately given by
\beq
\rho(\Nth, t) \approx \frac{\alpha v_e + \beta v_h}{\Gamma(A)} \exp\left(A \log \frac{\Nth}{2 \nu(t)} - \frac{\Nth}{2 \nu(t)} \right).
\label{time_resolution_asymptotic_unbounded}
\eeq
In this form we can see that a scaling of the threshold $\Nth \rightarrow \Nth \lambda$ corresponds to a shift of the time argument by $t\rightarrow t-\Delta t$, with $\Delta t = \frac{\log\lambda}{\alpha v_e + \beta v_h}$.
Figure \ref{Fig:time_response_function_comparison} indeed shows that the shape of the time response function changes significantly only for small $\Nth$. In the regime of large $\Nth$, only a shift in the time argument results.

\subsection{Time resolution from time response function}
Using the time response function from Section \ref{Sec:time_response_function}, we can compute the expected threshold-crossing time $\avg{\tth}$ and its second moment $\avg{\tth^2}$ can be computed as
\beq
\avg{\tth} = \int_0^{\infty} dt \, t \, \rho(\Nth, t) = \frac{1}{\alpha v_e + \beta v_h} \left[-\psi_0\left(A\right) + \psi_0\left(1 + \frac{\Nth-N^0}{2} + A \right) \right],
\label{result_tth}
\eeq
\beq
\avg{\tth^2} = \int_0^{\infty} dt \, t^2 \, \rho(\Nth, t) = \avg{\tth}^2 + \frac{1}{(\alpha v_e + \beta v_h)^2} \left[\psi_1(A) - \psi_1\left(1 + \frac{\Nth-N^0}{2} + A \right) \right],
\label{result_tth_tth}
\eeq
where the polygamma function $\psi_k(z)$ is defined in terms of derivatives of the gamma function as $\psi_k(z) = d^{k+1} \ln \Gamma(z) / dz^{k+1}$.

\begin{figure}[ht]
  \centering
  \includegraphics[width=10cm]{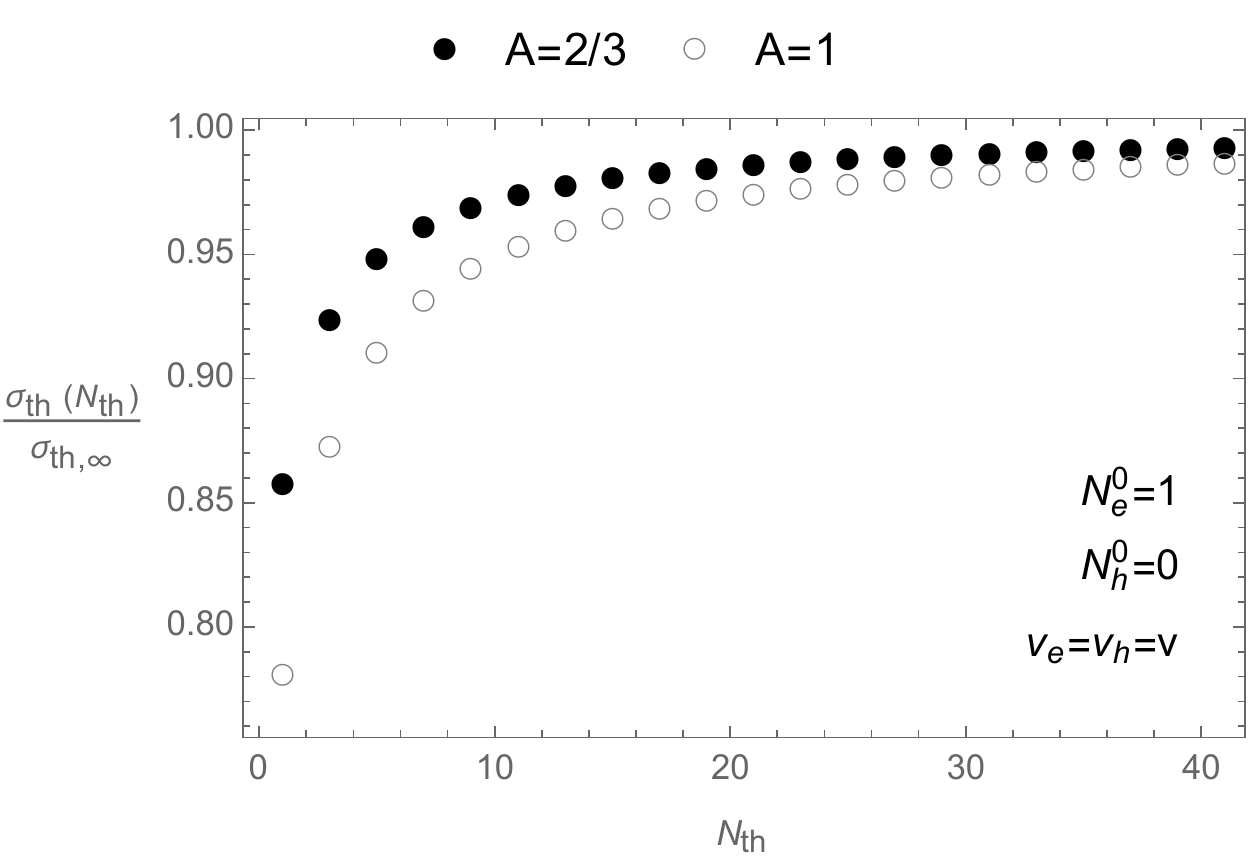}
  \caption{Convergence of the time resolution $\sth(\Nth)$ to its limiting value $\sthinf$ as $\Nth$ increases. The open circles correspond to $\beta = 0$, i.e.~a pure electron avalanche. The filled black circles are valid for an electron-hole avalanche with identical $\alpha+\beta$, but $\beta\neq 0$, i.e.~for a non-integer value for the avalanche parameter $A$.}
  \label{Fig:time_resolution_relative_comparison}
\end{figure}

From Equations \ref{result_tth} and \ref{result_tth_tth}, the standard deviation of the threshold crossing time, $\sth = \sqrt{\avg{\tth^2}-\avg{\tth}^2}$, can be obtained. We use this quantity as a measure for the intrinsic time resolution that can be achieved by a device that relies on the avalanche multiplication of some initially present charge. We find
\beq
\sth(\Nth) = \frac{1}{\alpha v_e + \beta v_h} \sqrt{ \psi_1(A) - \psi_1\left(1 + \frac{\Nth-N^0}{2} + A \right)}.
\eeq
The function $\psi_1(z)$ is monotonically decreasing with $z$. For large thresholds the second term thus becomes negligible compared to the first, and the time resolution $\sth$ saturates at $\sthinf = \lim_{\Nth\rightarrow\infty}\sth(\Nth)$. We get
\beq
\sthinf = \frac{\sqrt{\psi_1(A)}}{\alpha v_e + \beta v_h}.
\label{time_resolution_inf_unbounded}
\eeq
The rapid convergence $\sth(\Nth)\rightarrow\sthinf$, even for moderate thresholds, is shown in Figure \ref{Fig:time_resolution_relative_comparison}: the time resolution is essentially identical to its limiting value already for thresholds $\Nth$ of 30-40.

For the special case of an electron avalanche (i.e.~$\beta = 0$) initiated by $N_e^0$ electrons, we have
\beqs
\sthinf = \frac{\sqrt{\psi_1(N_e^0)}}{\alpha v_e} \approx \frac{1}{\alpha v_e} \frac{1}{\sqrt{N_e^0}},
\eeqs
where the approximation is valid for large $N_e^0$. This shows that the time resolution improves as $\sth \sim 1/\sqrt{N_e^0}$ as the primary charge increases. In case the avalanche is initiated by a single electron, $N_e^0 = 1$, we recover the result
\beqs
\sthinf = \frac{1}{\alpha v_e} \frac{\pi}{\sqrt{6}},
\eeqs
which was derived in \cite{rpc_time_resolution} in the context of resistive plate chambers.

\subsection{Time resolution for large thresholds}
\label{Sec:timeres_large_th}
For large thresholds, $\Nth\rightarrow\infty$, the time resolution can be computed in a physically more intuitive way that does not require the full time response function $\rho(\Nth, t)$. At late times, i.e.~for large avalanches, we expect the evolution of the avalanche to follow
\beq
\dNeh(t) \sim k_\infty \, e^{(\alpha v_e + \beta v_h) t} \quad \text{as} \quad t\rightarrow\infty.
\label{asymptotic_avalanche}
\eeq
Note that this is a relation between the two random variables $\dNeh(t)$ and $k_\infty$. The latter implements the fluctuations in the absolute size of the avalanche. The exponential scaling is factored out and made manifest. Whenever $\dNeh \gg N^0$, as is the case in the limit we consider here, a similar relation also holds for the total charge content of the avalanche, $N$.

The random variable describing the threshold-crossing time $\tth$ is then given in terms of $k_\infty$ by
\beqs
\tth = \frac{1}{\alpha v_e + \beta v_h} \log\left(\frac{\Nth}{k_\infty} \right) = \frac{\log\Nth - \log k_\infty}{\alpha v_e + \beta v_h},
\eeqs
and its standard deviation $\sthinf$ is
\beq
\sthinf = \frac{\sigma(\log k_\infty)}{\alpha v_e + \beta v_h} = \lim_{t\rightarrow\infty} \frac{\sigma(\log N)}{\alpha v_e + \beta v_h}.
\label{time_resolution_asymptotic_unbounded}
\eeq
This shows that the asymptotic time resolution is determined by the fluctuations of the logarithm of the number of charges in the avalanche.

The argument as presented above is very general and holds for any multiplication process that leads to (asymptotically) exponential growth. We now make the connection to the specific avalanche process described above in Section \ref{probability_distribution_unbounded}. At late times, the distribution $p(\dNeh, t)$ in Equation \ref{result_p_Neh} can be written as
\beqs
p(\dNeh, t) \approx \int_0^{\infty} dk_\infty \, \frac{k_\infty^{A-1} e^{-k_\infty}}{\Gamma(A)} \, \delta\left(\dNeh - k_\infty \, e^{(\alpha v_e + \beta v_h) t} \right) \quad \text{as} \quad t\rightarrow\infty.
\eeqs
We see explicitly that the distribution has support only for exponentially growing avalanches of the form of Equation \ref{asymptotic_avalanche}. The distribution of the factor $k_\infty$ can be directly read off: it follows a gamma distribution with a shape parameter of $A$ and a rate parameter of unity.
As is shown by an explicit calculation in Appendix \ref{Sec:asymptotic_logN_unbounded}, we have that
\beq
\lim_{t\rightarrow\infty} \sigma(\log N) = \sqrt{\psi_1(A)},
\label{logN_fluctuations_unbounded}
\eeq
and so we recover the expression for the time resolution already stated in Equation \ref{time_resolution_inf_unbounded}.


\section{Electron-hole avalanche in a thin semiconductor with an arbitrary electric field}
\label{Sec:results_bounded}

\begin{figure}[ht]
  \centering
  a)\quad \includegraphics[height=4cm]{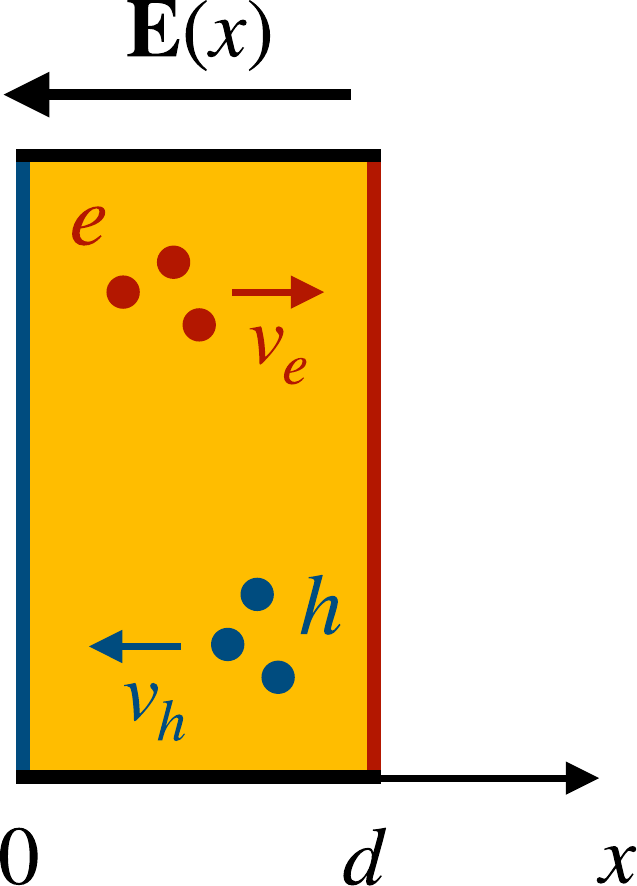}\qquad\qquad
  b)\quad \includegraphics[height=4cm]{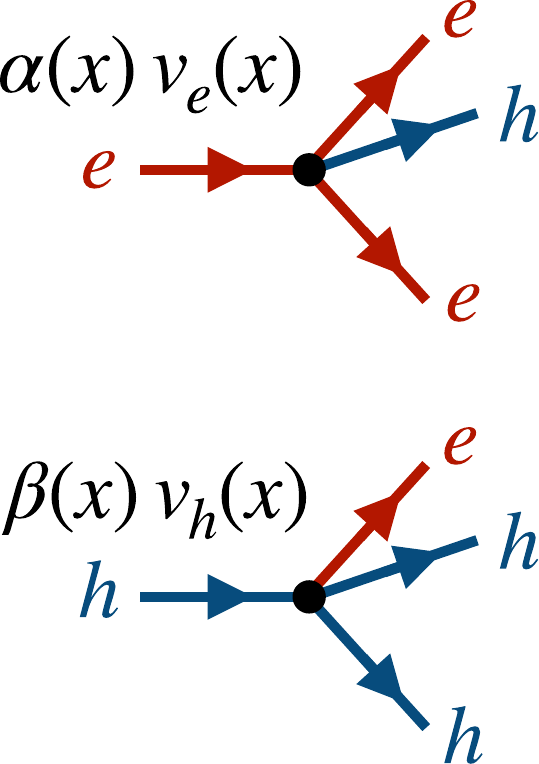}
  \caption{a) The avalanche develops in a semiconductor of thickness $d$ that is exposed to a position-dependent electric field $\mathbf{E}(x)$. We take electrons and holes to move with drift velocities $v_e(x)$ and $v_h(x)$. Electrons can leave the avalanche region through the boundary at $x=d$; holes cross the boundary at $x=0$. b) The probability for charge multiplication to occur now depends on the position $x$.}
  \label{Fig:bounded_domain}
\end{figure}

We now return to the discussion of the development of an electron-hole avalanche in a particle detector of the type described in Section \ref{Sec:introduction}. We consider a bounded one-dimensional region $0 \leq x \leq d$ in which the avalanche develops, as shown in Figure \ref{Fig:bounded_domain}a. Practically feasible doping profiles generally lead to a position-dependent electric field $\mathbf{E}(x)$ in the semiconductor upon the application of the bias voltage. To take this into account, we allow the Townsend coefficients and drift velocities to depend on the position, i.e.~we assume general profiles $\alpha(x)$, $\beta(x)$, $v_e(x)$ and $v_h(x)$.
The probability for a charge multiplication reaction to occur thus also depends on the position, as shown in Figure \ref{Fig:bounded_domain}b.

As explained in detail in Appendix \ref{Sec:formalism_1D}, this scenario presents significant technical challenges, which make the computation of explicit distributions such as $p(N, t)$ very difficult. In the following, we will thus focus on the discussion of expectation values of the spatial charge densities $n_e(x, t)$ and $n_h(x, t)$. These are now random variables indexed by $x$ and $t$. This will allow us to gain insight into the average spatial distribution of charges in the semiconductor, as well as their correlations across different positions. In all cases, we shall consider the evolution of the avalanche starting from well-defined initial charge densities at $t=0$, $n_e^0(x)$ and $n_h^0(x)$.

\subsection{Average development of the avalanche}
\label{Sec:first_moment_bounded}
We begin with the discussion of the average development of the avalanche. The quantities of interest here are the expectation values $\avg{n_e(x, t)}$ and $\avg{n_h(x, t)}$, i.e.~the average densities of electrons and holes at a specific position $x$. For brevity, we shall often suppress the explicit position- and time dependence.

As is shown in Appendix \ref{Sec:moments_equations_bounded}, these averages satisfy the following system of equations,
\beq
\frac{\partial}{\partial t}\avg{n_e(x)} + \frac{\partial}{\partial x} v_e(x)\avg{n_e(x)} = \alpha(x) v_e(x) \avg{n_e(x)} + \beta(x) v_h(x) \avg{n_h(x)},\label{equations_first_moment_bounded_start}
\eeq
\beq
\frac{\partial}{\partial t}\avg{n_h(x)} - \frac{\partial}{\partial x} v_h(x)\avg{n_h(x)} = \alpha(x) v_e(x) \avg{n_e(x)} + \beta(x) v_h(x) \avg{n_h(x)}.\label{equations_first_moment_bounded_end}
\eeq
These equations are the position-dependent analogue of Equation \ref{equations_first_moment_unbounded}: they immediately reduce to the latter for constant Townsend coefficients and drift velocities upon integration over $x$ and the assumption that the average densities vanish as $x\rightarrow\pm\infty$. They can also be interpreted as continuity equations for the average electron- and hole densities in the avalanche region.
The situation in Figure \ref{Fig:bounded_domain}a demands the boundary conditions $\avg{n_e(0, t)}=0$ and $\avg{n_h(d, t)}=0$ as well as the initial conditions $\avg{n_e(x, 0)} = n_e^0(x)$ and $\avg{n_h(x, 0)} = n_h^0(x)$.

For general field profiles $\mathbf{E}(x)$, Equations \ref{equations_first_moment_bounded_start}-\ref{equations_first_moment_bounded_end} can be solved numerically. Below, we present their analytical solutions for constant electric fields $\mathbf{E}$, i.e.~constant Townsend coefficients and drift velocities.

\subsubsection{Solution for constant electric field}

In this case, Equations \ref{equations_first_moment_bounded_start}-\ref{equations_first_moment_bounded_end} are equivalent to the following second order differential equation for $\avg{n_e}$,
\beq
\frac{\partial^2 \avg{n_e}}{\partial x^2} + \left(\frac{1}{v_e} - \frac{1}{v_h} \right) \frac{\partial^2 \avg{n_e}}{\partial x \partial t} - (\alpha -\beta) \frac{\partial \avg{n_e}}{\partial x} = \frac{1}{v_e v_h} \frac{\partial^2 \avg{n_e}}{\partial t^2} - \left(\frac{\alpha}{v_h} + \frac{\beta}{v_e} \right) \frac{\partial \avg{n_e}}{\partial t},
\label{n_e_second_order}
\eeq
together with the algebraic relation
\beq
\avg{n_h} = \frac{1}{\beta v_h}\left(\frac{\partial \avg{n_e}}{\partial t} + v_e \frac{\partial \avg{n_e}}{\partial x} - \alpha v_e \avg{n_e} \right).
\label{n_h_algeb_relation}
\eeq

To find the most general solution of Equation $\ref{n_e_second_order}$, we make the ansatz $\avg{n_e} = X(x) T(t)$. Inserting this into the equation and acting on it with $\partial_x \partial_t$ shows that either $X'/X$ or $T'/T$ must be constant. The former is incompatible with our boundary conditions. Focusing on the second case, we set $T' = \lambda_t T$, i.e.~$T(t) = C e^{\lambda_t t}$ with some constant $C$. We then get
\beq
\frac{X''}{X} + \left(\frac{1}{v_e} - \frac{1}{v_h} \right) \lambda_t \frac{X'}{X} - (\alpha - \beta) \frac{X'}{X} = \frac{1}{v_e v_h} \lambda_t^2 - \left(\frac{\alpha}{v_h} + \frac{\beta}{v_e} \right) \lambda_t.
\label{n_e_second_order_separated}
\eeq

\begin{figure}[tph]
  \centering
  a)\quad\includegraphics[width=7cm]{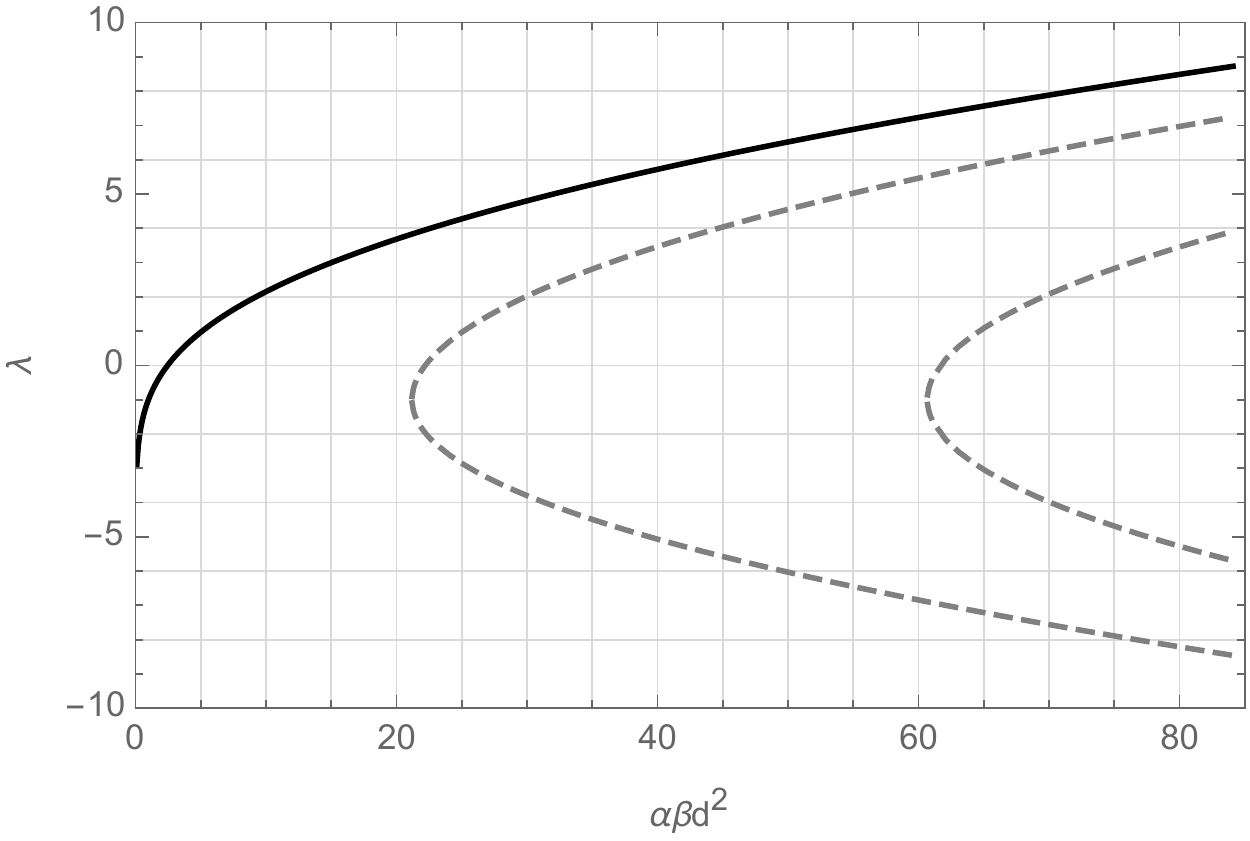}\quad
  b)\quad\includegraphics[width=7cm]{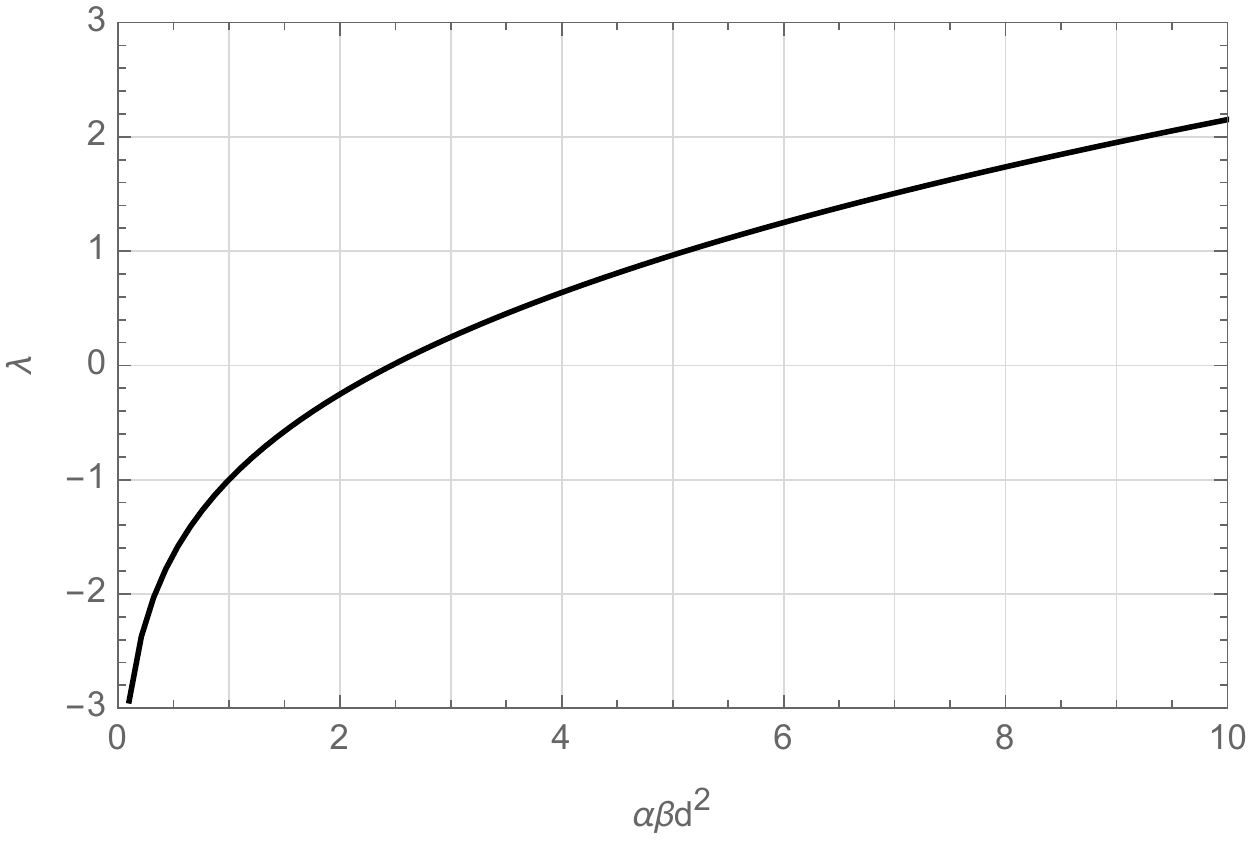}\\
  \caption{a) Real-valued eigenvalues that are solutions of Equation \ref{eigenvalue_equation}, as a function of the characteristic quantity $\alpha\beta d^2$. The solid black line indicates the largest eigenvalue. The dashed grey lines represent the remaining eigenvalues. b) The largest eigenvalue for small $\alpha\beta d^2$.}
  \label{Fig:eigenvalues_real}
\end{figure}

\begin{figure}[tph]
  \centering
  \includegraphics[width=9cm]{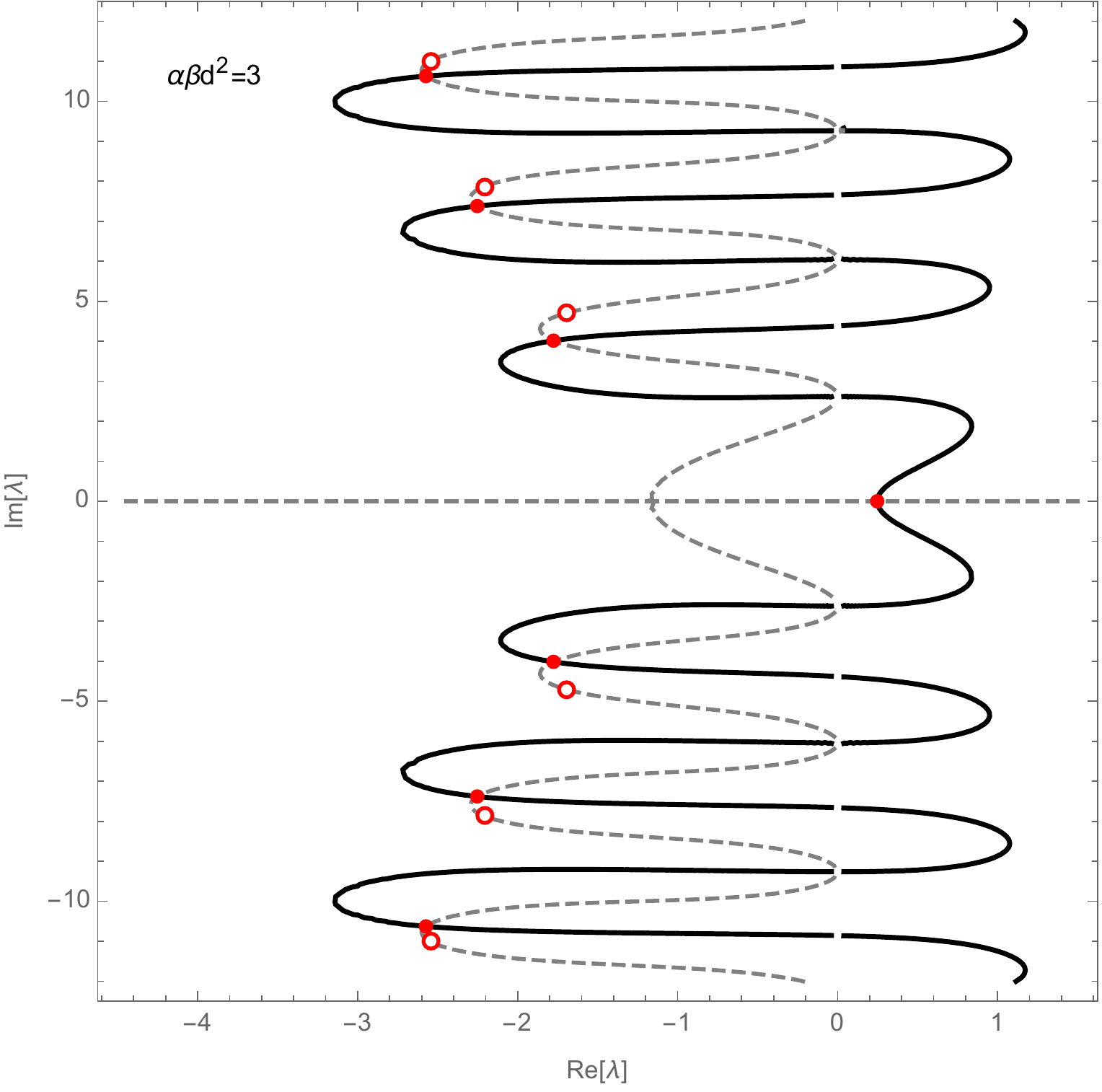}
  \caption{Illustration of the allowed eigenvalues for $\alpha\beta d^2=3$ in the complex plane. The solid black line corresponds to the contour $\realpart[\lambda + \bar\kappa \coth \bar\kappa]=0$ and the dashed grey line to $\imagpart[\lambda + \bar\kappa \coth \bar\kappa]=0$. The solutions to the eigenvalue equation are situated at their intersection points and are marked by filled red circles. The approximate initial values according to Equation \ref{initial_values_eigenvalues} are labelled by empty red circles.}
  \label{Fig:eigenvalues_complex}
\end{figure}

With the following definitions,
\bea
\gamma &=& \frac{\alpha+\beta}{2} + \frac{\lambda}{d},\label{gamma_def}\\
v^* &=& \frac{2 v_e v_h}{v_e+v_h},\nonumber\\
\lambda_t &=& \gamma v^*,\nonumber\\
a &=& \frac{v^*}{2}\left(\frac{\alpha}{v_h} - \frac{\beta}{v_e} \right) + \frac{v^*}{2d} \left(\frac{1}{v_h} - \frac{1}{v_e}\right) \lambda,\nonumber\\
\kappa &=& \frac{1}{d} \bar{\kappa} = \frac{1}{d} \sqrt{\lambda^2 - \alpha\beta d^2},\nonumber
\eea
we can write the solution of Equation \ref{n_e_second_order_separated} as
\bea
f_\lambda^e(x, t) &=& e^{\gamma v^* t} e^{a x} \sinh \kappa x,\label{efunc_e}\\
f_\lambda^h(x, t) &=& \frac{1}{\beta d} \frac{v_e}{v_h} e^{\gamma v^* t} e^{a x} \left[\lambda \sinh \kappa x + \bar\kappa \cosh \kappa x \right].\label{efunc_h}
\eea
The ``eigenvalues'' $\lambda$ for which these solutions satisfy all boundary conditions must satisfy the equation
\beq
\lambda + \bar\kappa \coth \bar\kappa = 0,
\label{eigenvalue_equation}
\eeq
which depends only on the characteristic combination $\alpha \beta d^2$. We will refer to the corresponding solutions $f_\lambda^e$ and $f_\lambda^h$ as ``eigenfunctions''. The eigenvalue equation has a finite number of real-valued solutions, as well as an infinite number of complex-valued solutions, which appear in complex-conjugate pairs. The latter can be efficiently obtained by solving Equation \ref{eigenvalue_equation} numerically, using
\beq
\lambda_n = -\frac{1}{2} \log\left(\frac{(2n+1)^2 \pi^2}{\alpha\beta d^2} \right) + i \frac{\pi}{2} (2 n + 1)\qquad\mathrm{for}\qquad n\in\mathbb{Z},
\label{initial_values_eigenvalues}
\eeq
as initial values. The number of real-valued solution depends on the value of $\alpha\beta d^2$. Their behaviour is shown in Figure \ref{Fig:eigenvalues_real}. For the case $\alpha\beta d^2=3$, Figure \ref{Fig:eigenvalues_complex} pictures in addition the complex-valued solutions along with the approximate initial values from Equation \ref{initial_values_eigenvalues}.

\paragraph{Inner product and orthogonality of eigenfunctions}
We can construct an inner product that makes the eigenfunctions in Equations \ref{efunc_e} and \ref{efunc_h} orthogonal for different eigenvalues. For arbitrary functions $f^e(x)$, $f^h(x)$, $g^e(x)$ and $g^h(x)$ defined on $x\in [0, d]$, we introduce
\beq
\begin{bmatrix}
f^e\\f^h
\end{bmatrix}
\cdot
\begin{bmatrix}
g^e\\g^h
\end{bmatrix}
:=
\int_0^d \left[\alpha v_e f^e(d - x) g^e(x) + \beta v_h f^h(d - x) g^h(x) \right].
\label{eigenfunctions_inner_product_definition}
\eeq
As is shown in Appendix \ref{Sec:eigenfunctions_orthogonality}, the eigenfunctions for different eigenvalues are orthogonal (but not orthonormal) w.r.t.~this inner product,
\beq
\begin{bmatrix}
f_{\lambda}^e(\cdot, t)\\f_{\lambda}^h(\cdot, t)
\end{bmatrix}
\cdot
\begin{bmatrix}
f_{\lambda'}^e(\cdot, t)\\f_{\lambda'}^h(\cdot, t)
\end{bmatrix}
=
\mathcal{N}(\lambda, t)\delta_{\lambda, \lambda'}.
\label{eigenfunctions_orthogonality_property}
\eeq
The normalisation $\mathcal{N}(\lambda)$ is given by
\beq
\mathcal{N}(\lambda, t) = - e^{2 \gamma v^* t} e^{a d} \frac{\alpha d}{2 \bar\kappa} \frac{v_e}{v_h} (v_e + v_h) (1 + \lambda) \sinh\bar\kappa.
\label{normalisation_eigenfunctions}
\eeq

\paragraph{General solution of the initial value problem} The most general solution of the initial value problem posed by Equations \ref{n_e_second_order}-\ref{n_h_algeb_relation} is given by a linear combination of the eigenfunctions,
\beq
\avg{n_e(x,t)} = \sum_{\lambda} C(\lambda) f_\lambda^e(x, t) \qquad\qquad \avg{n_h(x,t)} = \sum_{\lambda} C(\lambda) f_\lambda^h(x, t) \label{general_solution_ne_nh}.
\eeq
The coefficients $C(\lambda)$ are determined by the initial condition. Making use of the orthogonality property in Equation \ref{eigenfunctions_orthogonality_property}, we can write them explicitly as
\beq
C(\lambda) = \frac{1}{\mathcal{N}(\lambda, t = 0)}
\begin{bmatrix}
  f_\lambda^e(\cdot, t=0)\\
  f_\lambda^h(\cdot, t=0)
\end{bmatrix}
\cdot
\begin{bmatrix}
  n_e^0\\
  n_h^0
\end{bmatrix}.
\label{coeffs_general}
\eeq

If we start with $N_e^0$ electrons and $N_h^0$ holes at the position $x=x_0$, i.e.~with $n_e^0(x) = N_e^0 \, \delta(x-x_0)$ and $n_h^0(x) = N_h^0 \, \delta(x-x_0)$, the coefficients become
\beq
C(\lambda) = \frac{1}{\mathcal{N}(\lambda, t = 0)}\left[\alpha v_e N_e^0 f_e^\lambda(d - x_0, t=0) + \beta v_h N_h^0 f_h^\lambda(d - x_0, t=0) \right].
\label{coeffs_eh}
\eeq
A similar formula for the coefficient $C(\lambda_1)$ is mentioned in \cite{average_avalanche_comments} for the special case $\alpha = \beta$ and $v_e = v_h$.

The eigenvalue with the largest real component, labelled as $\lambda_1$, is of particular significance for the behaviour of the avalanche at late times. If $\gamma(\lambda_1) = \gamma_1 > 0$ and $C(\lambda_1)\neq 0$, the average grows exponentially on average. The device in which the avalanche forms is then said to operate above breakdown. Below breakdown, i.e.~for $\gamma_1 < 0$, the total charge created by the avalanche (and therefore the gain) is finite. The conditions under which breakdown occurs are discussed in \cite{mcintyre_breakdown, oldham_breakdown}.

\paragraph{Induced current and late-time behaviour} If we assume a constant weighting field throughout the avalanche region, Equation \ref{current_random_variable} determines the average current induced on the readout electrodes of a particle detector. It is given by
\beq
\avg{\Iind(t)} = e_0 \frac{E_w}{V_w} \left[v_e \avg{N_e(t)} + v_h \avg{N_h(t)} \right].
\label{ramo_shockley_average_current}
\eeq
The average total numbers of electrons and holes present in the avalanche region at time $t$, $\avg{N_e(t)}$ and $\avg{N_h(t)}$, are given by
\bea
\avg{N_e(t)} &=& \int_0^d dx \, \avg{n_e(x,t)} = \sum_{\lambda} C(\lambda) \int_0^d dx\, f^e_\lambda(x, t) = \sum_{\lambda} N_e(\lambda) e^{\gamma v^* t},\nonumber\\
\avg{N_h(t)} &=& \int_0^d dx \, \avg{n_h(x,t)} = \sum_{\lambda} C(\lambda) \int_0^d dx\, f^h_\lambda(x, t) = \sum_{\lambda} N_h(\lambda) e^{\gamma v^* t}.\nonumber
\eea
The coefficients $N_e(\lambda)$ and $N_h(\lambda)$ are
\bea
N_e(\lambda) &=& C(\lambda) \frac{\kappa + e^{a d} \left(a \sinh \bar\kappa - \kappa \cosh \bar\kappa \right)}{a^2-\kappa^2}\nonumber\\
N_h(\lambda) &=& C(\lambda) \frac{1}{\beta d} \frac{v_e}{v_h} \frac{\kappa\lambda - a \bar\kappa - \kappa e^{a d}\left( \lambda \cosh \bar\kappa + \bar\kappa\sinh\bar\kappa \right)}{a^2-\kappa^2}.\nonumber
\eea

\begin{figure}[ht]
  \centering
  a)\quad\includegraphics[height=5.1cm]{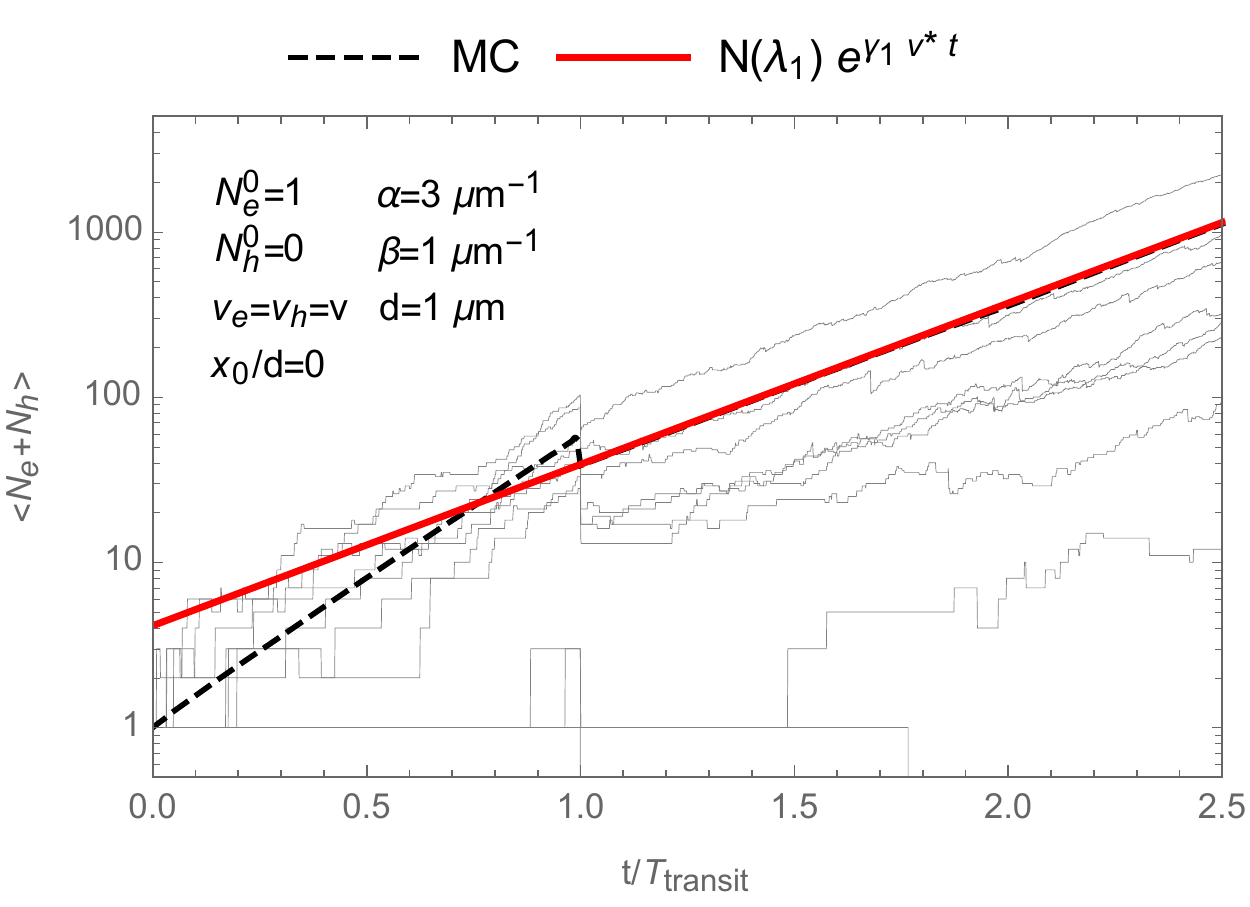}\quad
  b)\quad\includegraphics[height=5.1cm]{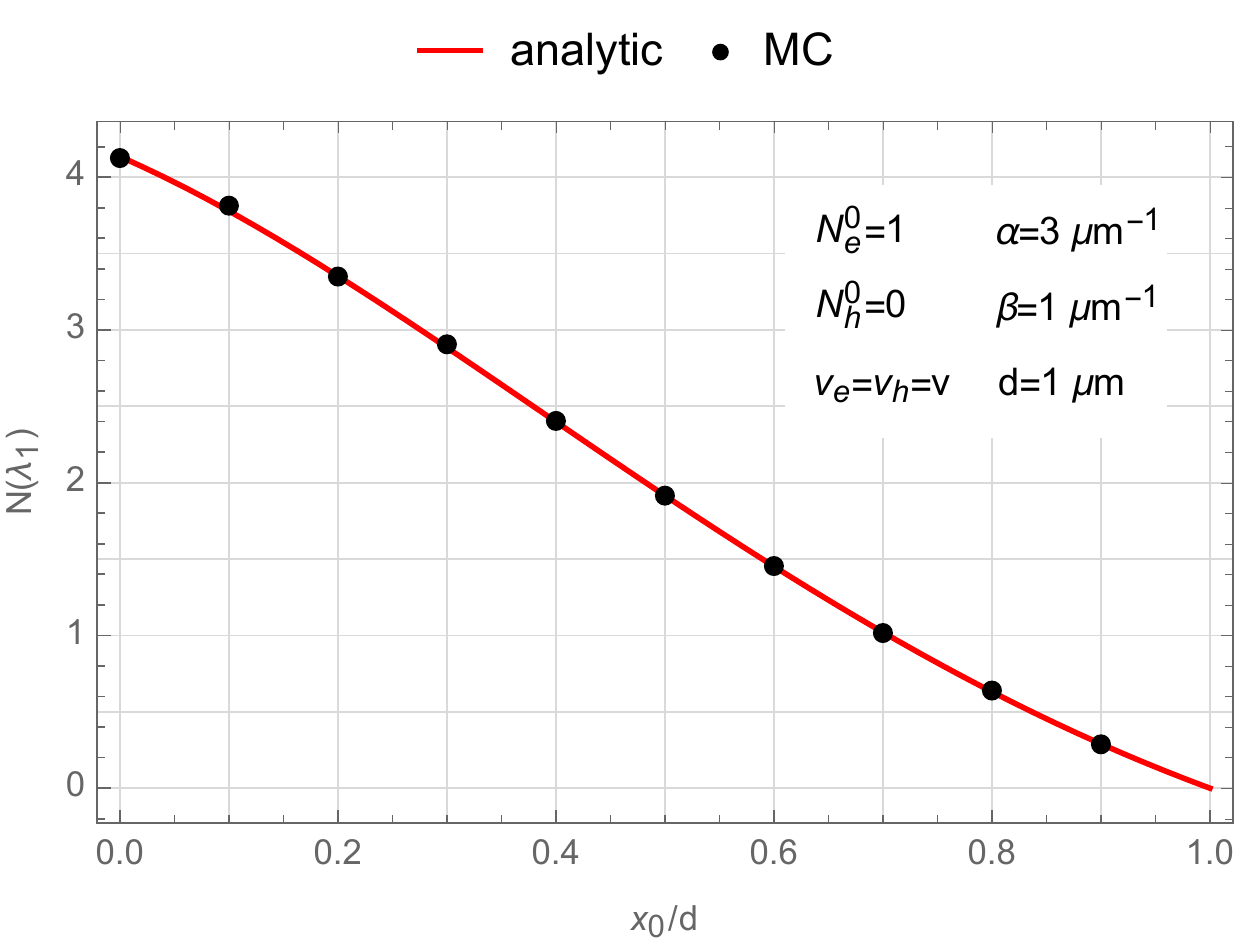}
  \caption{a) Evolution of the expected total number of charges $\avg{N_e + N_h}$ as a function of time for an avalanche initiated by a single electron placed at $x_0=0$. In the considered case where $v_e = v_h$, this quantity is directly proportional to the induced current. The dashed black line shows the average obtained from a MC simulation and the solid red line represents the fastest-growing component normalised according to the coefficient $N(\lambda_1)$. The thin grey lines correspond to individual avalanches. b) The coefficient $N(\lambda_1)$ as a function of the position $x_0$ of the avalanche-initiating electron.}
  \label{Fig:Ce0_comparison}
\end{figure}

For a device operated above breakdown, $N(\lambda_1) = N_e(\lambda_1) + N_h(\lambda_1)$ is related to the normalisation of the fastest-growing component that increases as $e^{\gamma_1 v^* t}$ and becomes dominant at late times. Figure \ref{Fig:Ce0_comparison} shows a comparison with results obtained from Monte Carlo (MC) simulations. The avalanche is initiated by a single electron placed at $x_0=0$. Already after one transit time ($\ttrans = d/v^*$), this approximation describes the average charge content with excellent precision. For a typical avalanche region with $d=1\,\mu$m that is implemented in silicon, the transit time is of the order of 10\,ps. Also visible in Figure \ref{Fig:Ce0_comparison}a is a change in the growth rate of the avalanche at $t = \ttrans$, as well as a drop in its average charge content. Both are caused by charges leaving the avalanche region across the boundary at $x=d$, which becomes visible only at $t = \ttrans$ for the chosen initial conditions.

\begin{figure}[ht]
  \centering
  \includegraphics[height=5.1cm]{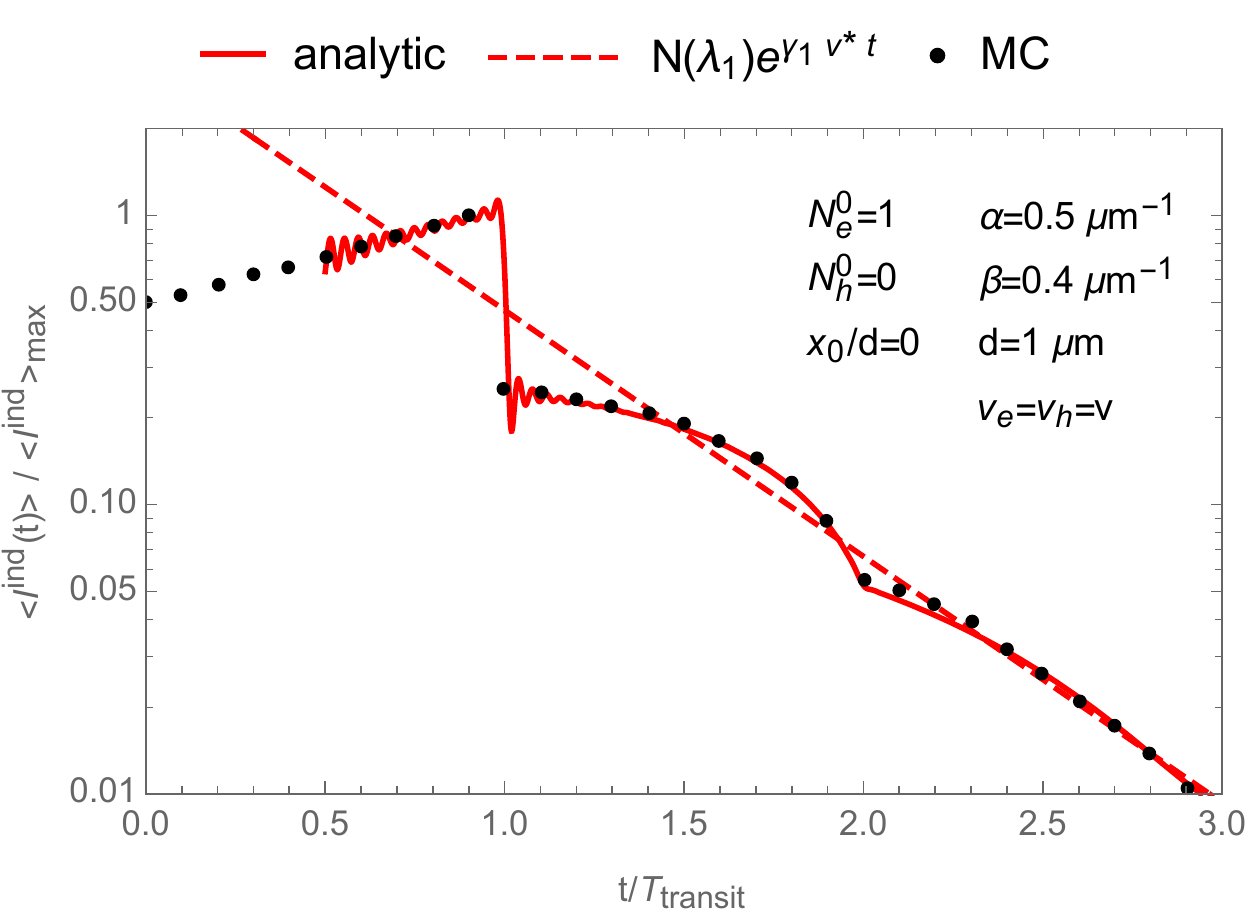}
  \caption{Average induced current $\avg{\Iind(t)}$ as given by Equation \ref{ramo_shockley_average_current}, scaled by its maximum, for a device operated below breakdown. The avalanche is initiated by a single electron placed at the left-hand side of the avalanche region. The black markers correspond to the current obtained from MC simulations, while the solid red line represents the analytical solution according to Equation \ref{ramo_shockley_average_current}, where the 50 most important eigenfunctions have been included. The asymptotic form of the current for late times, normalised by the coefficient $N(\lambda_1)$, is indicated by the dashed red line.}
  \label{Fig:APD_current}
\end{figure}

Below breakdown, $N(\lambda_1)$ can be used to compute the absolute size of the induced (decaying) current at late times, as is shown in Figure \ref{Fig:APD_current} for the same initial conditions. This is relevant for detectors such as avalanche photodiodes. In this regime, the predominant fraction of the total induced charge is generated within one transit time. Indeed, a significant fraction of all charge carriers leave the avalanche region at $t = \ttrans$. For $t\gg\ttrans$, the average induced current decays approximately exponentially with a time constant given by $\gamma_1 v^*$.

\subsection{Development of spatial correlations and fluctuations}
\label{Sec:development_spatial_correlations}
The spatial development of an electron-hole avalanche proceeds in a very different manner compared to an avalanche that is driven by electrons only. Since electrons and holes move in opposite directions, we can no longer expect the charge densities at different positions to be independent of each other (as is the case for an electron-only avalanche), but should rather anticipate nontrivial correlations. In turn, these will affect the fluctuations of the charge content of the avalanche, and thus of the induced current.

To study these, we look at the expectation values $\avg{n_e(x) n_e(y)}$, $\avg{n_h(x) n_h(y)}$ and $\avg{n_e(x) n_h(y)}$. As is shown in Appendix \ref{Sec:moments_equations_bounded}, they satisfy the following system of equations
\begin{multline}
\frac{\partial}{\partial t}\avg{n_e(x) n_e(y)} + \frac{\partial}{\partial x} v_e(x)\avg{n_e(x) n_e(y)} + \frac{\partial}{\partial y}v_e(y)\avg{n_e(x) n_e(y)} = \\
= \alpha(x) v_e(x) \delta(x-y) \avg{n_e(x)} + \beta(x) v_h(x) \delta(x-y) \avg{n_h(y)} + \\
+ \left[\alpha(x)v_e(x) + \alpha(y)v_e(y)\right] \avg{n_e(x) n_e(y)} + \beta(y) v_h(y) \avg{n_e(x) n_h(y)} + \beta(x) v_h(x) \avg{n_e(y) n_h(x)},\label{equations_second_moment_bounded_start}
\end{multline}
\begin{multline}
\frac{\partial}{\partial t}\avg{n_h(x) n_h(y)} - \frac{\partial}{\partial x}v_h(x)\avg{n_h(x) n_h(y)} - \frac{\partial}{\partial y}v_h(y)\avg{n_h(x) n_h(y)} = \\
= \alpha(x) v_e(x) \delta(x-y) \avg{n_e(y)} + \beta(x) v_h(x) \delta(x-y) \avg{n_h(y)} + \\
+ \left[\beta(x) v_h(x) + \beta(y) v_h(y)\right] \avg{n_h(x) n_h(y)} + \alpha(y) v_e(y) \avg{n_e(y) n_h(x)} + \alpha(x) v_e(x) \avg{n_e(x) n_h(y)},
\end{multline}
\begin{multline}
\frac{\partial}{\partial t}\avg{n_e(x) n_h(y)} + \frac{\partial}{\partial x}v_e(x)\avg{n_e(x) n_h(y)} - \frac{\partial}{\partial y}v_h(y)\avg{n_e(x) n_h(y)} = \\
= \alpha(x) v_e(x) \delta(x-y) \avg{n_e(y)} + \beta(x) v_h(x) \delta(x-y) \avg{n_h(y)} + \\
+ \alpha(y) v_e(y) \avg{n_e(x) n_e(y)} + \beta(x) v_h(x) \avg{n_h(x) n_h(y)} + \left[\alpha(x) v_e(x) + \beta(y) v_h(y)\right] \avg{n_e(x) n_h(y)}. \label{equations_second_moment_bounded_end}
\end{multline}
The relevant boundary conditions are $\avg{n_e(0,t) n_e(y, t)} = \avg{n_h(d, t) n_h(y, t)} = 0$ and $\avg{n_e(0, t) n_h(y, t)} \allowbreak = \allowbreak \avg{n_e(x, t) n_h(d, t)} \allowbreak = 0$ and the initial conditions are $\avg{n_e(x, 0) n_e(y, 0)} = n_e^0(x) n_e^0(y)$, $\avg{n_h(x, 0) n_h(y, 0)} = n_h^0(x) n_h^0(y)$ and $\avg{n_e(x, 0) n_h(y, 0)} = n_e^0(x) n_h^0(y)$.

These equations are the direct position-dependent analogues of the evolution equations for the moments $\avg{N_e^m N_h^n}$ encountered earlier: taking the Townsend coefficients and drift velocities to be constant and integrating over $x$ and $y$ immediately reproduces Equations \ref{equations_second_moment_unbounded_start}-\ref{equations_second_moment_unbounded_end}.

We are interested in computing the covariances
\beas
\cov[n_e(x), n_e(y)]&=&\avg{n_e(x)n_e(y)}-\avg{n_e(x)}\avg{n_e(y)},\\
\cov[n_h(x), n_h(y)]&=&\avg{n_h(x)n_h(y)}-\avg{n_h(x)}\avg{n_h(y)},\\
\cov[n_e(x), n_h(y)]&=&\avg{n_e(x)n_h(y)}-\avg{n_e(x)}\avg{n_h(y)},
\eeas
which encode information that is not already contained in the averages $\avg{n_e}$ and $\avg{n_h}$.

The diagonal components ($x=y$) of $\cov[n_e(x), n_e(y)]$ and $\cov[n_h(x), n_h(y)]$ correspond to the variances of the charge densities, i.e.~measure the magnitude of fluctuations at a certain location. Their off-diagonal entries ($x\neq y$) contain information about the correlations of the charge densities at different positions. Furthermore, we expect $\cov[n_e(x), n_h(y)]$ to be trivially nonzero if $x > y$: electron-hole pairs that are created at a common position at a time $t_c$ (and are thus positively correlated) move apart such that the electrons are always to the right of the holes for $t > t_c$. We thus make the ansatz
\bea
\cov[n_e(x), n_e(y)]&=&f_1^{ee}(x,t)\delta(x-y) + f_2^{ee}(x,y,t),\nonumber\\
\cov[n_h(x), n_h(y)]&=&f_1^{hh}(x,t)\delta(x-y) + f_2^{hh}(x,y,t),\nonumber\\
\cov[n_e(x), n_h(y)]&=&f_1^{eh}(x,y,t)\theta(x-y) + f_2^{eh}(x,y,t),\label{cov_eh_defn}
\eea
where the three functions $f_1^{ee}$, $f_1^{hh}$ and $f_1^{eh}$ make the expected features manifest. The remaining functions $f_2^{ee}$, $f_2^{hh}$ and $f_2^{eh}$ encapsulate additional, more indirect, effects. The functions $f_2^{ee}$ and $f_2^{hh}$ are symmetric in $x$ and $y$, while $f_2^{eh}$ does not have any symmetry property.

Since $n_e$ and $n_h$ are spatial densities, the functions $f_2^{ee}$ and $f_2^{hh}$ have dimensions of $L^{-2}$, while $f_1^{ee}$ and $f_1^{hh}$ have dimensions of $L^{-1}$. We have seen above that densities scale as $e^{\gamma_1 v^* t}$ for large avalanches. Purely on dimensional grounds, we therefore expect $f_2^{ee}$ and $f_2^{hh}$ to dominate at late times, while $f_1^{ee}$ and $f_1^{hh}$ are important at early times. As an explicit calculation shows, the same is in fact true for $f_1^{eh}$ and $f_2^{eh}$.

As a measure of the fluctuations of the avalanche around its average evolution, we consider again the ratios $\sigma(n_e(x))^2 / \avg{n_e(x)}^2$ and $\sigma(n_h(x))^2 / \avg{n_h(x)}^2$. These are the position-dependent versions of the relative fluctuations from Equation \ref{relfluc_late_times_unbounded}. Their limiting behaviour for large avalanches, i.e.~late times, will play an important role for the discussion of the time resolution in Section \ref{Sec:time_resolution_bounded} below. Making use of the scaling laws mentioned above, the relative fluctuations can be expressed entirely in terms of $f_2$ in this limit,
\beq
\frac{\sigma(n_e(x))^2}{\avg{n_e(x)}^2} = \frac{f_2^{ee}(x, x, t)}{\avg{n_e(x)}^2}, \qquad\qquad \frac{\sigma(n_h(x))^2}{\avg{n_h(x)}^2} = \frac{f_2^{hh}(x, x, t)}{\avg{n_h(x)}^2} \quad \text{as} \quad t\rightarrow\infty.
\label{local_relfluc_defn}
\eeq

To make progress towards this goal and to find evolution equations for the functions $f_1$ and $f_2$, we insert the above ansatz into the original Equations \ref{equations_second_moment_bounded_start}-\ref{equations_second_moment_bounded_end}. The functions $f_1^{ee}$, $f_1^{hh}$ follow the differential equations
\bea
\frac{\partial}{\partial t} f_1^{ee}(x,t) + \frac{\partial}{\partial x} v_e(x) f_1^{ee}(x,t) = \alpha(x) v_e(x) \avg{n_e(x,t)} + \beta(x) v_h(x) \avg{n_h(x)} + 2 \alpha(x) v_e(x) f_1^{ee}(x,t),\label{f1ee_equation}\\
\frac{\partial}{\partial t} f_1^{hh}(x,t) - \frac{\partial}{\partial x} v_h(x) f_1^{hh}(x,t) = \alpha(x) v_e(x) \avg{n_e(x,t)} + \beta(x) v_h(x) \avg{n_h(x)} + 2 \beta(x) v_h(x) f_1^{hh}(x,t),\label{f1hh_equation}
\eea
with the boundary conditions $f_1^{ee}(0, t) = f_1^{hh}(d, t) = 0$ and the initial conditions $f_1^{ee}(x, 0) = f_2^{hh}(x, 0) = 0$.

In the domain $x > y$, the function $f_1^{eh}$ is determined by
\beq
\frac{\partial}{\partial t} f_1^{eh}(x,y,t) + \frac{\partial}{\partial x} v_e(x) f_1^{eh}(x,y,t) - \frac{\partial}{\partial y} v_h(y) f_1^{eh}(x,y,t) = \left[\alpha(x) v_e(x) + \beta(y) v_h(y)\right] f_1^{eh}(x,y,t),\label{f1eh_equation}
\eeq
with the boundary condition
\beq
\left[v_e(x) + v_h(x)\right] f_1^{eh}(x,x,t) = \alpha(x) v_e(x) \avg{n_e(x)} + \beta(x) v_h(x) \avg{n_h(x)} + \alpha(x) v_e(x) f_1^{ee}(x,t) + \beta(x) v_h(x) f_1^{hh}(x,t)\label{f1eh_condition}
\eeq
and the initial condition $f_1^{eh}(x, y, 0)=0$.

Finally, the functions $f_2^{ee}$, $f_2^{hh}$ and $f_2^{eh}$ are governed by the following equations,
\begin{multline}
\frac{\partial}{\partial t} f_2^{ee}(x,y,t) + \frac{\partial}{\partial x} v_e(x) f_2^{ee}(x,y,t) + \frac{\partial}{\partial y} v_e(y) f_2^{ee}(x,y,t) = \left[\alpha(x) v_e(x) + \alpha(y) v_e(y)\right] f_2^{ee}(x,y,t) + \\
+ \beta(y) v_h(y) f_2^{eh}(x,y,t) + \beta(x) v_h(x) f_2^{eh}(y,x,t) + \left[ \beta(y) v_h(y) \theta(x-y) f_{1}^{eh}(x,y,t) + (x\leftrightarrow y)\right],\label{f2_system_start}
\end{multline}
\begin{multline}
\frac{\partial}{\partial t} f_2^{hh}(x,y,t) - \frac{\partial}{\partial x} v_h(x) f_2^{hh}(x,y,t) - \frac{\partial}{\partial y} v_h(y) f_2^{hh}(x,y,t) = \left[\beta(x) v_h(x) + \beta(y) v_h(y)\right] f_2^{hh}(x,y,t) + \\
+ \alpha(x) v_e(x) f_2^{eh}(x,y,t) + \alpha(y) v_e(y) f_2^{eh}(y,x,t) + \left[ \alpha(x) v_e(x) \theta(x-y) f_{1}^{eh}(x,y,t) + (x\leftrightarrow y) \right],\label{f2_system_middle}
\end{multline}
\begin{multline}
\frac{\partial}{\partial t} f_2^{eh}(x,y,t) + \frac{\partial}{\partial x} v_e(x) f_2^{eh}(x,y,t) - \frac{\partial}{\partial y} v_h(y) f_2^{eh}(x,y,t) = \alpha(y) v_e(y) f_2^{ee}(x,y,t) + \beta(x) v_h(x) f_2^{hh}(x,y,t) + \\
+ \left[\alpha(x) v_e(x) + \beta(y) v_h(y)\right] f_2^{eh}(x,y,t),\label{f2_system_end}
\end{multline}
with the boundary conditions $f_2^{ee}(x, 0, t) = f_2^{ee}(0, y, t) = 0$, $f_2^{hh}(x, d, t) = f_2^{hh}(d, y, t) = 0$, $f_2^{eh}(0, y, t) = f_2^{eh}(x, d, t) = 0$ and the initial conditions $f_2^{ee}(x, y, 0) = f_2^{hh}(x, y, 0) = f_2^{eh}(x, y, 0) = 0$. This is a system of three coupled differential equations, for which $f_1^{eh}$ plays the role of an external source.

Equations \ref{f1ee_equation}-\ref{f2_system_end} are equivalent to the original system in Equations \ref{equations_second_moment_bounded_start}-\ref{equations_second_moment_bounded_end}, but all participating functions and source terms are now continuous. For general field distributions $\mathbf{E}(x)$, they must be solved numerically. Below, we discuss the characteristics of their solutions for constant electric fields $\mathbf{E}$.

\subsubsection{Solution for constant electric field}

For position-independent Townsend coefficients and drift velocities, Equations \ref{f1ee_equation} and \ref{f1hh_equation} can be solved directly, and their solutions are
\beq
f_1^{ee}(x,t) = \int_0^t dt' \, e^{2\alpha v_e (t-t')} \left[\alpha v_e \avg{n_e(x-v_e(t-t'), t')} + \beta v_h \avg{n_h(x-v_e(t-t'),t')} \right],\label{f1ee_solution}
\eeq
and
\beq
f_1^{hh}(x,t) = \int_0^t dt' \, e^{2\beta v_h (t-t')} \left[\alpha v_e \avg{n_e(x+v_h(t-t'), t')} + \beta v_h \avg{n_h(x+v_h(t-t'),t')} \right].\label{f1hh_solution}
\eeq
These integrals can be computed analytically by inserting the expansions of $\avg{n_e}$ and $\avg{n_h}$ in terms of their eigenfunctions from Equation \ref{general_solution_ne_nh}. The resulting expressions are quite lengthy and listed in Appendix \ref{Sec:f1_analytic} for reference.

Equations \ref{f1eh_equation} and \ref{f1eh_condition} determine the function $f_1^{eh}$ only in the domain $x > y$. It is given by
\bea
f_1^{eh}(x,y,t) = \frac{1}{v_e + v_h} \exp\left( \frac{\alpha v_e + \beta v_h}{v_e + v_h} (x-y) \right) \left[ \alpha v_e \avg{n_e\left(\bar x, \tret \right)} + \beta v_h \avg{n_h\left(\bar x, \tret \right)} + \right. \nonumber \\ \left. + \alpha v_e f_1^{ee}\left( \bar x, \tret \right) + \beta v_h f_1^{hh}\left( \bar x, \tret \right) \right] \quad \text{for} \quad x > y.
\label{f1eh_analytic}
\eea
The retarded time $\tret$ and the position $\bar{x}$ are defined as
\beqs
\tret = t - \frac{x-y}{v_e + v_h},  \qquad\qquad  \bar x = \frac{v_h x + v_e y}{v_e + v_h}.
\eeqs
For constant Townsend coefficients and drift velocities, the source term entering into Equations \ref{f2_system_start} and \ref{f2_system_middle} for $f_2^{ee}$ and $f_2^{hh}$ can be expressed in terms of the function
\beqs
f_{1,\mathrm{sym}}^{eh}(x,y,t) = f_{1}^{eh}(x,y,t) \theta(x-y) + (x\leftrightarrow y).
\eeqs
It is the symmetric continuation of the function $f_1^{eh}$ into the region $x < y$.

The system of equations determining the functions $f_2^{ee}$, $f_2^{hh}$ and $f_2^{eh}$ deserves a bit more thought. First, we introduce the auxiliary function $f_2^{he}(x, y, t) = f_2^{eh}(y, x, t)$. Then, this system of equations is of the form
\beq
\left(\frac{\partial}{\partial t} + L_{x, y} \right) F_2(x, y, t) = g(x, y, t),
\label{original_inhomogeneous_system}
\eeq
where $F_2$ is the vector of functions $F_2 = (f_2^{ee}, f_2^{hh}, f_2^{eh}, f_2^{he})^T$ and $g$ is the source term built from $f_{1,\mathrm{sym}}^{eh}$, $g = (\beta v_h f_{1,\mathrm{sym}}^{eh}, \alpha v_e f_{1,\mathrm{sym}}^{eh}, 0, 0)^T$. $L_{x,y}$ is a linear operator that contains only spatial derivatives. If we define $\tilde F_2^{t'}(x, y, t)$ as the solution of the following homogeneous initial value problem,
\beq
\left(\frac{\partial}{\partial t} + L_{x, y} \right) \tilde F_2^{t'}(x, y, t) = 0, \qquad \tilde F_2^{t'}(x, y, t') = g(x, y, t'), \qquad t > t',\label{durhamel_initial_value_problem}
\eeq
then the solution of the original system in Equation \ref{original_inhomogeneous_system} (and therefore the solution for the functions $f_2$) is given by
\beq
F_2(x, y, t) = \int_0^t dt' \, \tilde F_2^{t'}(x, y, t).
\label{durhamel_convolution}
\eeq
It is easy to see that this is indeed a solution of the inhomogeneous system of equations. Acting on this expression with the operator $\partial_t + L_{x, y}$ yields, by Equation \ref{durhamel_initial_value_problem}
\beqs
\left(\frac{\partial}{\partial t} + L_{x, y}\right)\int_0^t dt' \, \tilde F_2^{t'}(x, y, t) = \tilde{F}_2^{t}(x, y, t) + \int_0^t dt' \, \underbrace{\left(\frac{\partial}{\partial t} + L_{x, y}\right) \tilde F_2^{t'}(x, y, t)}_{=0} = g(x, y, t).
\eeqs

\paragraph{Solution of the homogeneous system}
To be able to evaluate the integral in Equation \ref{durhamel_convolution}, we first need to solve the homogeneous system in Equation \ref{durhamel_initial_value_problem}. Written explicitly, it reads
\bea
\left(\frac{\partial}{\partial t} + v_e \frac{\partial}{\partial x} + v_e \frac{\partial}{\partial y} \right) f_2^{ee}(x,y,t) = 2 \alpha v_e f_2^{ee}(x,y,t) + \beta v_h f_2^{eh}(x,y,t) + \beta v_h f_2^{he}(x,y,t),\nonumber \\
\left(\frac{\partial}{\partial t} - v_h \frac{\partial}{\partial x} - v_h \frac{\partial}{\partial y} \right) f_2^{hh}(x,y,t) = 2 \beta v_h f_2^{hh}(x,y,t) + \alpha v_e f_2^{eh}(x,y,t) + \alpha v_e f_2^{he}(x,y,t),\nonumber\\
\left(\frac{\partial}{\partial t} + v_e \frac{\partial}{\partial x} - v_h \frac{\partial}{\partial y} \right) f_2^{eh}(x,y,t) = \alpha v_e f_2^{ee}(x,y,t) + \beta v_h f_2^{hh}(x,y,t) + (\alpha v_e+\beta v_h) f_2^{eh}(x,y,t),\nonumber\\
\left(\frac{\partial}{\partial t} - v_h \frac{\partial}{\partial x} + v_e \frac{\partial}{\partial y} \right) f_2^{he}(x,y,t) = \alpha v_e f_2^{ee}(x,y,t) + \beta v_h f_2^{hh}(x,y,t) + (\alpha v_e+\beta v_h) f_2^{he}(x,y,t).\nonumber
\eea
Inserting the ansatz
\bea
f_2^{ee}(x, y, t) = f_2^{e}(x, t) f_2^{e}(y, t), \quad f_2^{hh}(x, y, t) = f_2^{h}(x, t) f_2^{h}(y, t),\nonumber\\
f_2^{eh}(x, y, t) = f_2^{e}(x, t) f_2^{h}(y, t), \quad f_2^{he}(x, y, t) = f_2^{h}(x, t) f_2^{e}(y, t),\label{ansatz_f2_homogeneous}
\eea
we find that the two functions $f_2^{e}$ and $f_2^{h}$ must follow the equations,
\beqs
\frac{\partial f_2^{e}}{\partial t} + v_e \frac{\partial f_2^{e}}{\partial x} = \alpha v_e f_2^{e}(x,t) + \beta v_h f_2^{h}(x, t), \qquad \frac{\partial f_2^{h}}{\partial t} - v_h \frac{\partial f_2^{h}}{\partial x} = \alpha v_e f_2^{e}(x,t) + \beta v_h f_2^{h}(x, t).
\eeqs
These are identical to the equations for the expected densities $\avg{n_e}$ and $\avg{n_h}$ studied in Section \ref{Sec:first_moment_bounded}. Making use of the results found there, we can express the general solution for $f_2^{ee}$, $f_2^{hh}$ and $f_2^{eh}$ again as a linear combination of eigenfunctions. They are now indexed by two eigenvalues, $\lambda_a$ and $\lambda_b$. We define $F_{\lambda_a, \lambda_b} = (f^{ee}_{\lambda_a, \lambda_b}, f^{hh}_{\lambda_a, \lambda_b}, f^{eh}_{\lambda_a, \lambda_b}, f^{he}_{\lambda_a, \lambda_b})^T$ as the vector combining the eigenfunctions for the functions contained in $F_2$. Its entries are given in terms of the eigenfunctions $f^e_{\lambda}$ and $f^h_{\lambda}$ defined in Equations \ref{efunc_e} and \ref{efunc_h},
\beqs
f^{ee}_{\lambda_a, \lambda_b}(x, y, t) = f^{e}_{\lambda_a}(x, t) f^{e}_{\lambda_b}(y, t), \qquad f^{hh}_{\lambda_a, \lambda_b}(x, y, t) = f^{h}_{\lambda_a}(x, t) f^{h}_{\lambda_b}(y, t),
\eeqs
\beqs
f^{eh}_{\lambda_a, \lambda_b}(x, y, t) = f^{e}_{\lambda_a}(x, t) f^{h}_{\lambda_b}(y, t), \qquad f^{he}_{\lambda_a, \lambda_b}(x, y, t) = f^{h}_{\lambda_a}(x, t) f^{e}_{\lambda_b}(y, t).
\eeqs
The most general solution of the homogeneous system is then a linear combination of these eigenfunctions,
\beqs
\tilde{F}_2(x, y, t) = \sum_{\lambda_1, \lambda_2} \tilde{C}(\lambda_1, \lambda_2) F_{\lambda_1, \lambda_2}(x, y, t),
\eeqs
where the coefficients $\tilde{C}(\lambda_1, \lambda_2)$ are determined by the initial condition. We derive an explicit expression below for the problem in Equation \ref{durhamel_initial_value_problem}.

\paragraph{Inner product and orthogonality of eigenfunctions}
Analogously to before, we again define a convenient inner product. For two arbitrary vectors of functions $F = (f^{ee}, f^{hh}, f^{eh}, f^{he})^T$ and $G = (g^{ee}, g^{hh}, g^{eh}, g^{he})^T$, where each component function is defined on $(x,y)\in[0,d]\times[0,d]$, we introduce
\bea
F\cdot G
:=
\int_0^d dx \int_0^d dy \left[ \alpha^2 v_e^2 f^{ee}(d-x, d-y) g^{ee}(x, y) + \beta^2 v_h^2 f^{hh}(d-x, d-y) g^{hh}(x, y) + \right.\nonumber\\
  \left. + \alpha \beta v_e v_h f^{eh}(d-x, d-y) g^{eh}(x, y) + \alpha \beta v_e v_h f^{he}(d-x, d-y) g^{he}(x, y) \right].\nonumber
\eea
Following this definition, the inner product of two eigenfunctions $F_{\lambda_a, \lambda_b}$ and $F_{\lambda_a', \lambda_b'}$ for eigenvalues $\lambda_a, \lambda_b$ and $\lambda_a', \lambda_b'$ can be expressed in terms of the inner product defined in Equation \ref{eigenfunctions_inner_product_definition} as,
\beqs
{F}_{\lambda_a, \lambda_b}(t) \cdot
{F}_{\lambda_a', \lambda_b'}(t) =
\left(\begin{bmatrix} f^{e}_{\lambda_a} \\ f^{h}_{\lambda_a} \end{bmatrix} \cdot \begin{bmatrix} f^{e}_{\lambda_a'} \\ f^{h}_{\lambda_a'} \end{bmatrix}\right) \left(\begin{bmatrix} f^{e}_{\lambda_b} \\ f^{h}_{\lambda_b} \end{bmatrix} \cdot \begin{bmatrix} f^{e}_{\lambda_b'} \\ f^{h}_{\lambda_b'} \end{bmatrix}\right) = \mathcal{N}(\lambda_a, t) \mathcal{N}(\lambda_b, t) \delta_{\lambda_a,\lambda_a'} \delta_{\lambda_b,\lambda_b'}.
\eeqs
As before, this inner product makes eigenfunctions corresponding to different eigenvalues orthogonal to each other. The normalisation $\mathcal{N}(\lambda, t)$ is the one defined in Equation \ref{normalisation_eigenfunctions}.

\paragraph{Solution of the inhomogeneous system}
Making use of the orthogonality of the eigenfunctions, the coefficients $\tilde{C}(\lambda_a, \lambda_b, t')$ for the initial value problem in Equation \ref{durhamel_initial_value_problem} are thus given by
\bea
\tilde{C}(\lambda_a, \lambda_b, t') = \frac{1}{\mathcal{N}(\lambda_a, t')\mathcal{N}(\lambda_b, t')} \, {F}_{\lambda_a, \lambda_b}(t') \cdot \begin{bmatrix} \beta v_h f_{1,\mathrm{sym}}^{eh}(t') \\ \alpha v_e f_{1,\mathrm{sym}}^{eh}(t') \\ 0 \\ 0 \end{bmatrix}.
\label{coefficients_second_moment}
\eea
They combine with the eigenfunctions $F_{\lambda_a, \lambda_b}$ to produce the solution
\beqs
\tilde F_2^{t'}(x, y, t) = \sum_{\lambda_a, \lambda_b} \tilde{C}(\lambda_a, \lambda_b, t') F_{\lambda_a, \lambda_b}(x, y, t).
\eeqs
The full solution given by Equation \ref{durhamel_convolution} then reads
\beq
F_2(x, y, t) = \int_0^t dt' \, \tilde F_2^{t'}(x, y, t) = \sum_{\lambda_a, \lambda_b} \underbrace{\left(\int_0^t dt' \, \tilde{C}(\lambda_a, \lambda_b, t')\right)}_{C(\lambda_a, \lambda_b, t)} F_{\lambda_a, \lambda_b}(x, y, t). \label{F2_full_solution}
\eeq
We see that the solution of the inhomogeneous system has the same character as the solution of the homogeneous system, except that the coefficients multiplying the eigenfunctions now explicitly depend on time.

\paragraph{Late-time behaviour}
The behaviour of the solution $F_2$ as $t\rightarrow\infty$ is of particular interest, as it contains information about the asymptotic level of fluctuations of the avalanche. For late times, the fastest-growing component $F_{\lambda_1, \lambda_1}$ becomes dominant.
The corresponding coefficient $C(\lambda_1, \lambda_1, t)$ can be read off from Equation \ref{F2_full_solution}. It approaches a constant $C_{\infty}(\lambda_1, \lambda_1)$ like
\beqs
C(\lambda_1, \lambda_1, t) \sim C_{\infty}(\lambda_1, \lambda_1)\left( 1 - e^{-\gamma_1 v^* t} \right) \quad \text{as} \quad t\rightarrow\infty,
\eeqs
where the saturation timescale $T_{\mathrm{sat}}$ is now given by $T_{\mathrm{sat}}=(\gamma_1 v^*)^{-1}$, in complete analogy to the result obtained below Equation \ref{relfluc_late_times_unbounded}.

This shows that avalanche fluctuations at early times (and their interactions with the boundaries) are important to be able understand the behaviour of the avalanche at late times. Intuitively, this fact can already be appreciated from Figure \ref{Fig:practical_situation}b. The constant $C_{\infty}$ can be obtained from Equation \ref{coefficients_second_moment} as
\bea
C_{\infty}(\lambda_1, \lambda_1) = 2 \alpha \beta v_e v_h \int_0^{\infty} dt'\, \frac{1}{\mathcal{N}(\lambda_1, t')^2} \int_0^d dx \int_0^x dy \left[\alpha v_e f_{\lambda_1}^e(d-x,t') f_{\lambda_1}^e(d-y,t') + \right. \nonumber \\ \left. + \beta v_h f_{\lambda_1}^h(d-x,t') f_{\lambda_1}^h(d-y,t') \right] f_1^{eh}(x, y, t').
\label{C_infty_numerical_integral}
\eea
In this form, it is well-suited for numerical evaluation, where $f_1^{eh}$ is taken from Equation \ref{f1eh_analytic} and $\mathcal{N}$ from Equation \ref{normalisation_eigenfunctions}. This integral also exists analytically, but the explicit expression is too lengthy to be of much practical use.

The late-time solution, valid on timescales that are long compared to $T_{\mathrm{sat}}$, thus reads
\beq
F_2(x, y, t) \approx C_{\infty}(\lambda_1, \lambda_1) F_{\lambda_1, \lambda_1}(x, y, t) \quad \text{as} \quad t\rightarrow\infty.
\label{F2_late_time_solution}
\eeq
As mentioned below Equation \ref{cov_eh_defn}, the covariances are given entirely in terms of the functions $f_2$ in this limit,
\bea
\cov[n_e(x), n_e(y)] &\approx& f_2^{ee}(x, y, t) \quad \text{as}\quad t\rightarrow\infty,\nonumber\\
\cov[n_h(x), n_h(y)] &\approx& f_2^{hh}(x, y, t) \quad \text{as}\quad t\rightarrow\infty,\nonumber\\
\cov[n_e(x), n_h(y)] &\approx& f_2^{eh}(x, y, t) \quad \text{as}\quad t\rightarrow\infty.\nonumber
\eea
Keeping in mind that $f_2^{ee}(x, y, t) \approx C_{\infty}(\lambda_1, \lambda_1) f^e_{\lambda_1}(x, t) f^e_{\lambda_1}(y, t)$ at late times, we see that
\beq
\lim_{t\rightarrow\infty} \frac{\cov[n_e(x), n_e(y)]}{\sigma(n_e(x))\sigma(n_e(y))} = \lim_{t\rightarrow\infty} \frac{f_2^{ee}(x, y, t)}{\sqrt{f_2^{ee}(x, x, t)}\sqrt{f_2^{ee}(y, y, t)}}=1.
\label{correlation_ee}
\eeq
Similarly, we find that
\beq
\lim_{t\rightarrow\infty} \frac{\cov[n_h(x), n_h(y)]}{\sigma(n_h(x))\sigma(n_h(y))} = \lim_{t\rightarrow\infty} \frac{f_2^{hh}(x, y, t)}{\sqrt{f_2^{hh}(x, x, t)}\sqrt{f_2^{hh}(y, y, t)}}=1,
\label{correlation_hh}
\eeq
and
\beq
\lim_{t\rightarrow\infty} \frac{\cov[n_e(x), n_h(y)]}{\sigma(n_e(x))\sigma(n_h(y))} = \lim_{t\rightarrow\infty} \frac{f_2^{eh}(x, y, t)}{\sqrt{f_2^{ee}(x, x, t)}\sqrt{f_2^{hh}(y, y, t)}}=1.
\label{correlation_eh}
\eeq

The above results have important implications. Equations \ref{correlation_ee}-\ref{correlation_eh} correspond to the Pearson correlations between electron- and hole densities at positions $x$ and $y$: at late times, the charge distributions inside the avalanche region become maximally correlated across space. This also implies that the absolute numbers of electrons and holes become maximally Pearson-correlated at late times,
\beqs
\lim_{t\rightarrow\infty} \rho(N_e, N_h) = \lim_{t\rightarrow\infty} \frac{\cov[N_e, N_h]}{\sigma(N_e)\sigma(N_h)} = 1.
\eeqs

By the same argument, we see that the relative fluctuations of the charge densities defined in Equation \ref{local_relfluc_defn} become independent of the position $x$ at late times, and identical to the relative fluctuation of the total charge content,
\beqs
\lim_{t\rightarrow\infty} \frac{\sigma(n_e(x))^2}{\avg{n_e(x)}^2} = \lim_{t\rightarrow\infty} \frac{\sigma(N_e)^2}{\avg{N_e}^2} = \lim_{t\rightarrow\infty} \frac{f_2^{ee}(x, x, t)}{\avg{n_e(x,t)}^2}, \qquad \lim_{t\rightarrow\infty} \frac{\sigma(n_h(x))^2}{\avg{n_h(x)}^2} = \lim_{t\rightarrow\infty} \frac{\sigma(N_h)^2}{\avg{N_h}^2} = \lim_{t\rightarrow\infty} \frac{f_2^{hh}(x, x, t)}{\avg{n_h(x,t)}^2}.
\eeqs
This demonstrates again that, at late times, the only degree of freedom of the charge densities $n_e(x)$ and $n_h(x)$ lies in their overall normalisations: once maximal spatial correlation is achieved, the fluctuations become identical for all $x$. The calculation also shows that the relative fluctuations for electrons and holes in fact become identical at late times,
\beq
\lim_{t\rightarrow\infty} \frac{\sigma(N)^2}{\avg{N}^2} = \lim_{t\rightarrow\infty} \frac{\sigma(N_e)^2}{\avg{N_e}^2} = \lim_{t\rightarrow\infty} \frac{f_2^{ee}(x, x, t)}{\avg{n_e(x,t)}^2} = \lim_{t\rightarrow\infty} \frac{\sigma(N_h)^2}{\avg{N_h}^2} = \lim_{t\rightarrow\infty} \frac{f_2^{hh}(x, x, t)}{\avg{n_h(x,t)}^2}.
\label{relfluc_bounded}
\eeq
This is in direct correspondence to Equation \ref{relfluc_late_times_unbounded} for the case of an avalanche developing in an infinitely large semiconductor. For a constant electric field, the relative fluctuation is conveniently computed as
\beq
\lim_{t\rightarrow\infty} \frac{\sigma(N)^2}{\avg{N}^2} = \frac{C_{\infty}(\lambda_1, \lambda_1)}{C(\lambda_1)^2},
\eeq
where $C_\infty(\lambda_1, \lambda_1)$ is taken from Equation \ref{C_infty_numerical_integral} and $C(\lambda_1)$ from Equation \ref{coeffs_general}. 

The total instantaneous charge content of the avalanche, and thus its relative fluctuation, is not generally an experimentally accessible quantity. Of direct relevance for the achievable time resolution is the relative fluctuation of the induced current $\Iind(t)$ at late times, i.e.~the ratio $\lim_{t\rightarrow\infty} \sigma(\Iind(t))^2 / \avg{\Iind(t)}^2$. Taking the form of the current from Equation \ref{current_random_variable} and making use of Equation \ref{relfluc_bounded}, we find that it again limits to
\beq
\lim_{t\rightarrow\infty} \frac{\sigma(\Iind(t))^2}{\avg{\Iind(t)}^2} = \lim_{t\rightarrow\infty} \frac{\sigma(N)^2}{\avg{N}^2}.
\label{relfluc_current_bounded}
\eeq

Note that the strategy for the solution of Equations \ref{f2_system_start}-\ref{f2_system_end} and in particular the ansatz in Equation \ref{ansatz_f2_homogeneous} also works in the general case of position-dependent Townsend coefficients and drift velocities. For late times, the position-dependence of $f_2^{ee}$, $f_2^{hh}$ and $f_2^{eh}$ thus continues to be given by products of the eigenfunctions of Equations \ref{equations_first_moment_bounded_start}-\ref{equations_first_moment_bounded_end} (which must now be found numerically). Maximal correlation occurs regardless of the shape of the field profile $\mathbf{E}(x)$. Also Equations \ref{relfluc_bounded} and \ref{relfluc_current_bounded} continue to hold in the general case and allow to compute the relative fluctuations from the numerical solutions for $f_2^{ee}$ and $f_2^{hh}$.

\subsection{Time resolution}
\label{Sec:time_resolution_bounded}
We focus on the discussion of the time resolution for large thresholds placed on the induced current $\Iind(t)$, which is the case of practical relevance. Our discussion here parallels the treatment given in Section \ref{Sec:timeres_large_th} for infinitely large semiconductors and constant electric fields.

Before quantifying the time resolution itself, it is worthwhile to point out a conceptual difference to the case studied there. An avalanche developing in a bounded semiconductor will generally diverge only with a finite probability (or ``efficiency'') $\epsilon < 1$. Correspondingly, a fraction $1-\epsilon$ of all avalanches will end once all charges have escaped across the boundaries. The (generally unknown) probability distribution $p(N_e, N_h, t)$ for this case has the structure
\beq
p(N_e, N_h, t) = \left[1-\epsilon(t)\right] \, \delta_{N_e, 0} \delta_{N_h, 0} + \epsilon(t) \, p_S(N_e, N_h, t),
\label{general_pNeNh_efficiency}
\eeq
where $1 - \epsilon(t)$ is the fraction of all avalanches that have already ended at time $t$, and the distribution $p_S(N_e, N_h, t)$ describes the ``starters'', i.e.~those avalanches that have not (yet) ended. Once an avalanche has ended, it can never diverge again, i.e.~$\epsilon(t)$ is monotonically falling with $t$. Its limiting value $\epsilon = \lim_{t\rightarrow\infty} \epsilon(t)$ can be easily computed as shown in \cite{mcintyre_breakdown, oldham_breakdown}. For our purposes, the efficiency $\epsilon$ is merely an external input parameter like the Townsend coefficients $\alpha$ and $\beta$. It depends on the applied electric field and on the position $x_0$ of the charge carrier initiating the avalanche.

For the computation of the time resolution, therefore, we must restrict all expectation values to only include those avalanches that actually diverge (and therefore cross the threshold). We use a subscript $S$ to indicate that only the starters are considered. In complete analogy to the steps leading to Equation \ref{time_resolution_asymptotic_unbounded}, the time resolution for large thresholds is given by
\beq
\sthinf = \lim_{t\rightarrow\infty} \frac{\sigma(\log \Iind(t))_S}{\gamma_1 v^*}.
\eeq
However, no closed system of evolution equations exists for the exact evaluation of $\sigma(\log \Iind)_S$, even in the case of a constant electric field (see the discussion in Appendices \ref{Sec:asymptotic_logN_unbounded} and \ref{Sec:asymptotic_logN_bounded}). Instead, we establish an approximate expression for $\sthinf$, building on the intuition gained in Section \ref{Sec:results_unbounded}.

We saw in Section \ref{Sec:moments_unbounded} that in the simple case of an infinitely extended semiconductor and a constant electric field, the avalanche parameter $A$ is directly related to the relative fluctuations through $\lim_{t\rightarrow\infty} \sigma^2(N)/\avg{N}^2 = A^{-1}$. For this case, Equation \ref{logN_fluctuations_unbounded} then establishes a direct link between the relative fluctuation of the quantity on which a threshold is placed, and the corresponding time resolution. The exact form of this relationship depends on the shape of the underlying probability distribution $p(N)$.

For the general case considered here, we do not have access to the shape of this distribution, and so the time resolution can not be directly connected to the relative fluctuation of the current, $\sigma(\Iind)^2 / \avg{\Iind}^2$, as computed in Section \ref{Sec:development_spatial_correlations}. However, under the approximation that this relationship is in fact similar to that derived in Section \ref{Sec:moments_unbounded}, the time resolution for large thresholds is given by
\beq
\sthinf \approx \frac{\sqrt{\psi_1(\Aeff)}}{\gamma_1 v^*},
\label{time_resolution_approx_bounded}
\eeq
where the effective avalanche parameter $\Aeff$ is defined as
\beqs
\frac{1}{\Aeff} := \lim_{t\rightarrow\infty} \frac{\sigma(\Iind)^2_S}{\avg{\Iind}_S^2} = \lim_{t\rightarrow\infty} \frac{\sigma(N)^2_S}{\avg{N}_S^2}.
\eeqs

To evaluate this expression further, we note the following consequence of Equation \ref{general_pNeNh_efficiency}. The expectation values $\avg{\mathcal{O}}_S$ and $\avg{\mathcal{O}}$ are related through
\beqs
\avg{\mathcal{O}}_S = \sum_{N_e, N_h} \mathcal{O}(N_e, N_h) \, p_S(N_e, N_h, t) = \frac{1}{\epsilon(t)} \sum_{N_e, N_h} \mathcal{O}(N_e, N_h) \, p(N_e, N_h, t) = \frac{1}{\epsilon(t)} \avg{\mathcal{O}}
\eeqs
for any observable $\mathcal{O}$ that is homogeneous in $N_e$ and $N_h$, i.e.~for which $N_e = N_h = 0 \implies \mathcal{O}(N_e, N_h) = 0$.
This is true in particular for $\avg{N}$ and $\avg{N^2}$, and so, for late times, $\avg{N} = \epsilon \avg{N}_S$ and $\avg{N^2} = \epsilon \avg{N^2}_S$. The effective avalanche parameter can therefore be computed in terms of unconditional averages as
\beq
\frac{1}{\Aeff} = \lim_{t\rightarrow\infty} \epsilon \left(1 + \frac{\sigma(N)^2}{\avg{N}^2}\right) - 1 = \epsilon \left(1 + \frac{C_{\infty}(\lambda_1, \lambda_1)}{C(\lambda_1)^2} \right) - 1,
\eeq
where the last equality is valid only for constant electric fields. For the general case, $\sigma(N)^2/\avg{N}^2$ can be taken from the numerical solution as explained in Equation \ref{relfluc_bounded}.

Inspecting the expression for the time resolution in Equation \ref{time_resolution_approx_bounded}, we see that its overall scale is set by the inverse of the average asymptotic growth rate of the avalanche, $\gamma_1 v^*$. It can be computed from Equation \ref{gamma_def} and Figure \ref{Fig:eigenvalues_real} and generally accounts for the predominant part of the material- and field-dependence of the time resolution.

\begin{figure}[p]
  \centering
  \includegraphics[width=11cm]{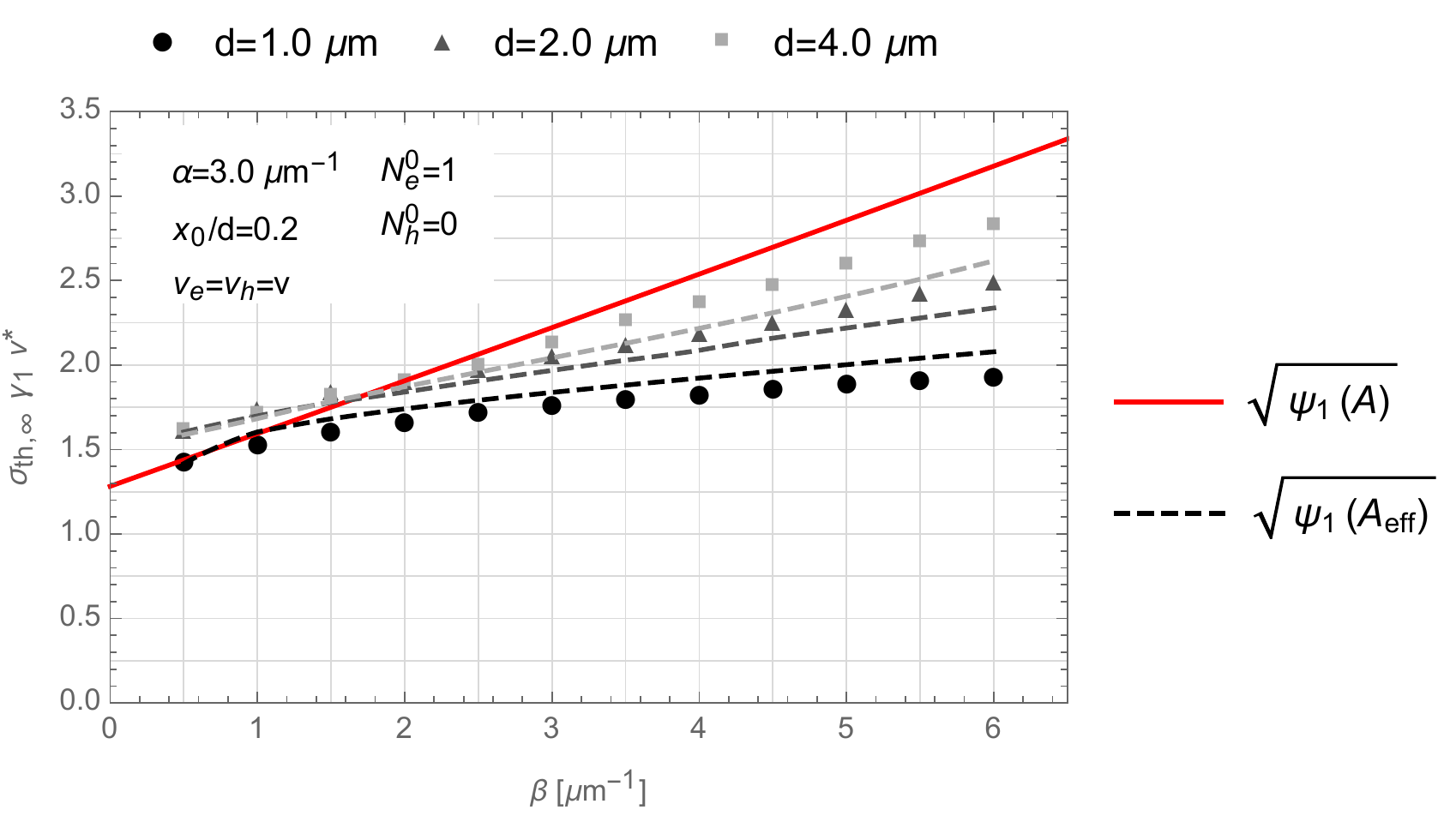}
  \caption{The black and grey markers show the quantity $\sthinf \gamma_1 v^*$ obtained from MC simulations as a function of the Townsend coefficient $\beta$ and the thickness $d$ of the semiconductor. The avalanche is initiated by a single electron placed at $x_0 / d = 0.2$. The dashed lines correspond to the approximation from Equation \ref{time_resolution_approx_bounded}. Neglecting the finite size of the semiconductor and setting $\Aeff = A$ leads to the solid red line.}
  \label{Fig:MC_timeres_beta_sweep}
\end{figure}

\begin{figure}[p]
  \centering
  \includegraphics[width=11cm]{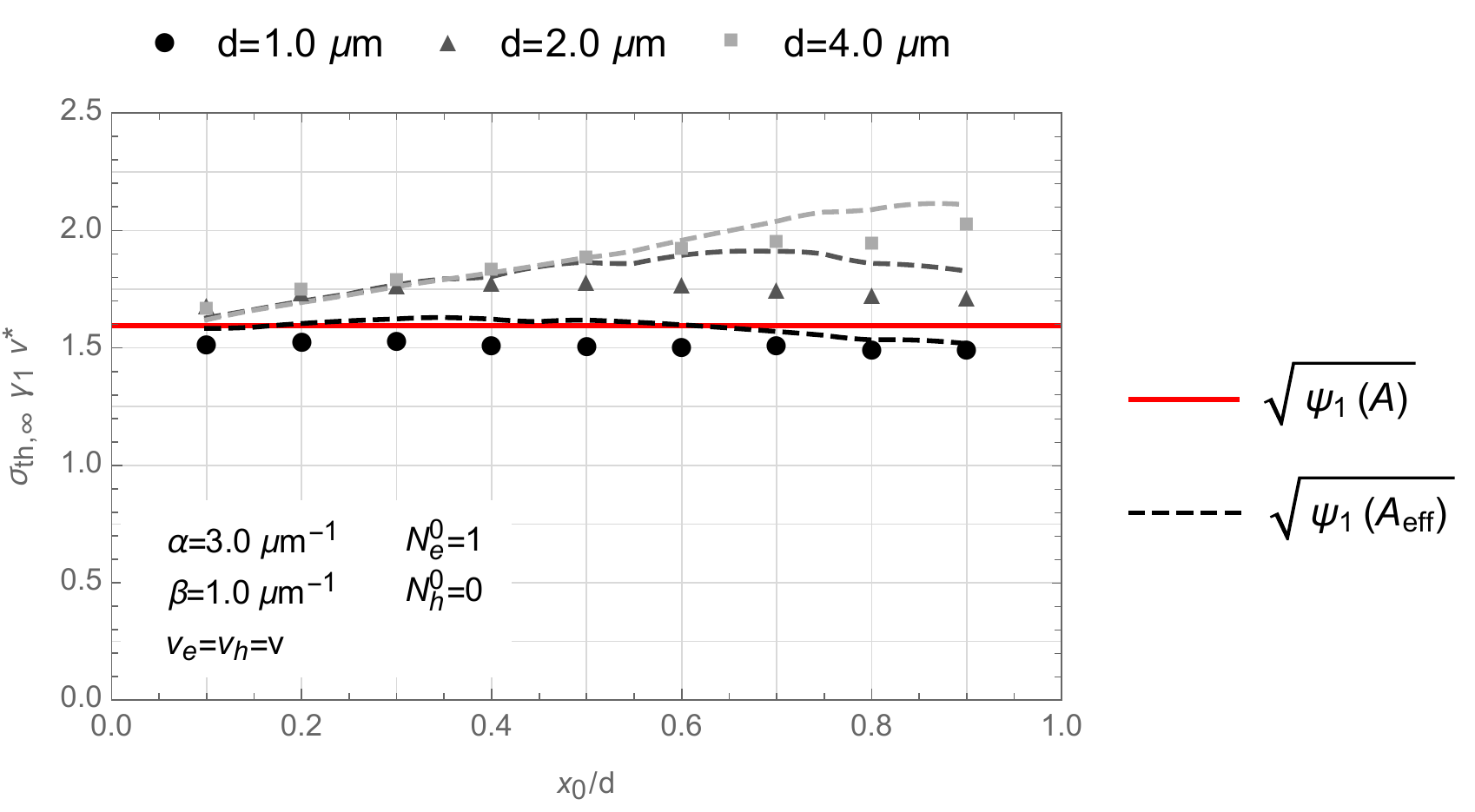}
  \caption{The black and grey markers show the quantity $\sthinf \gamma_1 v^*$ obtained from MC simulations as a function of the position $x_0$ of the initiating electron and the thickness $d$ of the semiconductor. The Townsend coefficients are fixed to $\alpha = 3\,\mathrm{\mu m}^{-1}$ and $\beta = 1\,\mathrm{\mu m}^{-1}$. The dashed lines correspond to the approximation from Equation \ref{time_resolution_approx_bounded}. Neglecting the finite size of the semiconductor and setting $\Aeff = A$ leads to the solid red line.}
  \label{Fig:MC_timeres_x0_sweep}
\end{figure}

Scaling out this trivial dependency, the dimensionless quantity $\sthinf \gamma_1 v^*$ measures the effect of avalanche fluctuations on the time resolution. Figure \ref{Fig:MC_timeres_beta_sweep} summarises the results from a series of MC simulations that show the evolution of this coefficient with the Townsend parameter $\beta$ and the thickness $d$ of the semiconductor.
Its dependence on the material properties is usually rather weak. The approximation made in Equation \ref{time_resolution_approx_bounded}, $\sthinf \gamma_1 v^* \approx \sqrt{\psi_1(\Aeff)}$, is accurate to within about 10\% across a wide range of the ratio $\alpha / \beta$. For avalanches that are initiated by an electron, it becomes exact in the limit $\alpha \gg \beta$. This case is relevant e.g.~for detectors based on silicon.
As shown in Figure \ref{Fig:MC_timeres_x0_sweep}, $\sthinf \gamma_1 v^*$ depends only weakly on the position $x_0$ of the charge that triggers the avalanche. Variations are limited to about $30\%$ as $x_0$ traverses the medium.

Neglecting the finite thickness of the semiconductor altogether, i.e.~taking $\Aeff = A$ generally reproduces the correct scaling of $\sthinf \gamma_1 v^*$ with $\beta$. Of course, this approximation does not capture the residual dependence on the thickness $d$ and the initial position $x_0$.

\section{Conclusions}

We have given a comprehensive description of electron-hole avalanches developing in a thin semiconductor that is exposed to a strong electric field, a structure that is commonly used for the amplification and detection of a small amount of initial charge.

Starting from a series of differential equations, we describe the average evolution of the avalanche as well as its fluctuations around this average. For general position-dependent electric fields, these equations can be efficiently solved with numerical methods. For constant fields, we give their analytical solutions, including compact expressions for the average current induced on the readout electrodes of a particle detector.

We also study the achievable time resolution for devices that operate above the breakdown limit, e.g.~SPADs and SiPMs. The time resolution depends on the average growth rate of the avalanche as well as a numerical coefficient of order unity which encodes fluctuations of the avalanche. The latter can be approximated with an accuracy of about 10\% using our methods and shows only a relatively weak dependence on the strength of the electric field and the thickness of the semiconductor.


\section{Acknowledgements}
PW acknowledges support by the Science and Technology Facilities Council (STFC) under grant ST/S505638/1 and through a Buckee scholarship awarded by Merton College, Oxford.

\FloatBarrier

\begin{appendices}

\section{Avalanches as stochastic many-body systems}
\label{Sec:formalism}

Under the assumptions detailed in Section \ref{Sec:assumptions}, the charge avalanche turns into a Markov process. That is, a full description of the avalanche at a certain time $t_0$ (the current ``state'' of the avalanche) contains enough information to allow for the computation of the state of the avalanche at any future time $t > t_0$.

A charge avalanche evolves stochastically, and so a complete description of its state is equivalent to a specification of the probability $p(\mathcal{A})$ with which a certain avalanche configuration $\mathcal{A}$ occurs. By ``avalanche configuration'' we mean a sufficiently detailed description of a concrete realisation of an avalanche. In practice, this can take various forms. In the most general (and most complex) case, $\mathcal{A}$ contains the spatial positions of all participating charges (or, equivalently, their spatial densities). In many situations, however, merely keeping track of the total number of charges is sufficient. We will consider both of these situations in what follows.

The probability $p(\mathcal{A})$ will generally change with time: we write $p(\mathcal{A}, t)$. For a Markov process, the equation describing this time evolution is of the form
\beq
\frac{d}{dt} p(\mathcal{A}, t) = \sum_{\mathcal{A}'} \left[ T(\mathcal{A}'\rightarrow\mathcal{A}) \, p(\mathcal{A}', t) - T(\mathcal{A}\rightarrow\mathcal{A'}) \, p(\mathcal{A}, t) \right].
\label{master_equation_general}
\eeq

The two terms on the right-hand side correspond to the two ways in which the probability $p(\mathcal{A}, t)$ can change in the small time interval $[t, t+dt]$. First, an avalanche with a configuration $\mathcal{A}'$ can evolve into $\mathcal{A}$, thereby increasing $p(\mathcal{A}, t)$ by an amount proportional to $p(\mathcal{A}', t)$. This is captured by the first (positive) term in Equation \ref{master_equation_general}, where $T(\mathcal{A}'\rightarrow\mathcal{A})$ labels the rate at which this transition occurs. The second (negative) term represents avalanches that are evolving away from the configuration $\mathcal{A}$ into another configuration $\mathcal{A}'$, thereby reducing $p(\mathcal{A}, t)$ by an amount proportional to itself.

The set of all transition rates $T(\mathcal{A}\rightarrow\mathcal{A'})$ are collectively referred to as the ``transition matrix''. The transition matrix elements encode the dynamics of the avalanche process. However, writing them down explicitly can be very complicated, in particular if the description of the avalanche configuration $\mathcal{A}$ is very detailed. It is therefore important to set up an efficient bookkeeping mechanism to render the calculations manageable.

In the remainder of this Appendix, we introduce such a formalism, based on the occupation-number representation familiar from quantum mechanics \cite{stochastic_mechanics_1, stochastic_mechanics_2, stochastic_mechanics_3, quantum_techniques}. Note that this does in no way imply that we treat the avalanche as a quantum system. Rather the opposite: the way in which many-body quantum mechanics is conventionally formulated provides a very convenient framework for the description of a general Markov process.

Section \ref{Sec:formalism_0D} applies these ideas to the case of an electron-hole avalanche evolving in an unbounded semiconductor that is exposed to a constant electric field. Section \ref{Sec:formalism_1D} then treats the significantly more complicated case of a thin semiconductor and a general position-dependent electric field.

\subsection{Electron-hole avalanche in an infinite semiconductor with a constant electric field}
\label{Sec:formalism_0D}
For a position-independent electric field $\mathbf{E}$, the Townsend coefficients $\alpha$ and $\beta$ and the drift velocities $v_e$ and $v_h$ become constants. Then, the spatial distribution of the charge carriers in the avalanche is irrelevant and the avalanche configuration is fully specified by the total numbers of electrons ($N_e$) and holes ($N_h$) in the medium.

\subsubsection{Avalanche configurations and state vectors}
As in many-body quantum mechanics, we represent such an avalanche configuration by a vector in a Fock space that we denote as $\ket{\mathcal{A}} = \ket{N_e, N_h}$. The ``empty'' avalanche that does not contain any charges is consequently written as $\ket{0, 0} = \ket{0}$. These vectors serve as a basis, from which more complicated states can be assembled.

To describe the evolution of an avalanche, we routinely need to create basis vectors with different electron and hole contents, e.g.~$\ket{\mathcal{A}} = \ket{N_e, N_h} \rightarrow \ket{\mathcal{A}'} = \ket{N_e', N_h'}$.
To this end, we introduce creation and annihilation operators for electrons ($\cre$, $\ane$) and holes ($\crh$, $\anh$) that act on these states in the way familiar from quantum mechanics. The defining properties of these operators are encoded in their commutation relations (for two operators $\hat{A}$ and $\hat{B}$, their commutator is $[\hat{A}, \hat{B}] := \hat{A}\hat{B} - \hat{B}\hat{A}$). They read
\beq
    [\cre, \crh] = [\ane, \anh] = [\ane, \crh] = [\anh, \cre] = 0,
    \label{commutation_relations_cran_1}
\eeq
and
\beq
    [\ane, \cre] = [\anh, \crh] = 1.
    \label{commutation_relations_cran_2}
\eeq
The constant appearing on the right-hand side of Equation \ref{commutation_relations_cran_2} is chosen by convention and its precise value is not important. The fact that it is nonzero is of great physical relevance: there is one more way of first adding a particle to an arbitrary avalanche and then removing one, than doing it the other way around.

Using the creation and annihilation operators, we can also define number operators for electrons and holes,
\beqs
\Nume = \cre\ane, \qquad \Numh = \crh\anh.
\eeqs
Equations \ref{commutation_relations_cran_1} and \ref{commutation_relations_cran_2} lead to the following commutation relations for the number operators,
\beq
    [\Nume, \cre] = \cre, \qquad\qquad [\Numh, \crh] = \crh, \qquad\qquad [\Nume, \ane] = -\ane, \qquad\qquad [\Numh, \anh] = -\anh.
    \label{commutation_relations_num}
\eeq
All commutators that combine an operator involving electrons with an operator involving holes, e.g.~$[\Nume, \crh]$, vanish.

Making use of the creation operators, we identify the basis states as
\beqs
\ket{N_e, N_h} = \left(\cre\right)^{N_e} \left(\crh\right)^{N_h} \ket{0},
\eeqs
i.e.~as the empty avalanche, to which the correct numbers of electrons and holes have been added. That this assertion is in fact consistent with the definitions of $\crg$, $\ang$ and $\Numg$ is easy to see. Applying the commutation relations in Equation \ref{commutation_relations_num} shows that the states $\ket{N_e, N_h}$ indeed contain a well-defined number of charges, as measured by the number operators,
\beqs
\Nume \ket{N_e, N_h} = N_e \ket{N_e, N_h}, \qquad\qquad \Numh \ket{N_e, N_h} = N_h \ket{N_e, N_h}.
\eeqs
The action of the creation operators on the basis states is trivially given by
\beqs
\cre \ket{N_e, N_h} = \ket{N_e+1, N_h}, \qquad\qquad \crh \ket{N_e, N_h} = \ket{N_e, N_h+1},
\eeqs
while the relations in Equation \ref{commutation_relations_num} describe the action of the annihilation operators as
\beqs
\ane \ket{N_e, N_h} = N_e \ket{N_e-1, N_h}, \qquad\qquad \anh \ket{N_e, N_h} = N_h \ket{N_e, N_h-1},
\eeqs
and in particular,
\beqs
\ane\ket{0,0} = 0, \qquad \anh\ket{0,0} = 0.
\eeqs

The basis states $\ket{N_e, N_h}$ describe specific configurations $\mathcal{A}$ of the avalanche. We represent a generic ``state'' $\ket{\psi(t)}$ of the avalanche as the vector corresponding to the sum of all allowed configurations, weighted by their respective probabilities $p(\mathcal{A}, t) = p(N_e, N_h, t)$,
\beq
\ket{\psi(t)} = \sum_{N_e = 0}^{\infty} \sum_{N_h=0}^{\infty} \, p(N_e, N_h, t) \ket{N_e, N_h}.
\label{superposition}
\eeq
This notion of a general state will turn out to be very useful for the computation of expectation values. Equation \ref{superposition} can be thought of as the analogue of superposition in quantum mechanics. An avalanche which attains a particular configuration $\ket{N_e, N_h}$ with probability one is thus represented by the corresponding basis vector, i.e.~$\ket{\psi} = \ket{N_e, N_h}$.

\subsubsection{Dual vectors and inner product} To every vector $\ket{N_e, N_h} = \left(\cre\right)^{N_e} \left(\crh\right)^{N_h} \ket{0}$ we can associate a dual vector $\bra{N_e, N_h} = \bra{0} {\ane}^{N_e} {\anh}^{N_h}$. This allows us to denote the inner product of two vectors $\ket{N_e, N_e}$ and $\ket{N_e', N_h'}$ as $\braket{N_e', N_h' | N_e, N_h}$. With respect to this inner product, the adjoints of the operators $\ane$ and $\anh$ are $\cre$ and $\crh$---a fact already anticipated by the notation. We define the vector representing the empty avalanche to be normalised as $\braket{0|0}=1$. Then, using the commutation relations in Equations \ref{commutation_relations_cran_1} and \ref{commutation_relations_cran_2}, we can evaluate the inner product between two basis vectors to give
\beqs
\braket{N_e', N_h' | N_e, N_h} = \bra{0} {\ane}^{N_e'} {\anh}^{N_h'} \left(\cre\right)^{N_e} \left(\crh\right)^{N_h} \ket{0} = N_e! \, N_h! \, \delta_{N_e, N_e'} \, \delta_{N_h, N_h'}.
\eeqs
This shows that the basis states $\ket{N_e, N_h}$ are orthogonal (but not orthonormal) w.r.t.~this inner product.

Given a general state $\ket{\psi(t)}$ like in Equation \ref{superposition}, we can recover the probability $p(N_e, N_h, t)$ by taking the inner product of $\ket{\psi(t)}$ with the basis state $\bra{N_e, N_h}$,
\beq
p(N_e, N_h, t) = \frac{\braket{N_e, N_h | \psi(t)}}{\braket{N_e, N_h | N_e, N_h}}.
\label{probability_reconstruction}
\eeq

\subsubsection{Observables and expectation values}
\label{Sec:expectation_values_unbounded}
Given a general state $\ket{\psi(t)} = \sum p(N_e, N_h, t) \ket{N_e, N_h}$, an important task is to find the expectation value of an observable $\mathcal{O}(N_e, N_h)$ that depends on $N_e$ and $N_h$, i.e.~to compute the sum
\beq
\avg{\mathcal{O}(N_e, N_h)} := \sum_{N_e = 0}^{\infty} \sum_{N_h = 0}^{\infty} \mathcal{O}(N_e, N_h) \, p(N_e, N_h, t).
\label{general_expectation_value}
\eeq
Important observables include the number of electrons and holes, i.e.~$\mathcal{O}(N_e, N_h)=N_e$ and $\mathcal{O}(N_e, N_h)=N_h$, as well as the total charge in the avalanche, $\mathcal{O}(N_e, N_h)=N_e+N_h$.

To this end, we introduce a special dual vector $\avgbra$,
\beqs
\avgbra = \bra{0} e^{\ane + \anh} = \bra{0} \sum_{n = 0}^{\infty} \frac{(\ane + \anh)^n}{n!},
\eeqs
which has the important property of being a left-eigenstate of the creation operators. In other words, we have
\beqs
\avgbra \cre = \avgbra, \qquad\qquad \avgbra \crh = \avgbra.
\eeqs
Its overlap with the empty state is $\avgbraket{0} = \braket{0 | \mathds{1} + \ldots | 0} = 1$. From this, it follows that also $\avgbraket{N_e, N_h} = 1$.

To compute the expectation in Equation \ref{general_expectation_value}, we turn the observable $\mathcal{O}(N_e, N_h)$ into an operator by inserting the number operators into its arguments: $\mathcal{O}(\Nume, \Numh)$. Acting with this operator on the state $\ket{\psi(t)}$ and taking the inner product with $\avgbra$ computes the expectation $\avg{\mathcal{O}(N_e, N_h)}$,
\bea
\avgbraket{\mathcal{O}(\Nume, \Numh) | \psi(t)} = \sum p(N_e, N_h, t) \avgbraket{\mathcal{O}(\Nume, \Numh) | N_e, N_h} =\nonumber\\
= \sum p(N_e, N_h, t) \mathcal{O}(N_e, N_h) \avgbraket{N_e, N_h} = \avg{\mathcal{O}(N_e, N_h)}.
\label{observable_expectation}
\eea
This makes explicit that expectation values are linear functions of the state vector, and thus of the probabilities. This is to be contrasted with quantum mechanics, where the expectation value of an operator $\mathcal{O}$ is $\avg{\mathcal{O}} = \braket{\psi|\mathcal{O}|\psi}$ and thus depends quadratically on the amplitudes.

An important special case of Equation \ref{observable_expectation} occurs for $\mathcal{O} = \mathds{1}$, which shows the normalisation of $p(N_e, N_h, t)$ as actual probabilities,
\beqs
\avg{\mathds{1}} = 1 = \sum p(N_e, N_h, t).
\eeqs

\subsubsection{Time evolution}
\label{Sec:time_evolution_unbounded}
The remaining aspect is to implement the time evolution of the stochastic system, i.e.~to describe the time dependence of $\ket{\psi(t)}$, or equivalently, of $p(N_e, N_h, t)$. To this end, we introduce the time translation operator $\uhat(dt) = \mathds{1} + dt \ham$ which acts on the state $\ket{\psi(t)}$ as $\uhat(dt)\ket{\psi(t)} = \ket{\psi(t+dt)}$. The operator $\ham$ encodes the dynamics of the system, and for this reason we refer to it as the Hamiltonian.
From the definition of the time translation operator, we obtain the evolution equation for the state $\ket{\psi}$ as
\beq
\frac{d}{dt}\ket{\psi(t)} = \ham\ket{\psi(t)}.
\label{schroedinger_equation}
\eeq
This is just the general Equation \ref{master_equation_general} for a Markov process, where the Hamiltonian operator $\ham$ now encodes the set of all transition matrix elements $T(N_e, N_h \rightarrow N_e', N_h')$. We will make this connection more explicit in Section \ref{Sec:hamiltonian_unbounded} below, where we derive the Hamiltonian for an electron-hole avalanche. By virtue of the property in Equation \ref{probability_reconstruction}, this equation is equivalent to an evolution equation for the probability $p(N_e, N_h, t)$. Taking the inner product of Equation \ref{schroedinger_equation} with $\frac{\bra{N_e, N_h}}{\braket{N_e, N_h | N_e, N_h}}$ gives
\beq
\frac{d}{dt} p(N_e, N_h, t) = \frac{d}{dt} \frac{\braket{N_e, N_h | \psi(t)}}{\braket{N_e, N_h | N_e, N_h}} = \frac{\braket{N_e, N_h | \ham | \psi(t)}}{\braket{N_e, N_h | N_e, N_h}}.
\label{schroedinger_equation_probability}
\eeq

Equation \ref{schroedinger_equation} also gives rise to evolution equations for the expectation values of arbitrary observables $\mathcal{O}(N_e, N_h)$. Making use of Equation \ref{observable_expectation}, we find
\beq
\frac{d}{dt}\avg{\mathcal{O}(N_e, N_h)} = \frac{d}{dt} \avgbraket{\mathcal{O}(\Nume, \Numh) | \psi(t)} = \avgbraket{\mathcal{O}(\Nume, \Numh)\ham|\psi(t)}.
\label{evolution_equation_expectation_1}
\eeq
For the special case $\mathcal{O} = \mathds{1}$, we obtain the relation
\beqs
0 = \avgbraket{\ham|\psi(t)}.
\eeqs
Since $\ham\ket{\psi(t)}\neq 0$, we have the condition
\beqs
\avgbra\ham = 0
\eeqs
that any valid Hamiltonian must satisfy. Therefore, Equation \ref{evolution_equation_expectation_1} can also be written in the form
\beq
\frac{d}{dt}\avg{\mathcal{O}(N_e, N_h)} = \avgbraket{\left[\mathcal{O}(\Nume, \Numh), \ham \right]|\psi(t)}.
\label{evolution_equation_expectation_2}
\eeq
We shall find that this evolution equation is extremely useful. It can be used to extract important information about the system even in complicated situations where Equation \ref{schroedinger_equation} itself can no longer be solved directly.

\subsubsection{Hamiltonian for electron-hole avalanche}
\label{Sec:hamiltonian_unbounded}
In order to derive the form of the Hamiltonian for an electron-hole avalanche, it is useful to expand on the connection between Equation \ref{schroedinger_equation} and the general form of the time evolution for a Markov process in Equation \ref{master_equation_general}.

To this end, we write the time evolution operator $\uhat(dt)$ in terms of the basis states $\ket{N_e, N_h}$,
\beq
\uhat(dt) = \sum_{N_e, N_h} \sum_{N_e', N_h'} p(N_e, N_h \rightarrow N_e', N_h'; dt) \frac{\ket{N_e', N_h'}\bra{N_e, N_h}}{\braket{N_e, N_h | N_e, N_h}} ,
\label{time_evolution_basis_expansion}
\eeq
The operator matrix elements $p(N_e, N_h \rightarrow N_e', N_h'; dt)$ in this basis correspond to the probabilities that a transition $\ket{N_e, N_h} \rightarrow \ket{N_e', N_h'}$ happens within a small time interval $dt$. This is seen by acting with $\uhat(dt)$ on a general state vector $\ket{\psi(t)}$,
\bea
\uhat(dt)\ket{\psi(t)} &=& \sum_{N_e, N_h} \sum_{N_e', N_h'} p(N_e, N_h \rightarrow N_e', N_h'; dt) \, p(N_e, N_h, t) \ket{N_e', N_h'} \nonumber\\
&=& \sum_{N_e', N_h'} \left( \sum_{N_e, N_h} p(N_e, N_h \rightarrow N_e', N_h'; dt) \, p(N_e, N_h, t) \right) \ket{N_e', N_h'} \nonumber\\
&=& \sum_{N_e, N_h} p(N_e, N_h, t + dt) \ket{N_e, N_h},\nonumber
\eea
where we have relabelled $N_e' \rightarrow N_e$, $N_h' \rightarrow N_h$ in the last line. This shows that the operator $\uhat(dt)$ as expanded in Equation \ref{time_evolution_basis_expansion} indeed effects an infinitesimal time translation $t \rightarrow t + dt$ when applied to any state. The probabilities $p(N_e, N_h \rightarrow N_e', N_h'; dt)$ are related to the transition matrix elements as follows,
\bea
p(N_e, N_h \rightarrow N_e', N_h'; dt) &=& dt \cdot T(N_e, N_h \rightarrow N_e', N_h') \qquad \mathrm{for} \qquad (N_e, N_h) \neq (N_e', N_h'),\nonumber\\
p(N_e, N_h \rightarrow N_e, N_h; dt) &=& 1 - \sum_{\substack{(N_e', N_h')\neq\\(N_e, N_h)}} p(N_e, N_h \rightarrow N_e', N_h'; dt)\nonumber\\
&=& 1 - dt \sum_{\substack{(N_e', N_h')\neq\\(N_e, N_h)}} T(N_e, N_h \rightarrow N_e', N_h').\nonumber
\eea

We make use of this correspondence to write down the Hamiltonian for the electron-hole avalanche. It is particularly convenient to work with the expansion in Equation \ref{time_evolution_basis_expansion}. Within a small time interval $dt$, the avalanche configuration $\ket{N_e, N_h}$ can evolve into the configuration $\ket{N_e+1, N_h+1}$ (with a probability of order $\mathcal{O}(dt)$) or remain itself (with a probability of order unity). Transitions to other states are possible within $dt$, but with probabilities of order at least $\mathcal{O}(dt^2)$.
A transition occurs if any one of the $N_e$ electrons creates an electron-hole pair, or if any one of the $N_h$ holes creates an electron-hole pair. Since all of these processes occur independently, the corresponding probability is
\beqs
p(N_e, N_h \rightarrow N_e+1, N_h+1; dt) = \alpha v_e N_e dt + \beta v_h N_h dt.
\eeqs
Similarly, the probability for the avalanche not to leave the state $\ket{N_e, N_h}$ within $dt$ is
\beqs
p(N_e, N_h \rightarrow N_e, N_h; dt) = 1 - p(N_e, N_h \rightarrow N_e+1, N_h+1; dt) = 1 - \alpha v_e N_e dt - \beta v_h N_h dt.
\eeqs
We can thus express the time evolution operator $\uhat(dt)$ as
\bea
\uhat(dt) = \sum_{N_e, N_h} (\alpha v_e N_e dt + \beta v_h N_h dt) \frac{\ket{N_e + 1, N_h + 1}\bra{N_e, N_h}}{\braket{N_e, N_h | N_e, N_h}} + \nonumber \\
+ \sum_{N_e, N_h} (1 - \alpha v_e N_e dt - \beta v_h N_h dt) \frac{\ket{N_e, N_h}\bra{N_e, N_h}}{\braket{N_e, N_h | N_e, N_h}}.\nonumber
\eea
Writing $\ket{N_e+1, N_h+1}$ as $\cre\crh\ket{N_e, N_h}$ and $N_e \ket{N_e, N_h}$ as $\Nume \ket{N_e, N_h}$, this simplifies to
\bea
\uhat(dt) &=& \left[ \mathds{1} + \alpha v_e dt (\cre\crh \Nume - \Nume) + \beta v_h dt (\cre\crh \Numh - \Numh) \right] \underbrace{\sum_{N_e, N_h} \frac{\ket{N_e, N_h}\bra{N_e, N_h}}{\braket{N_e, N_h | N_e, N_h}}}_{\mathds{1}} \nonumber \\
&=& \mathds{1} + dt \left[\alpha v_e (\cre\crh\Nume - \Nume) + \beta v_h (\cre\crh\Numh - \Numh) \right].\label{uhat_0D}
\eea
We can now read off the Hamiltonian $\ham$, which is
\beq
\ham = \alpha v_e (\cre\crh\Nume - \Nume) + \beta v_h (\cre\crh\Numh - \Numh).
\label{hamiltonian_unbounded}
\eeq

\begin{figure}[ht]
  \centering
  a) \raisebox{0.5\height}{\includegraphics[height=1.25cm]{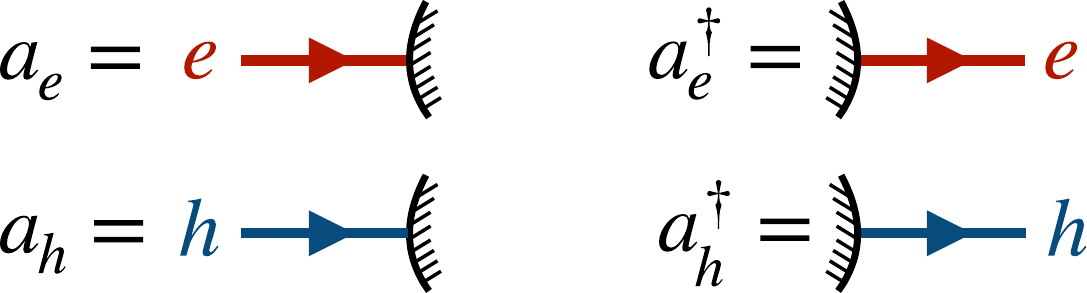}}\qquad
  b) \includegraphics[height=2.5cm]{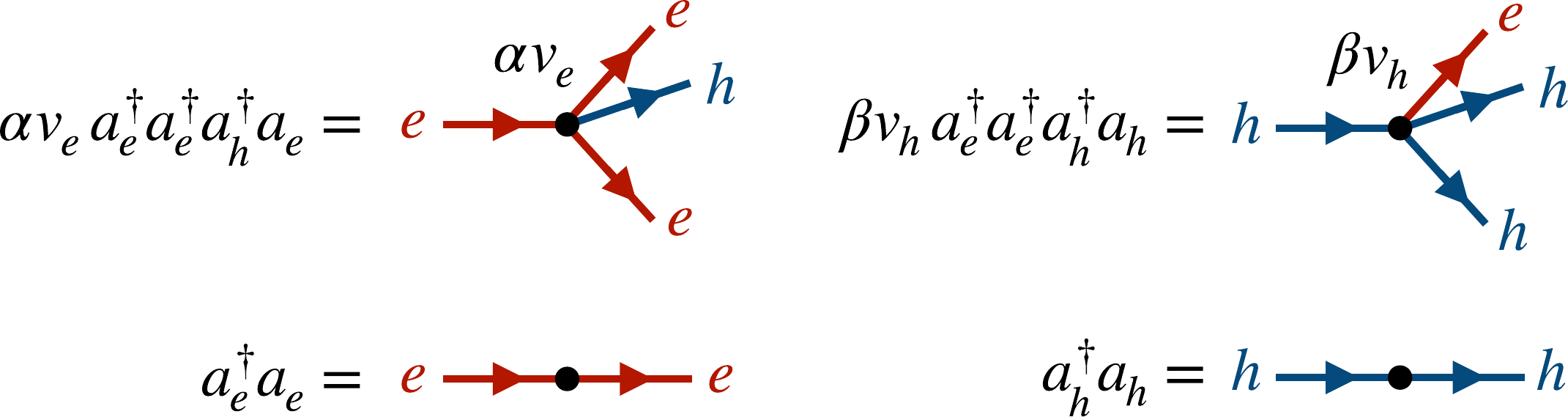}
  \caption{a) Representation of annihilation operators ``absorbing'' a charge, and of creation operators ``emitting'' a charge. b) Diagrams corresponding to the four terms in the Hamiltonian in Equation \ref{hamiltonian_unbounded}.}
  \label{Fig:hamiltonian_diagrammatic}
\end{figure}

We have derived the Hamiltonian by starting from the transition probabilities $p(N_e, N_h \rightarrow N_e', N_h'; dt)$. One could have also started from a ``microscopic'' point of view, i.e.~by considering the interactions that each charge can undergo. We shall not expand on this picture here, but merely hint at the related diagrammatic representation of Equation \ref{hamiltonian_unbounded}. We represent the creation and annihilation operators as lines leaving or entering any given diagram, as shown in Figure \ref{Fig:hamiltonian_diagrammatic}a. The representation of the four terms in the Hamiltonian, expressed in terms of $\crg$ and $\ang$ are shown in Figure \ref{Fig:hamiltonian_diagrammatic}b: they correspond exactly to all possible interactions available to a charge carrier. The probability per unit time for an interaction to happen appears as a numerical coefficient in the Hamiltonian and plays the role of a coupling constant.

\subsubsection{Evolution equation for $p(N_e, N_h)$}
\label{Sec:derivation_evolution_equation_p_Ne_Nh}
We make use of Equation \ref{schroedinger_equation_probability}. Inserting the explicit form of the Hamiltonian from Equation \ref{hamiltonian_unbounded}, we find,
\beqs
\alpha v_e \frac{\bra{N_e, N_h} \cre\crh\Nume - \Nume \ket{\psi(t)}}{\braket{N_e, N_h | N_e, N_h}} = \alpha v_e \left[(N_e - 1) p(N_e - 1, N_h - 1, t) - N_e p(N_e, N_h, t) \right],
\eeqs
\beqs
\beta v_h \frac{\bra{N_e, N_h} \cre\crh\Numh - \Numh \ket{\psi(t)}}{\braket{N_e, N_h | N_e, N_h}} = \beta v_h \left[(N_h - 1) p(N_e - 1, N_h - 1, t) - N_h p(N_e, N_h, t) \right].
\eeqs
Putting these terms together immediately leads to Equation \ref{evolution_equation_p_Ne_Nh}.

\subsubsection{Evolution equations for moments}
\label{Sec:derivation_evolution_equation_moments}
We use Equation \ref{evolution_equation_expectation_2} to obtain evolution equations for the moments $\avg{N_e^m N_h^n}$. The required commutation relation $\left[ \Nume^m \Numh^n, \ham \right]$ can be evaluated by using the relations
\beqs
[\Nume^m, \cre] = \cre \sum_{k = 0}^{m-1} \binom{m}{k} \Nume^k, \qquad\qquad [\Numh^m, \cre] = \crh \sum_{k = 0}^{m-1} \binom{m}{k} \Numh^k.
\eeqs
Using these results, we obtain the following commutator with the Hamiltonian in Equation \ref{hamiltonian_unbounded},
\beqs
\left[ \Nume^m \Numh^n, \ham \right] = \cre\crh \sum_{(l, k) \in S} \binom{m}{l}\binom{n}{k} \left(\alpha v_e \Nume^{l+1} \Numh^k + \beta v_h \Nume^l \Numh^{k+1} \right),
\eeqs
where $S = \{(l, k) | 0 \leq l \leq m, 0 \leq k \leq n \} \setminus (m, n)$. This directly gives Equation \ref{differential_equation_general_moment}.

\subsubsection{Asymptotic value for $\sigma(\log N)$}
\label{Sec:asymptotic_logN_unbounded}

The time resolution for large thresholds, $\sthinf$, can be computed directly from the Hamiltonian in Equation \ref{hamiltonian_unbounded}, without the need to access the time response function $\rho$. We saw in Equation \ref{time_resolution_asymptotic_unbounded} that $\sthinf$ is directly related to $\sigma(\log N) = \sqrt{\avg{\log^2 N}-\avg{\log N}^2}$ at late times.

However, the Hamiltonian in Equation \ref{hamiltonian_unbounded} does not admit closed evolution equations for the moments $\avg{\log N}$ and $\avg{\log^2 N}$. To see this, note that e.g.~the evolution equation for the observable $\log \Nume$ will contain terms like $\frac{d}{dt}\avg{N_e} = \avgbra [\log \Nume, \ham] \ket{\psi} \ni \alpha v_e \avg{\log\left(1 + \frac{1}{N_e} \right) N_e}$. Solving such an equation would require an ad-hoc closure scheme.
However, expanding the right-hand side for large $N_e$ shows that the equation does become closed for large avalanches, where $\avg{\log\left(1 + \frac{1}{N_e} \right) N_e} \approx 1$. This suggests that we might find an observable which is not generally identical to $\log \Numg$, but which limits to this quantity for large avalanches, and which does obey a closed evolution equation.

To this end, we consider the observables
\beq
\hat{f}_1 = \frac{1}{\alpha v_e + \beta v_h} \, \psi_0\left(\frac{\alpha v_e \Nume + \beta v_h \Numh}{\alpha v_e + \beta v_h} \right),
\label{f1_definition}
\eeq
and
\beq
\hat{f}_2 = \frac{1}{2 (\alpha v_e+\beta v_h)^2} \left[ \psi_0\left(\frac{\alpha v_e \Nume + \beta v_h \Numh}{\alpha v_e+\beta v_h} \right)^2 + \psi_1\left(\frac{\alpha v_e \Nume + \beta v_h \Numh}{\alpha v_e+\beta v_h} \right) \right],
\label{f2_definition}
\eeq
with the polygamma functions $\psi_k$. They indeed limit to $\log N$ for large avalanches,
\beqs
\lim_{N_e, N_h \rightarrow \infty} \hat{f}_1 = \frac{1}{\alpha v_e + \beta v_h} \left[\log \Numg + C_0 \right],\qquad \lim_{N_e, N_h \rightarrow \infty} \hat{f}_2 = \frac{1}{2(\alpha v_e+\beta v_h)^2} \left[\left(\log \Numg + C_0 \right)^2 \right],
\eeqs
where $C_0$ is an irrelevant constant that will drop out. This means that
\beq
\lim_{t\rightarrow\infty} 2 \avg{f_2} - \avg{f_1}^2 = \lim_{t\rightarrow\infty} \frac{1}{\alpha v_e+\beta v_h}\left[\avg{\log^2 N} - \avg{\log N}^2 \right] = \sthinf^2.
\label{f1_f2_limit}
\eeq
The commutators of $\hat{f}_1$ and $\hat{f}_2$ with the Hamiltonian are trivial,
\beqs
\left[\hat{f}_1, \ham \right] = \cre\crh, \qquad\qquad \left[\hat{f}_2, \ham \right] = \cre\crh\hat{f}_1,
\eeqs
and so are their time evolution equations,
\beqs
\frac{d}{dt}\avg{f_1}=1, \qquad\qquad \frac{d}{dt}\avg{f_2}=\avg{f_1}.
\eeqs
Together with Equation \ref{f1_f2_limit}, this shows that
\beqs
\frac{d}{dt}\left(2\avg{f_2} - \avg{f_1}^2 \right) = 0 \implies 2\avg{f_2} - \avg{f_1}^2 = \mathrm{const.} = \sthinf^2.
\eeqs
The asymptotic time resolution thus becomes
\beqs
\sthinf = \sqrt{2\avg{f_2} - \avg{f_1}^2}\Big\rvert_{t=0} = \frac{\sqrt{\psi_1(A)}}{\alpha v_e + \beta v_h},
\eeqs
as stated in Equation \ref{logN_fluctuations_unbounded}.

\subsection{Electron-hole avalanche in a thin semiconductor with an arbitrary electric field}
\label{Sec:formalism_1D}
We now proceed to including the spatial development of the avalanche. This is important in case the avalanche develops in a bounded region where the electric field $\mathbf{E}(x)$ depends on the position $x$. This leads to position-dependent Townsend coefficients $\alpha(x)$ and $\beta(x)$ and drift velocities $v_e(x)$ and $v_h(x)$.

\subsubsection{State vectors}
In this case, an arbitrary configuration $\mathcal{A}$ of the avalanche is labelled by the spatial densities of electrons and holes, $n_e(x)$ and $n_h(x)$. The number of electrons in a small interval $[x, x+dx]$ is then $n_e(x) dx$ and the number of holes in the same region is $n_h(x) dx$. The basis vectors spanning the space of all avalanche configurations are then $\ket{\mathcal{A}} = \ket{n_e(x), n_h(x)}$.

Again, we introduce creation and annihilation operators to create different avalanche configurations. These operators now add or remove a charge at position $x$, i.e.~they pick up a position index. As before, their action is defined by the commutation relations
\beq
    [\cre(x), \crh(y)] = [\ane(x), \anh(y)] = [\ane(x), \crh(y)] = [\anh(x), \cre(y)] = 0,
    \label{commutation_relations_local_cran_1}
\eeq
\beq
    [\ane(x), \cre(y)] = [\anh(x), \crh(y)] = \delta(x-y).
    \label{commutation_relations_local_cran_2}
\eeq
Using the creation operators, we can again build up the basis vectors starting from the empty configuration $\ket{0}$,
\beqs
\ket{n_e(x), n_h(x)} = \prod_{\{x\}} \cre(x)^{n_e(x) dx} \,\, \crh(x)^{n_h(x) dx} \ket{0}.
\eeqs
We also have the number density operators
\beqs
\nume(x) = \cre(x) \ane(x), \qquad\qquad \numh(x) = \crh(x) \anh(x).
\eeqs
Using relations \ref{commutation_relations_local_cran_1} and \ref{commutation_relations_local_cran_2}, we see that the number operators obey the following commutation relations
\beqs
[\nume(x), \cre(y)] = \cre(x) \delta(x-y), \qquad\qquad [\numh(x), \crh(y)] = \crh(x) \delta(x-y),
\eeqs
\beqs
[\nume(x), \ane(y)] = -\ane(x) \delta(x-y), \qquad\qquad [\numh(x), \anh(y)] = -\anh(x) \delta(x-y).
\eeqs
Their action on the basis vectors is thus given by
\beqs
\nume(y) \ket{n_e(x), n_h(x)} = n_e(y) \ket{n_e(x), n_h(x)}, \qquad\qquad \numh(y) \ket{n_e(x), n_h(x)} = n_h(y) \ket{n_e(x), n_h(x)},
\eeqs
as expected.

A general state $\ket{\psi(t)}$ is again built as a linear combination of basis vectors, weighted by the probability to find the corresponding avalanche configuration, $p(\mathcal{A}, t)$. Since $\mathcal{A}$ is labelled by the functions $n_e(x)$ and $n_h(x)$, this probability is now formally a functional, $p(\mathcal{A}, t) = p[n_e(x), n_h(x), t]$, where we use square brackets to denote the functional dependency. A generic state $\ket{\psi(t)}$ is then given by the functional integral
\beqs
\ket{\psi(t)} = \int Dn_e(x) Dn_h(x) \, p[n_e(x), n_h(x), t] \ket{n_e(x), n_h(x)}.
\eeqs

\subsubsection{Translations}
\label{Sec:translation_operators}
The movement of charge carriers through the semiconductor is central to the formation of the avalanche in position space. We introduce dedicated operators $\transe$ and $\transh$ that implement the effects of spatial translations on the basis vectors $\ket{n_e(x), n_h(x)}$ separately for electrons and holes.

Under a general translation $T_\Delta$, the position of a charge carrier changes as
\beq
T_\Delta: x \mapsto y = x + \Delta(x).
\label{general_translation_defn}
\eeq
The translation is characterised by the function $\Delta(x)$ that describes the magnitude of the (generally position-dependent) shift. The translation operators thus depend on $\Delta(x)$ in a functional way. We write $\transe[\Delta(x)]$ and $\transh[\Delta(x)]$.

Under a translation of the type of Equation \ref{general_translation_defn}, a generic density $n(x)$ transforms as
\beq
T_\Delta n(y) dy = n(x) dx \implies T_\Delta n(y) = \frac{n(x)}{1 + \frac{d\Delta(x)}{dx}},
\label{density_transformation_translation}
\eeq
where we have used the shorthand notation $T_\Delta n(x)$ to represent the transformed density. For constant shifts, $\Delta(x)=\Delta$, the Jacobian factor $\left(1+\frac{d\Delta}{dx}\right)^{-1}$ vanishes and the density gets translated as $y = x + \Delta$.

The action of the translation operators $\transe[\Delta(x)]$ and $\transh[\Delta(x)]$ on the state $\ket{n_e(x), n_h(x)}$ is thus defined as
\bea
\transe[\Delta(x)] \ket{n_e(x), n_h(x)} &:=& \ket{T_\Delta n_e(x), n_h(x)},\label{translation_e_action_state}\\
\transh[\Delta(x)] \ket{n_e(x), n_h(x)} &:=& \ket{n_e(x), T_\Delta n_h(x)}.\label{translation_h_action_state}
\eea
The transformation in Equation \ref{density_transformation_translation} is homogeneous, which means that the empty state is invariant under translations, $\transe \ket{0} = \ket{0}$ and $\transh \ket{0} = \ket{0}$, as expected.

Equivalently, we can describe the translation operators by giving their commutation relations with the number density operators $\nume(x)$ and $\numh(x)$. Using the definitions in Equations \ref{translation_e_action_state}-\ref{translation_h_action_state}, we find
\bea
\left[\nume(y), \transe[\Delta]\right] &=& \transe[\Delta] \left(\frac{\nume(x)}{1+\frac{d\Delta}{dx}} - \nume(y) \right),\label{commutation_trans_e}\nonumber\\
\left[\numh(y), \transh[\Delta]\right] &=& \transh[\Delta] \left(\frac{\numh(x)}{1+\frac{d\Delta}{dx}} - \numh(y) \right).\label{commutation_trans_h}\nonumber
\eea

The case of infinitesimal translations deserves a special mention. A typical case would have $\Delta(x) = v(x) dt$, i.e.~corresponds to a position-dependent drift velocity acting for a short time interval $dt$. The above commutation relations specialise to
\bea
\left[\nume(x), \transe[v_e(x) dt] \right] &=& -dt \, \transe[v_e(x) dt] \, \frac{d}{dx} v_e(x) \nume(x),\label{commutation_trans_e_infinitesimal}\nonumber\\
\left[\numh(x), \transh[-v_h(x) dt] \right] &=& dt \, \transh[-v_h(x) dt] \, \frac{d}{dx} v_h(x) \numh(x).\label{commutation_trans_h_infinitesimal}\nonumber
\eea
In these expressions, the action of a derivative on a generic operator $\opg(x)$ is to be understood as
\beqs
dx \frac{d}{dx}\opg(x) := \opg(x+dx) - \opg(x),
\eeqs
which is inspired by the fact that, for smooth expectations,
\beqs
\avg{\frac{d}{dx} \opg(x)} = \frac{d}{dx} \avg{\opg(x)}.
\eeqs

\subsubsection{Global observables}
Starting from the number density operators $\nume(x)$ and $\numh(x)$, we can also construct operators that are sensitive to the total number of charges in a finite region $\reg$,
\beq
\Nume(\reg) = \int_{\reg} dx \, \nume(x), \qquad\qquad \Numh(\reg) = \int_{\reg} dx \, \numh(x).
\label{global_number_operators_defs}
\eeq
Their commutation relations with the creation and annihilation operators are given by
\beqs
[\Nume(\reg), \cre(x)] = \cre(x) \cdot \indicator_{\reg}(x), \qquad\qquad [\Numh(\reg), \crh(x)] = \crh(x) \cdot \indicator_{\reg}(x),
\eeqs
\beqs
[\Nume(\reg), \ane(x)] = -\ane(x) \cdot \indicator_{\reg}(x), \qquad\qquad [\Numh(\reg), \anh(x)] = -\anh(x) \cdot \indicator_{\reg}(x),
\eeqs
where $\indicator_{\reg}$ is the indicator function of $\reg$, i.e.~$\indicator_{\reg}(x) = 1$ if $x\in\reg$ and $0$ otherwise.

\subsubsection{Expectation values}
Just like the probability $p(\mathcal{A}, t)$, also observables are now generally functionals of the densities $n_e(x)$ and $n_h(x)$, we write~$\mathcal{O}[n_e(x), n_h(x)]$. Nevertheless, the computation of their expectation values proceeds in exactly the same way as in Section \ref{Sec:expectation_values_unbounded}. We introduce again the dual vector $\avgbra$
\beqs
\avgbra = \bra{0} e^{\int dx \, \ane(x) + \int dx \, \anh(x)},
\eeqs
which is a left-eigenstate not only of $\cre(x)$ and $\crh(x)$, but also of $\transe(\Delta)$ and $\transh(\Delta)$,
\beqs
\avgbra \cre(x) = \avgbra, \qquad\qquad \avgbra \crh(x) = \avgbra,
\eeqs
\beqs
\avgbra \transe(\Delta) = \avgbra, \qquad\qquad \avgbra \transh(\Delta) = \avgbra.
\eeqs
In particular, the important property $\avgbraket{n_e(x), n_h(x)} = 1$ holds as before.

To compute the expectation value $\avg{\mathcal{O}[n_e(x), n_h(x)]}$, we again replace the arguments with the number density operators, apply the resulting operator to the state $\ket{\psi(t)}$ and take the inner product with $\avgbra$. This gives
\bea
\avgbraket{\mathcal{O}[\nume(x), \numh(x)]|\psi(t)} = \int Dn_e(x) Dn_h(x) \, p[n_e(x), n_h(x), t] \, \avgbraket{\mathcal{O}[\nume(x), \numh(x)]| n_e(x), n_h(x)} =\nonumber
\\= \int Dn_e(x) Dn_h(x) \, p[n_e(x), n_h(x), t] \, \mathcal{O}[n_e(x), n_h(x)] = \avg{\mathcal{O}(n_e(x), n_h(x))}.\nonumber
\eea

\subsubsection{Time evolution operator for electron-hole avalanche}
We again introduce the time evolution operator $\uhat(dt) = \mathds{1} + dt \ham$ that produces an infinitesimal time evolution of the state vector $\ket{\psi(t)}$. The time evolution of the state $\ket{\psi(t)}$ is still given by Equation \ref{schroedinger_equation}. It now takes the form of a functional differential equation. The time evolution equation for the expectation value $\avg{\mathcal{O}[n_e(x), n_h(x)]}$ in Equation \ref{evolution_equation_expectation_2} generally becomes a partial differential equation.

In the present case, $\uhat(dt)$ is easier to write down than the Hamiltonian $\ham$ itself. Since $[\opg, \mathds{1}] = 0$, we see that $dt \left[\opg, \ham \right] = \left[\opg, \uhat(dt) \right]$ for an arbitrary observable $\opg$, and so both operators essentially contain the same amount of information.

We start from Equation \ref{uhat_0D} for the position-independent situation and generalise this expression to the general case considered here. Since charges at different coordinates multiply independently, it is easy to account for the position-dependence itself: we need to replace $\Nume \rightarrow \nume(x)$, $\Numh \rightarrow \numh(x)$, $\alpha \rightarrow \alpha(x)$, $\beta \rightarrow \beta(x)$, $v_e \rightarrow v_e(x)$, $v_h \rightarrow v_h(x)$ and integrate over $x$. In the language of Figure \ref{Fig:hamiltonian_diagrammatic}, this introduces a separate interaction vertex for each position $x$. The translation operators introduced in Section \ref{Sec:translation_operators} can be used to implement the movement of electrons and hole in opposite directions, and so the final expression for $\uhat(dt)$ reads
\bea
\uhat(dt) &=& \transe[v_e(x) dt] \transh[-v_h(x) dt] \Big[ \mathds{1} + \nonumber\\
  &+& dt \int dx \, \alpha(x) v_e(x) \left( \cre(x)\crh(x)\nume(x) - \nume(x) \right) + \nonumber\\
  &+& dt \int dx \, \beta(x) v_h(x) \left( \cre(x)\crh(x)\numh(x) - \numh(x) \right) \Big].\nonumber
\eea

\subsubsection{Evolution equations for moments}
\label{Sec:moments_equations_bounded}
To derive evolution equations for the first moments $\avg{n_e(x)}$ and $\avg{n_h(x)}$, we require the commutation relations (dropping contributions of order $\mathcal{O}(dt^2)$ or higher)
\bea
\left[\nume(x), \uhat(dt)\right] &=& dt\, \transe[v_e(x) dt] \transh[-v_h(x) dt] \left[-\frac{d}{dx} v_e(x) \nume(x) + \cre(x)\crh(x) \left[ \alpha(x) v_e(x) \nume(x) + \beta(x) v_h(x) \numh(x) \right] \right]\nonumber\\
\,\left[\numh(x), \uhat(dt)\right] &=& dt\, \transe[v_e(x) dt] \transh[-v_h(x) dt] \left[\frac{d}{dx} v_h(x) \numh(x) + \cre(x)\crh(x) \left[ \alpha(x) v_e(x) \nume(x) + \beta(x) v_h(x) \numh(x) \right] \right]\nonumber
\eea
Extracting the term of order $dt$ and computing the average $\avgbra \cdot \ket{\psi}$ immediately leads to Equations \ref{equations_first_moment_bounded_start}-\ref{equations_first_moment_bounded_end}.

For the second moments $\avg{n_e(x) n_e(y)}$, $\avg{n_e(x) n_h(y)}$ and $\avg{n_h(x) n_h(y)}$, the necessary commutators read (again dropping terms of order $\mathcal{O}(dt^2)$ or higher)
\bea
\left[\nume(x)\nume(y), \uhat(dt)\right] &=& dt \, \transe[v_e(x) dt] \transh[-v_h(x) dt] \left[-\frac{d}{dy} v_e(y)\nume(x)\nume(y) - \frac{d}{dx} v_e(x) \nume(y) \nume(x) + \right. \nonumber\\
&+& \alpha(y) v_e(y) \, \cre(y)\crh(y) \left[\delta(x-y) \nume(y) + \nume(x) \nume(y)\right] + \alpha(x) v_e(x) \, \cre(x)\crh(x) \nume(x) \nume(y) + \nonumber\\
&+& \left. \beta(y) v_h(y) \, \cre(y)\crh(y) \left[\delta(x-y) \numh(y) + \nume(x)\numh(y) \right] + \beta(x) v_h(x) \, \cre(x)\crh(x) \numh(x) \nume(y) \right], \nonumber
\eea
\bea
\left[\nume(x)\numh(y), \uhat(dt)\right] &=& dt \, \transe[v_e(x) dt] \transh[-v_h(x) dt] \left[\frac{d}{dy} v_h(y)\nume(x)\numh(y) -\frac{d}{dx} v_e(x)\numh(y)\nume(x) + \right. \nonumber\\
&+& \alpha(y) v_e(y) \, \cre(y)\crh(y) \left[\delta(x-y) \nume(y) + \nume(x)\nume(y) \right] + \alpha(x) v_e(x) \, \cre(x)\crh(x) \nume(x) \numh(y) + \nonumber\\
&+& \left. \beta(y) v_h(y) \, \cre(y)\crh(y) \left[\delta(x-y) \numh(y) + \nume(x) \numh(y) \right] + \beta(x) v_h(x) \, \cre(x)\crh(x) \numh(x)\numh(y)\right], \nonumber 
\eea
\bea
\left[\numh(x)\numh(y), \uhat(dt)\right] &=& dt \, \transe[v_e(x) dt] \transh[-v_h(x) dt] \left[\frac{d}{dy} v_h(y)\numh(x)\numh(y) + \frac{d}{dx} v_h(x)\numh(y)\numh(x) + \right. \nonumber\\
&+& \alpha(y) v_e(y) \, \cre(y)\crh(y) \left[\delta(x-y) \nume(y) + \numh(x) \nume(y) \right] + \alpha(x) v_e(x) \, \cre(x)\crh(x) \nume(x)\numh(y) + \nonumber\\
&+& \left. \beta(y) v_h(y) \, \cre(y)\crh(y) \left[\delta(x-y) \numh(y) + \numh(x)\numh(y) \right] + \beta(x) v_h(x) \, \cre(x)\crh(x) \numh(x)\numh(y) \right], \nonumber
\eea
which directly give rise to Equations \ref{equations_second_moment_bounded_start}-\ref{equations_second_moment_bounded_end}.

\subsubsection{Asymptotic value for $\sigma(\log N)$}
\label{Sec:asymptotic_logN_bounded}
Already the quantity $\sigma(\log N)$ does not admit a closed system of evolution equations. To see this, we take over the definitions for $\hat{f}_1$ and $\hat{f}_2$ from Equations \ref{f1_definition} and \ref{f2_definition}, but replace $\Nume$ and $\Numh$ with their analogues from Equation \ref{global_number_operators_defs}. For constant electric fields $\mathbf{E}$, the evolution equation for $2 \avg{f_2} - \avg{f_1}^2$ in the presence of boundaries reads
\bea
\frac{d}{dt}\left(2 \avg{f_2} - \avg{f_1}^2 \right) = -2 \alpha v_e \, \cov \left[\log \frac{\alpha v_e N_e + \beta v_h N_h}{\alpha v_e + \beta v_h}, \frac{v_e n_e(d)}{\alpha v_e N_e + \beta v_h N_h}\right] - \nonumber \\ - 2 \beta v_h \, \cov \left[\log \frac{\alpha v_e N_e + \beta v_h N_h}{\alpha v_e + \beta v_h}, \frac{v_h n_h(0)}{\alpha v_e N_e + \beta v_h N_h}\right]\nonumber.
\eea
For efficiencies $\epsilon\approx 1$, the left-hand side becomes proportional to $\sigma(\log N)_S$ at late times. Compared to Equation \ref{f1_f2_limit}, the right-hand side does not vanish, but implements the effects induced by the boundary. It essentially depends on the degree of correlation between the charges currently present in the avalanche region, and the relative rate at which charges disappear across the boundaries. Note that this equation is not closed, i.e.~the right-hand side contains variables whose evolution is not determined by the equation itself.

\section{Orthogonality of eigenfunctions}
\label{Sec:eigenfunctions_orthogonality}
\noindent Making use of the definition of the inner product in Equation \ref{eigenfunctions_inner_product_definition}, we find for $\lambda\neq\lambda'$
\bea
\mathcal{I}(\lambda, \lambda') &=&
\begin{bmatrix}
f^{\lambda}_e(\cdot, t)\\f^{\lambda}_h(\cdot, t)
\end{bmatrix}
\cdot
\begin{bmatrix}
f^{\lambda'}_e(\cdot, t)\\f^{\lambda'}_h(\cdot, t)
\end{bmatrix}
= \nonumber
\\
&=&
\int_0^d dx \, \left[ \alpha v_e f^{\lambda}_e(d-x, t) f^{\lambda'}_e(x, t) + \beta v_h f_h^{\lambda}(d-x, t) f_h^{\lambda'}(x, t) \right] = \nonumber\\ &=&
\frac{e^{(\gamma + \gamma')v^* t}}{2 d \beta}\frac{v_e}{v_h}\frac{v_e+v_h}{\lambda-\lambda'}\left[\bar\kappa' \left(\lambda \sinh \bar\kappa + \bar\kappa \cosh \bar\kappa \right) - e^{-\frac{v_e-v_h}{v_e+v_h}(\lambda-\lambda')} \bar\kappa \left(\lambda' \sinh \bar\kappa' + \bar\kappa' \cosh \bar\kappa' \right) \right].\nonumber
\eea
Since both $\lambda$ and $\lambda'$ are valid eigenvalues and satisfy Equation \ref{eigenvalue_equation}, we have
\bea
\lambda + \bar\kappa \coth \bar\kappa &=& 0,\nonumber\\
\lambda' + \bar\kappa' \coth \bar\kappa' &=& 0,\nonumber
\eea
and thus $\mathcal{I}(\lambda, \lambda') = 0$, as claimed.

\section{Analytic expressions for $f_1^{ee}$ and $f_1^{hh}$}
\label{Sec:f1_analytic}

Inserting the expansions given in Equation \ref{general_solution_ne_nh} into Equations \ref{f1ee_solution} and \ref{f1hh_solution} and performing the integration, the functions $f_1^{ee}$ and $f_1^{hh}$ can be expressed as
\beqs
f_1^{ee}(x, t) = \sum_{\lambda} C(\lambda) f_{\lambda}^{ee}(x, t),
\eeqs
and
\beqs
f_1^{hh}(x, t) = \sum_{\lambda} C(\lambda) f_{\lambda}^{hh}(x, t).
\eeqs

The functions $f_{\lambda}^{ee}$ and $f_{\lambda}^{hh}$ are given by
\beqs
f_{\lambda}^{ee}(x,t) = \frac{e^{\gamma v^* t} e^{a x}}{d (\alpha+\beta) - 2\lambda}
\begin{cases}
  2\bar\kappa \exp\left[x \frac{v^*}{2}\left(\frac{1}{v_e} + \frac{1}{v_h} \right)\left(\alpha - \frac{\lambda}{d}\right)\right] - 2\bar\kappa \cosh\kappa x - d(\alpha-\beta) \sinh\kappa x & t \geq \frac{x}{v_e}\\[1em]
  \begin{aligned}[b]
    \exp\left[-t\frac{v^*}{2}\left(1+\frac{v_e}{v_h}\right)\frac{\lambda}{d}\right]\left[-e^{t \frac{v_e\lambda}{d}}\left(2\bar\kappa\cosh\kappa x + d(\alpha-\beta)\sinh\kappa x \right) + \right. \\ \left. + e^{t v_e \alpha} \left(2\bar\kappa\cosh\left[\kappa(x - t v_e)\right] + d(\alpha-\beta)\sinh\left[\kappa(x-t v_e)\right] \right) \right]
  \end{aligned} & t < \frac{x}{v_e}, \nonumber
  \end{cases}
\eeqs
and
\beqs
f_{\lambda}^{hh}(x, t) = \frac{1}{\beta d}\frac{v_e}{v_h}\frac{e^{\gamma v^* t}e^{a x}}{d(\alpha+\beta)-2\lambda}
\begin{cases}
  \begin{aligned}[b]
    2 \alpha\beta d^2 e^{d\beta-\lambda} \sinh\bar\kappa \,\exp\left[-x \frac{v^*}{2}\left(\frac{1}{v_e}+\frac{1}{v_h}\right)\left(\beta - \frac{\lambda}{d}\right)\right] - \\
    - 2\alpha\beta d^2 \sinh\kappa x + (d(\alpha-\beta)+2\lambda)(\bar\kappa\cosh\kappa x + \lambda \sinh \kappa x)
  \end{aligned}  & t \geq \frac{d-x}{v_h} \\[1em]
  \begin{aligned}[b]
    (d(\alpha-\beta)+2\lambda)(\bar\kappa\cosh\kappa x+\lambda\sinh\kappa x) - 2\alpha\beta d^2 \sinh\kappa x + \\
    + \exp\left[t \frac{v^*}{2}\left(1+\frac{v_h}{v_e}\right)\left(\beta-\frac{\lambda}{d}\right)\right] \left[2 \alpha\beta d^2 \sinh\left[\kappa(x+t v_h)\right] - \right.\\
    \left. -(d(\alpha-\beta)+2\lambda)(\bar\kappa\cosh\left[\kappa(x+t v_h)\right] + \lambda\sinh\left[\kappa(x+t v_h)\right]) \right]
  \end{aligned} & t < \frac{d-x}{v_h}. \nonumber
  \end{cases}
\eeqs
  
\end{appendices}

\FloatBarrier

\end{document}